\documentclass[submission, Phys,amsmath]{SciPost}

\usepackage{mathrsfs} \usepackage{amssymb} \usepackage{graphicx,color} \usepackage{bbm} \usepackage{multirow}\usepackage{ulem}
\usepackage{bbm} \usepackage{mathrsfs} \usepackage{commath}\usepackage{tikz}
\usetikzlibrary{shapes,arrows,positioning,calc}\usepackage{stmaryrd}\usepackage{bm}
\usepackage{mhchem,chemformula}
\usepackage{url}

\hypersetup{
  colorlinks=true,
  linkcolor=blue,
  filecolor=magenta,
  citecolor=blue,
  urlcolor=blue,
}

\def\vS{\Vec{S}}

\newcommand{\eq}[1]{Eq.~(\ref{#1})}
\newcommand{\beq}{\begin{equation}} \newcommand{\eeq}{\end{equation}}
\newcommand{\beqn}{\begin{eqnarray}} \newcommand{\eeqn}{\end{eqnarray}}
\newcommand{\bmat}{\begin{mathdisplay}} \newcommand{\emat}{\end{mathdisplay}}
 
\newcommand{\Tr}{\mbox{Tr}}
\newcommand{\cD}{{\cal D}}

\newcommand{\red}[1]{\textcolor{black}{#1}}
\renewcommand{\Vec}[1]{{\bf #1}}

\newcommand{\inti}{\int_{-\infty}^{\infty}}

\newcommand{\bs}{\blacksquare}

\newcommand{\bc}{\begin{center}}
\newcommand{\ec}{\end{center}}

\begin{document}

\begin{center}{\Large \textbf{
      From complex to simple : hierarchical
      free-energy landscape renormalized in deep neural networks
}}\end{center}

\begin{center}
Hajime Yoshino\textsuperscript{1,2*}
\end{center}

\begin{center}
{\bf 1} Cybermedia Center, Osaka University, Toyonaka, Osaka 560-0043, Japan
\\
{\bf 2} Graduate School of Science, Osaka University, Toyonaka, Osaka 560-0043, Japan
\\
* yoshino@cmc.osaka-u.ac.jp
\end{center}

\begin{center}
\today
\end{center}

\section*{Abstract}
         {\bf
           We develop a statistical mechanical approach based on the replica method to study the design space of deep and wide neural networks constrained to meet a large number of training data. Specifically, we analyze the configuration space of the synaptic weights and neurons in the hidden layers in a simple feed-forward perceptron network for two scenarios:  a setting with random inputs/outputs and a teacher-student setting. By increasing the strength of constraints,~i.e. increasing the number of training data, successive 2nd order glass transition (random inputs/outputs) or 2nd order crystalline transition (teacher-student setting) take place layer-by-layer starting next to the inputs/outputs boundaries going deeper into the bulk with the thickness of the solid phase growing logarithmically with the data size.
           This implies the typical storage capacity of the network grows exponentially fast with the depth.
           In a deep enough network, the central part remains in the liquid phase.
           We argue that in systems of finite width N, the weak bias field can remain in the center and plays the role of a symmetry-breaking field that connects the opposite sides of the system. The successive glass transitions bring about a hierarchical free-energy landscape with ultrametricity, which evolves in space: it is most complex close to the boundaries but becomes renormalized into progressively simpler ones in deeper layers. These observations provide clues to understand why deep neural networks operate efficiently. Finally, we present some numerical simulations of learning which reveal spatially heterogeneous glassy dynamics truncated by a finite width $N$ effect.
}

\vspace{10pt}
\noindent\rule{\textwidth}{1pt}
\tableofcontents\thispagestyle{fancy}
\noindent\rule{\textwidth}{1pt}
\vspace{10pt}

\section{Introduction}

Machine learning by deep neural networks (DNN) is successful in numerous applications \cite{lecun2015deep}. However, it remains challenging to understand why DNNs actually work so well. Given the enormous parameter space, which is typically orders of magnitude larger than that of the data space, and the flexibility of non-linear functions used in DNNs, it is not very surprising that they can express complex data  \cite{cybenko1989approximation}. What is surprising is that such extreme machines can be put under control. On one hand, one would naturally fear that learning such a huge number of parameters would be extremely time-consuming because the fitness landscape is presumably quite complex with many local traps.  Moreover, over-fitting or poor generalization ability seems unavoidable in such over-parametrized machines. We would not dare to fit a data set of $10$ points by a $100$ the order polynomial, which does not make sense usually. Quite unexpectedly, these issues seem to be somehow resolved in practice and such extreme machines turned out to be very useful. Thus it is a very interesting scientific problem to uncover what is going on in DNNs \cite{zhang2016understanding,carleo2019machine}. This is also important in practice because we wish to use DNNs not merely as mysterious black boxes but control/design them in rational ways.  In the present paper we develop a statistical mechanical approach based on the replica method to obtain some insights into these issues.

In this paper, we investigate a class of simple machines made of feed-forward networks of layered perceptrons whose depth is $L$ and the width is $N$ (see Fig.~\ref{fig_multilayer_network}). Such a machine is parametrized by a configuration of synaptic weights in the hidden layers. \red{We consider the coupling between adjacent layers are {\it global}
in the sense that all neurons in the $l$-th layer are connected to all neurons in $l+1$-th layer.}
For a given pair of inputs/outputs patterns imposed on the input and output layers, there can be different realizations of the synaptic weights that match the same constraints. We call each of them as a 'solution'. Following the work of Gardner \cite{gardner1988space,gardner1989three} for the single perceptron, we consider statistical mechanics of the design space of the neural network which is compatible with a large number $M=\alpha N$ patterns of training data, in the large width  $N \to \infty$ limit with fixed $\alpha$.
For the choice of the training data, we consider two simple scenarios: 1)  pairs of purely random inputs/outputs patterns 2) teacher-student setting - pairs of random input and the corresponding output of a teacher machine with random synaptic weights are handed over to a student machine.

From a broader perspective, the setting 1) can be viewed as a random constraint satisfaction problem (CSP) \cite{mezard2009information,zdeborova2016statistical},
which is deeply related to the physics of glass transitions and jamming \cite{berthier2011theoretical,charbonneau2014fractal,parisi2020theory}. 
In the context of neural networks, it is a standard setting to study the storage capacity \cite{gardner1988space,engel2001statistical}.
If $\alpha$ is small so that the constraint is weak enough, it is natural to expect
that the phase space looks like that of a liquid: 
there are so many realizations of machines compatible with a given set of constraints that essentially all solutions are continuously connected.
Increasing $\alpha$ the system becomes more constrained so that
the volume of the solution space shrinks and ultimately
vanishes at some critical value $\alpha_{\rm j}$.
This is an SAT/UNSAT (jamming) transition and $\alpha_{\rm j}$ defines the storage capacity.
Interestingly, before reaching $\alpha_{\rm j}$, the solution space can become clustered into mutually disconnected islands.
This is a glass transition and it accompanies some type of replica symmetry breaking (RSB) \cite{parisi1979infinite, MPV87}.
Recently non-trivial glass transitions accompanying continuous replica symmetry breaking, which imply the emergence of
hierarchical free-energy landscape and ultrametricity\cite{mezard1984nature,MPV87,nemoto1987numerical,nemoto1988metastable},
and common jamming critically as that of the hard-spheres \cite{charbonneau2014fractal}
were found in a family of CSPs including a single perceptron problem
\cite{franz2016simplest,franz2017universality} and a family of vectorial spin models \cite{yoshino2018}.
Understanding the nature of such glass transitions and jamming is a fundamental problem in CSPs since it is intimately related to the efficiency of
algorithms to solve CSPs. In the context of DNN, it is certainly important to understand the characteristics of the free-energy landscape to understand the efficiency of various learning algorithms for DNNs \cite{baity2019comparing,franz2019jamming,geiger2019jamming}.

On the other hand the setting 2), is a statistical inference (SI) problem.
While the constraint satisfaction problems are related to the physics of glass transitions and jamming,
solving a statistical inference problem can be said to be equivalent to searching
of a hidden (planted) crystalline state \cite{zdeborova2016statistical}.
In the context of neural networks, it is a standard setting to study {\it learning} \cite{gardner1989three,engel2001statistical}.
As $\alpha$ becomes sufficiently large, the synaptic weights of the
student machine starts to become closer to those of the teacher machine.
If this happens, the student machine starts to generalize: the probability that the student machine yields the same output as the teacher machine
for a test data (not used during training), increases with $\alpha$.
Although very simple, this setting will provide useful insights into the generalization ability of DNNs.

The present work is following the standard statistical mechanical approach to machine learning \cite{engel2001statistical}.
Extension of it to deeper neural networks has remained challenging. 
Our key strategy is to regard a DNN, not as a system of
long-ranged interaction between the input and output through a highly convoluted non-linear mappings
but rather as a system with short-ranged interactions between adjacent layers.
This is enabled by the {\it internal representation} \cite{monasson1995weight},
in which one takes into account not only 'bonds' (synaptic weights)
but also 'spins' (neurons) in the hidden layers 
as dynamical variables which are constrained to satisfy proper inputs/outputs
relations at each perceptron embedded in the hidden layers.
Representing the states of a neuron associated with $M$-patterns
as $M$-component vectorial spins, the system can be represented as a network of dynamical variables
with a large number of components with dense connections to each other.

The system is almost disorder-free in the sense that the 'quenched disorder' is present only on the boundaries
so that one would fear that the usual replica theory for the single perceptron \cite{gardner1988space} cannot be easily extended for DNNs.
However, the replica approach is not merely a trick to take the average over the quenched disorder. 
The recent progress on the exact replica theory in disorder-free systems like simple hard-spheres 
\cite{kurchan2012exact,kurchan2013exact,charbonneau2014exact,charbonneau2014fractal,parisi2020theory}  in the large dimensional limit
and disorder-free glassy spin systems
\cite{yoshino2018} have proved that spontaneous replica symmetry breaking (RSB) exist in such systems which become manifest by considering infinitesimal symmetry breaking field explicitly
as pointed out by Parisi and Virasoro \cite{parisi1989mechanism}.
One can even study spontaneous glass transitions of multiple degrees of freedoms
such as translational and orientational degrees of freedom in aspherical particulate systems by the same approach\cite{yoshino2018translational}.

The above observations motivate us to investigate the interior of the DNN
through local glass or crystalline order parameters, for both the spins (neurons) and bonds (synaptic weights), which are allowed to vary over the space.
We formulate a replica theory to analyze the design space of the deep perceptron network
analyzing a free-energy expressed as a functional of the space-dependent, local order parameters.
\red{For simplicity, we limit ourselves within a tree-approximation which neglects effects
  of interaction loops along the $z$-axis. Thus our theory is inevitably a mean-field approximation of the original problem,
  which does not faithfully take into account 1 dimensional fluctuation along the $z$-axis (See Fig.~\ref{fig_multilayer_network}).
  Nevertheless we believe our theory captures important aspects of the DNN.
  In a sense the present work may be regarded as a Ginzburg-Landau type theory for the DNN.
  Consideration of loop-corrections would improve the quantitative accuracy of the microscopic mean-field theory but we leave it for future studies.
}

The main result of the present paper is that the solutions 
of the glass/crystalline order parameters of over-parametrized DNNs
become quite heterogeneous in space (along $z$-axis).
In both settings 1) and 2), the amplitude of the order parameters
close to the inputs/outputs boundaries
become finite and take higher values as the strength of the constraint $\alpha$ increases
while the amplitudes decay down to $0$ going deeper into the bulk.
Moreover, in the case of setting 1) random inputs/outputs,
even the pattern of the replica symmetry breaking
(RSB) varies in space: it is most complex close to the boundaries
with $k$(+continuous)-RSB, which becomes $k-1$(+continuous)-RSB
in the next layer, ... down to a replica symmetric ($0$ RSB) state in the central part.
The thickness $\xi$ of the region around the boundaries
where the glass/crystalline order parameters
become finite roughly scales as $\xi \propto \ln \alpha$.
This implies the storage capacity of the network $\alpha_{\rm j}(L)$ for {\it typical} instances
grows exponentially fast with the depth $L$,
while the  worst case scenarios \cite{vapnik2015uniform} would predict linear growth with $L$.

Thus if the network is deep enough $L > \xi$, the central part of a typical network
remains in the liquid phase: there are so many possibilities left in the central
part all of which meet the same constraints imposed at the boundaries.
The heterogeneous profile of the order parameters should have important implications on how DNNs work.

The organization of the paper is the following.
In sec.~\ref{sec-model} we define the deep neural network model studied in the present paper
and explain the two scenarios : 1) random inputs/outputs and 2) teacher-student setting.
In sec.~\ref{sec-replica} we formulate a replica theory to perform statistical mechanical analyses of the design space of
the deep neural network \red{within a tree-approximation}.  In sec.~\ref{sec-replica-random-inputs-outputs} and sec.~\ref{sec-replica-teacher-student}
we study the cases of 1) random inputs/outputs and 2) teacher-student settings respectively using the replica theory.
In sec.~\ref{sec-simulation}, we present some results of numerical simulations to examine the theoretical predictions.
Finally in sec.~\ref{sec-conclusions}, we conclude the paper and present some outlook.
In the appendices \ref{appendix-replicated-free-energy}, \ref{appendix-rsb-random-inputs-outputs} and
\ref{appendix-rsb-teacher-student} we present some details of the theoretical formulation.

\section{Model}
\label{sec-model}

\subsection{Multi-layer feed-forward network}
\label{subsec-multilayer-feed-forward-network}

\begin{figure}[h]
  \bc
  \includegraphics[width=\textwidth]{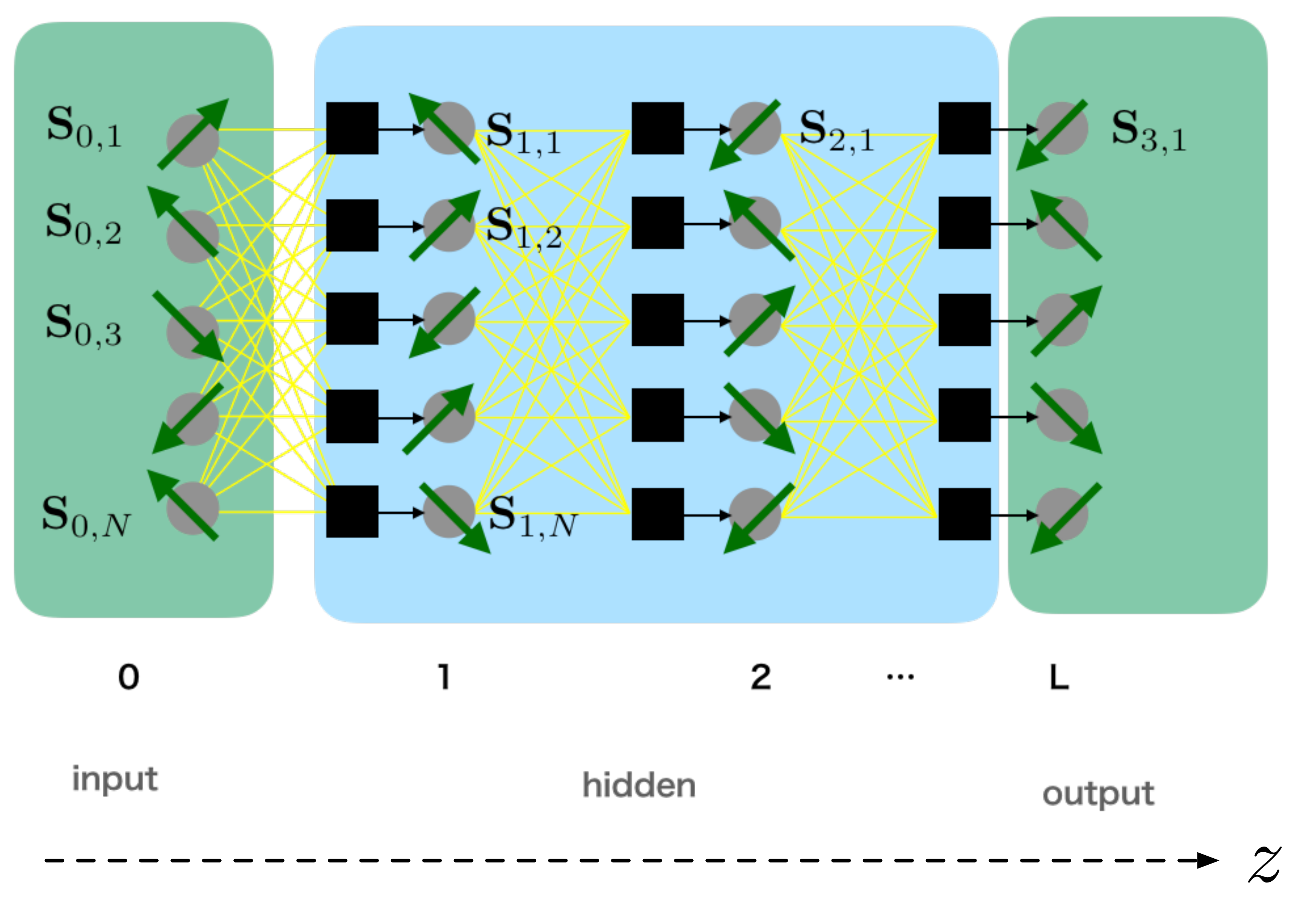}
  \ec
  \caption{ A simple multi-layer perceptron network
    of depth $L$ and width $N$. In this example
    the depth is $L=3$.
    Each arrow represents a $M$-component vector spin ${\bf S}_{i}=(S_{i}^{1},S_{i}^{2},\ldots,S_{i}^{M})$
    with its component $S_{i}^{\mu}=\pm 1$ representing the state of a 'neuron' in the $\mu$-th pattern.
  }
   \label{fig_multilayer_network}
 \end{figure}

 We consider a simple multi-layer neural network (See Fig.~\ref{fig_multilayer_network})
 which consists of an input layer ($l=0$), output layer
 ($l=L$) and hidden layers ($l=1,2,\ldots,L-1$). Each layer consists of $i=1,2,\ldots,N$ neurons
 ${\bf S}_{l,i}$, each of which consists of $M$-component Ising spins ${\bf S}_{l,i}=(S^{1}_{l,i},S^{2}_{l,i},\ldots,S^{M}_{l,i})$ with $S^{\mu}_{l,i}=\pm 1$.
 Here the label $\mu=1,2,\ldots,M$ is used to distinguish
 different firing patterns of the neurons (spins).
The spins in the inputs/outputs layers represent 'data' provided by
external sources.
We follow the notation of \cite{yoshino2018} to represent
a factor node, which is a perceptron here, as $\bs$.
We consider a feed-forward network of $N_{\bs}=NL$ perceptrons.
A perceptron $\bs$ receives $N$ inputs from the outputs of perceptrons $\bs(k)$ $(k=1,2,\ldots,N)$
in the previous layer, weighted by ${\bf J}_{\bs}=(J_{\bs}^{1},J_{\bs}^{2},\ldots,J_{\bs}^{N})$. Its output
${\bf S}_{\bs}$ is given by,
 \beq
 S^{\mu}_{\bs}={\rm sgn} \left(\frac{1}{\sqrt{N}}\sum_{k=1}^{N} J_{\bs}^{k}S^{\mu}_{\bs(k)}\right) \qquad \mu=1,2,\ldots,M.
 \label{eq-perceptron}
\eeq
We assume that the synaptic weights $J_{\bs}^{k}$ take real numbers normalized such that,
\beq
\sum_{k=1}^{N}(J_{\bs}^{k})^{2}=N.
\label{eq-J-normalization}
\eeq 
For simplicity, we call the variable for
the neurons ${\bf S}_{l,i}$'s  as 'spins', 
and  the synaptic weights $J_{\bs}^{k}$s
as 'bonds' in the present paper.

We will consider random input data of size
 \beq
 N_{\rm data}=NM=N^{2} \alpha
 \label{eq-number-of-data}
 \eeq
while the number of parameters is
\beq
N_{\rm parameter}=N_{\bs} N =N^{2}L.
 \label{eq-number-of-parameters}
 \eeq
Here we introduced a parameter
 \beq
 \alpha\equiv
 \frac{M}{N}.
  \label{eq-alpha}
  \eeq
  The task of {\it learning} is to design the synaptic weights
  $J_{\bs}^{k}$ to build a mapping (function) between the imposed
  random input data and output data, which can be completely
  different, by a network of width $N$ and depth $L$.

   We will consider the limit $N,M \to \infty$  with fixed $\alpha$.
 This scaling is known for
 the single perceptron \cite{gardner1988space}
 and we will find that it continues to be the key parameter for
 the bigger network much like the inverse temperature for condensed matters.
 Apparently the system is over parametrized $N_{\rm parameter} > N_{\rm input}$
 if it is deep enough $L > \alpha$. ( Actually our results imply that typical storage capacity
 grows exponentially with the depth $L$ as we see later so that
 the system is essentially over-parametrized if $L > \ln \alpha$.
 ( see sec. \ref{section_more_glass_transitions}))
  \red{We note that there might be other possible scalings different from
    \eq{eq-alpha}. For example studies on some types of
    two-layer perceptron networks
    suggest other scaling such as $M=\alpha N^{2}$ is also possible (see Chap 12 of \cite{engel2001statistical}). However, in the present paper we will limit ourselves to the scaling of the form \eq{eq-alpha}.}

The trajectories of such highly non-linear mapping as \eq{eq-perceptron} along  the random deep network is known to be highly chaotic \cite{sompolinsky1988chaos,poole2016exponential}: small differences in the input data lead to rapid decorrelation of the resulting trajectories. This feature is considered as responsible for the high expressive power of DNNs \cite{poole2016exponential}. Similarly, small changes made on the weights $J_{\bs}^{k}$ also leads to chaotic decorrelation of trajectories \cite{okazaki-yoshino}. But then we immediately face the obvious question: how the high generalization ability observed in DNNs can be explained when the system is so chaotic?
In the present paper we construct a statistical mechanics point of view to answer such questions.
Out of the set of all possible realizations of random deep networks, which typically give chaotic dynamics,
we focus on a substantially smaller
sub-manifold of it in which all trajectories (accidentally) meet the externally imposed boundary conditions put at the two opposite ends.
This {\it selection} (learning) may have significant consequences on the properties of the resultant ensemble.

Following the pioneering work by Gardner \cite{gardner1988space,gardner1989three} we consider the volume of the design space
of the system associated with a given set of inputs/outputs patterns represented by ${\bf S}_{0}$ and ${\bf S}_{L}$, which  can be expressed as,
\beqn
 V\left({\bf S}_{0},{\bf S}_{L}\right)&=& e^{N M {\cal S}\left({\bf S}_{0},{\bf S}_{l}\right)}
 =\left(\prod_{\bs} {\rm Tr}_{{\bf J}_{\bs}}\right)
\left( \prod_{\bs\backslash {\rm output}}{\rm Tr}_{{\bf S}_{\bs}}  \right)
\prod_{\mu=1}^{M}
\prod_{\bs}e^{-\beta V(r^{\mu}_{\bs})}
\label{eq-gardner-volume-DNN}
\qquad
\eeqn
where
\beq
e^{-\beta V(r)}=\theta(r)
\label{eq-hardcore}
\eeq
and we introduced the 'gap',
\beq
r^{\mu}_{\bs} \equiv
S^{\mu}_{\bs}
\sum_{i=1}^{N}\frac{J_{\bs}^{i}}{\sqrt{N}}S^{\mu}_{\bs(i)}
\label{eq-gap}
\eeq
The trace over the spin and bond configurations can be written explicitely as,
\beq
    {\rm Tr}_{\bf S} =\prod_{\mu=1}^{M}\sum_{S^{\mu}=\pm 1}
    \label{eq-spin-trace}
    \eeq
    and
    \beq
        {\rm Tr}_{\bf J} =\int_{-\infty}^{\infty}
\prod_{j=1}^{N}
dJ^{j}
\delta\left(\sum_{k=1}^{N}(J^{k})^{2}-N \right)
=N \int_{-i \infty}^{i \infty} \frac{d \lambda}{2\pi}e^{N\lambda}
\prod_{j=1}^{N} \int_{-\infty}^{\infty}dJ^{j}e^{-\lambda (J^{j})^{2}}
    \label{eq-bond-trace}
\eeq

Note that in \eq{eq-gardner-volume-DNN} summations are took
not only over the bonds (synaptic weights) but also over the
spins (neurons) in the hidden layers. This is the internal
representation \cite{monasson1995weight} which allows us
to avoid viewing the system as a system with
long-ranged interaction between the input and output through a highly convoluted non-linear mappings but rather as a system with short-ranged interactions
between adjacent layers.
From a physicist's point of view, this is far more convenient.
Indeed we can now write the effective Hamiltonian of the system
as
\beq
    {\cal H}_{\rm eff}=\sum_{\mu=1}^{M}\sum_{\bs}
    V\left(
S^{\mu}_{\bs}
\sum_{i=1}^{N}\frac{J_{\bs}^{i}}{\sqrt{N}}S^{\mu}_{\bs(i)}\right)
\label{eq-effective-hamiltonian}
\eeq
This simple  trick works because of the simple '${\rm sgn}$' activation function \eq{eq-perceptron} we consider in the present paper.
Let us emphasize here that both the spins $S_{\bs s}^{\mu}$
and bonds $J_{\bs}^{i}$ are dynamical variables,
except for the spins on the boundaries $l=0,L$ which are frozen.

Now our task is to analyze the equilibrium statistical mechanics of the system of many variables with the effective Hamiltonian \eq{eq-effective-hamiltonian}. In this point of view, we can forget about the 'feed-forwardness' of the original dynamical representation \eq{eq-perceptron}.
\red{If we imagine an inifinitely deep network without terminals or a network with
  the periodic boundary condition, one can regard the system
  as a globably homogneous $1(+\infty)$ dimensional system.
  With the boundaries, some inhomogeneity should emerge close to the boundaries.
}

The problem at our hands is similar to the statistical mechanics of an assembly of hard-spheres.
Each of the configurations which meet the hard-core constraint \eq{eq-hardcore} represents a
valid trajectory (more precisely a set of $M$ perceptron trajectories all of which meet the  corresponding inputs/outputs boundary conditions)
of the original feed-forward problem. Like in the statistical mechanics of hard-spheres \cite{parisi2020theory}, everything  that matters here is the entropy effect. For instance, we can expect that assembly of trajectories which consists of many nearby valid trajectories (which
meet the same inputs/outputs boundary conditions) have richer (local) entropy
so that they make important contributions to the total entropy.
This corresponds to the notion of 'free-volume' of an assembly of hard-spheres \cite{sastry1998free}.
Such an equilibrium statistical mechanics may not be merely academic.
Indeed the standard schemes of deep learning involve Stochastic Gradient
Descent (SGD) algorithms \cite{lecun2015deep,2018arXiv181000004Y}
which explores the solution space of DNNs in a stochastic way.
There also trajectories with richer local entropy would appear more often during the sampling.
In this paper we will find often that the analogy with the
physics of hard-sphere glass \cite{parisi2020theory}
is very useful to understand our results in physical terms.

\subsection{Two scenarios for inputs/outputs patterns}
\label{subsec-two-scenarios}

For the input and output patterns ${\bf S}_{0}$ and ${\bf S}_{L}$,
we consider the following two scenarios.

\subsubsection{Random inputs/outputs}
\label{subsubsect-random-inputs-outputs}

As the simplest setting, we consider the case of completely random inputs/outputs patterns,
which is the standard setting to study the storage capacity of the perceptrons\cite{gardner1988space,engel2001statistical}.
More precisely all  components of
${\bf S}_{0,i}=(S^{1}_{0,i},S^{2}_{0,i},\ldots,S^{M}_{0,i})$ and
${\bf S}_{L,i}=(S^{1}_{L,i},S^{2}_{L,i},\ldots,S^{M}_{L,i})$
for $i=1,2,\ldots,N$ 
are assumed to be iid random variables which take Ising values $\pm 1$.
As we noted in the introduction, this setting can be regarded as a random constraint
satisfaction problem (CSP).

\subsubsection{Teacher-student setting}
\label{subsubsect-teacher-student}

As a complementary approach, we consider the teacher-student setting,
which is a standard setting to study statistical inference problems \cite{zdeborova2016statistical}. We consider two machines : a teacher machine and student machine and assume that they have exactly the same architecture, i.e. the same width $N$ and the depth $L$.

We assume that the teacher is a 'quenched-random teacher': the set of the synaptic weights
$\{(J_{\bs}^{k})_{\rm teacher}\}$  of the teacher machine are iid random variables
which obey the normalization \eq{eq-J-normalization}.
Such a teacher machine is subjected to a set of random inputs, which are iid random variables,
${\bf S}_{0,i}=(S^{1}_{0,i},S^{2}_{0,i},\ldots,S^{M}_{0,i})$
for ($i=1,2,\ldots,N)$ and produces the corresponding set of outputs,
\beq
({\bf S}_{L,i})_{\rm teacher}=((S^{1}_{L,i})_{\rm teacher},(S^{2}_{L,i})_{\rm teacher},\ldots,(S^{M}_{L,i})_{\rm teacher})
\eeq
The task of the student machine is to try to
infer the synaptic weights $\{(J_{\bs}^{k})_{\rm teacher}\}$
of the teacher machine, by adjusting its own
synaptic weights $\{(J_{\bs}^{k})_{\rm student}\}$ such that
it successfully reproduces all the outputs of the teacher
$({\bf S}_{L,i})_{\rm teacher}$ starting from the same input data of the teacher.

Note that the student is given the full information of the input
${\bf S}_{0,i}$ and the output of the teacher $({\bf S}_{L,i})_{\rm teacher}$ plus
full information on the architecture of the teacher.
In the context of statistical inference,
this is an idealized situation called as Bayes optimal case
\cite{zdeborova2016statistical} and we limit ourselves to this
in the present paper for simplicity.

\section{Replica theory}
\label{sec-replica}

Now let us formulate a replica approach to study the solution space of the deep neural network.
To study the case of random inputs/outputs (sec. \ref{subsubsect-random-inputs-outputs})
we consider $n$ replicas $a=1,2,\ldots,n$ which are independent machines subjected to the common set of
inputs ${\bf S}_{0,i}$ and outputs ${\bf S}_{L,i}$ for $i=1,2,\ldots,N$.
For the case of the teacher-student setting (sec. \ref{subsubsect-teacher-student})
we consider $n=1+s$ replicas, with the replica $a=0$ to represent the teacher machine
and other replicas $a=1,2,\ldots,s$ to represent the replicas of the student.


\subsection{Order parameters}
\label{sec-order-parameters}

For the setting with random inputs/outputs, which is a constraint satisfaction problem, we anticipate that that the solution space exhibits clustering (glass transition) as we noted in the introduction.
Thus it is natural to consider order parameters that detect the glass transitions.
Given the dense connections of the network, we naturally introduce 'local' glass order parameters (see \cite{yoshino2018}),
\beq
Q_{ab,\bs}=\frac{1}{N}\sum_{i=1}^{N}(J_{\bs}^{i})^{a}(J_{\bs}^{i})^{b}
\qquad q_{ab,\bs}=\frac{1}{M}\sum_{\mu=1}^{M}(S_{\bs}^{\mu})^{a}(S_{\bs}^{\mu})^{b}
\label{eq-glass-order-parameters}
\eeq
Note that the normalization condition for the bonds \eq{eq-J-normalization} and the spins (which take Ising values $\pm 1$)
implies $Q_{aa,\bs}=q_{aa,\bs}=1$.

For the teacher-student setting, we continue to use the above order parameters
for $a=0,1,2,\ldots,s$ replicas where $0$-th replica is for the teacher machine.
Thus $Q_{0a}=Q_{a0}$ and $q_{0a}=q_{a0}$ for $a=1,2,\ldots,s$
represent the overlap between the teacher machine and student machines.

There are two comments regarding some trivial symmetries left in the system. First, the system is symmetric under permutations of the labels put on the data $\mu=1,2,\ldots,M$. The labels put on different replicas could be permuted differently. In the 2nd equation of \eq{eq-glass-order-parameters} it is assumed that all replicas follow the same labels breaking this permutation symmetry. Second, the system is symmetric under permutations of perceptrons $\bs$ within the same layer and the permutations could be done differently on different replicas. In \eq{eq-glass-order-parameters}, this permutation symmetry is also broken. Note that solutions with other permutations regarding the two symmetries mentioned above give exactly the same free-energy so that one choice is enough.

\subsection{Replicated Gardner volume}

The Gardner's volume \eq{eq-gardner-volume-DNN} fluctuates depending on the
realizations of the boundaries ${\bf S}_{0}$ and ${\bf S}_{L}$.
In the present paper we wish to analyze the {\it typical} behavior for
stochastic realizations of the boundaries.
To this end we consider the replicated phase space volume (the Gardner volume),
\beqn
 V^{n}\left({\bf S}_{0},{\bf S}_{L}\right)
&=&e^{N M {\cal S}_{n}\left({\bf S}_{0},{\bf S}_{l}\right)} \nonumber \\
& = &\prod_{a=1}^{n}
\left(\prod_{\bs} {\rm Tr}_{{\bf J}^{a}_{\bs}}\right)
\left( \prod_{\bs\backslash {\rm output}}{\rm Tr}_{{\bf S}^{a}_{\bs}}  \right)
\left\{ \prod_{\mu,\bs,a}
e^{-\beta V(r_{\bs,a}^{\mu})}
\right \}
\label{eq-replicated-gardner-volume}
\eeqn
with
\beq
r^{\mu}_{\bs,a} \equiv
(S^{\mu}_{\bs})^{a}
\sum_{i=1}^{N}\frac{(J_{\bs}^{i})^{a}}{\sqrt{N}}(S^{\mu}_{\bs(i)})^{a}
\label{eq-gap-replica}
\eeq
The {\it typical} behavior can be studied
by considering the $n \to 0$ limit \cite{parisi2008large}, i.~.e.
$\left. \partial_{n} \overline{V^{n}({\bf S}_{0},{\bf S}_{L})}^{{\bf S}(0),{\bf S}(L)} \right |_{n=0}$
where the overline represents the average over the different realizations of the boundaries
(see below for the details.)

\begin{figure}[h]
  \bc
  \includegraphics[width=0.5\textwidth]{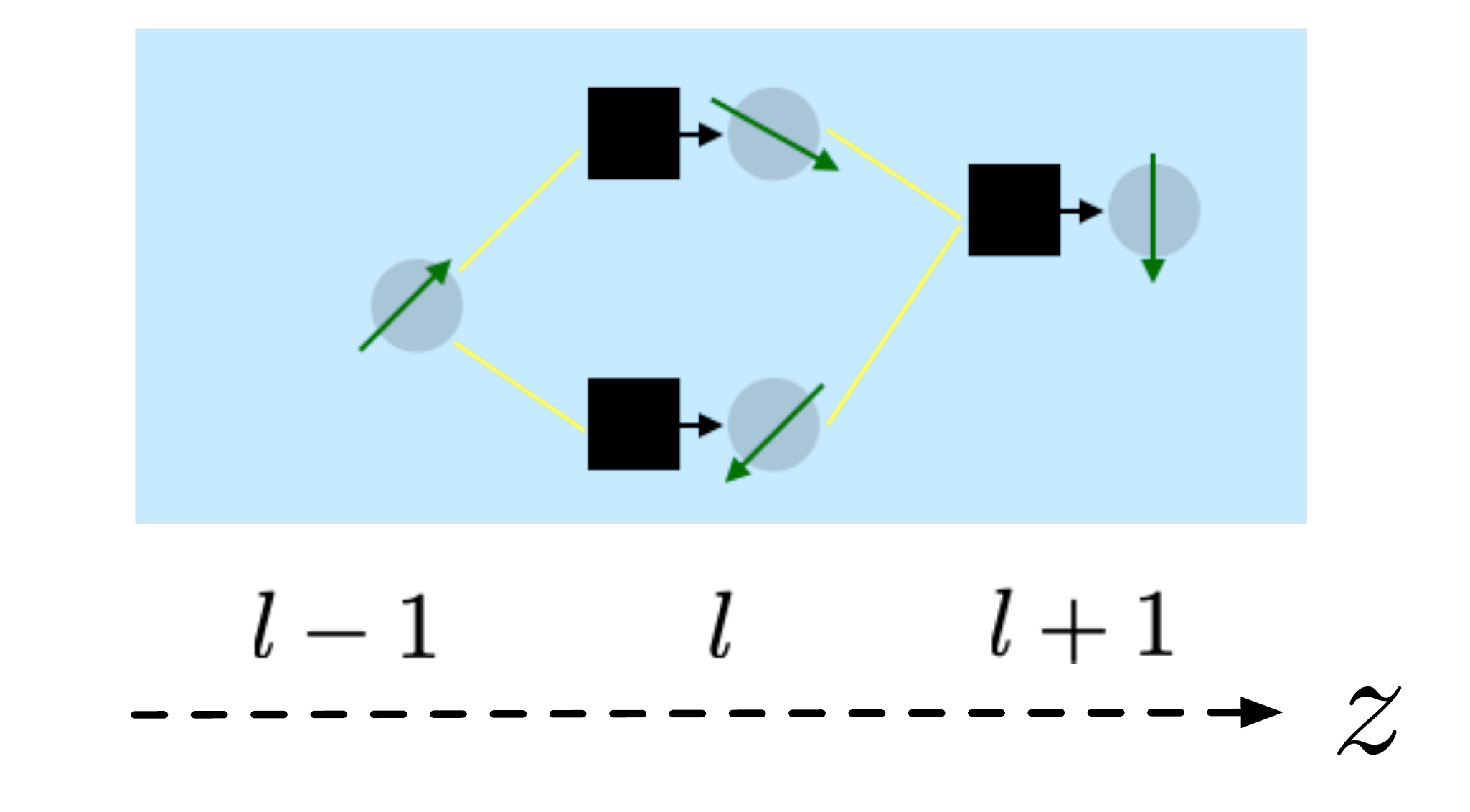}
  \ec
  \caption{ \red{A loop of interactions in a DNN
      extended over 3 layers, through 3 perceptrons
      and 4 bonds.
       We neglect effects
    of such loops (and more extended ones) in our theory.}
  }
   \label{fig_loop}
\end{figure}

As shown in appendix \ref{appendix-replicated-free-energy},
following similar steps as in \cite{yoshino2018},
we obtain the replicated free-entropy functional
 $s_{n}(\{Q_{\bs},q_{\bs}\})$ in terms of the order parameters $Q_{\bs}$ and $q_{\bs}$ defined in \eq{eq-glass-order-parameters}
in the limit $N,M \to \infty$ with fixed $\alpha=M/N$.
\red{For simplicity, we limit ourselves to
  a tree-approximation which neglects the effects of interaction-loops
  along the $z$-axis such as the one shown in Fig.~\ref{fig_loop}.
  The tree approximation has two essential problems: 1)
  it cannot describe faithfully 1-dimensional fluctuations
  along $z$-axis 2) it misses microscopic details 
  close to the boundaries
  where we naturally expect inhomogeneities.
  Especially it fails to capture the difference of
  the two opposite boundaries. In principle, this accidental symmetry
  can be removed taking into account
  loop-corrections. Indeed the loop
  shown in Fig.~\ref{fig_loop}
  is not symmetric with respect to the interchange
  of the left and right hand sides.
}

Given the structure of the network (see Fig.~\ref{fig_multilayer_network}), it is natural to assume that order parameters are uniform
within each layer $l=0,1,2,\ldots,L$,
\beq
Q_{ab,\bs}=Q_{ab}(l) \qquad q_{ab,\bs}=q_{ab}(l),
\label{eq-layerwise-orderparameter}
\eeq

To represent the quenched boundaries, we impose the boundary conditions on the inputs/outputs layers by simply putting
$q_{ab}(0)=q_{ab}(L)=1$ (see below).

The above general formulation can be adapted for the two scenarios introduced in sec.~\ref{subsec-two-scenarios}
as follows,
\begin{itemize}
\item Random inputs/outputs
  
  In the case of random inputs/outputs (sec. \ref{subsubsect-random-inputs-outputs}) we consider the free-energy functional,
\beq
\frac{-\beta F[\{{\hat Q}(l),{\hat q}(l)\}]}{NM}=\frac{\left. \partial_{n} \overline{V^{n}({\bf S}_{0},{\bf S}_{L})}^{{\bf S}_{0},{\bf S}_{L}}\right |_{n=0}}{NM}=
\left.  \partial_{n}s_{n}[\{{\hat Q}(l),{\hat q}(l)\}]\right |_{n=0}.
\label{eq-F-random-inputs-outputs}
\eeq
The presence of the imposed random inputs/outputs can be specified by providing values of $q_{ab}(0)$ and $q_{ab}(L)$.
Since all replicas are subjected to the same inputs and outputs, we can simply set,
\beq
q_{ab}(0)=q_{ab}(L)=1.
\label{eq-boundary-qab}
\eeq
As we discuss later we will also consider the case of fluctuating boundary conditions.

\item Teacher-student setting

  In the case of the teacher-student setting (sec. \ref{subsubsect-teacher-student}) we consider instead the so called Franz-Parisi potential \cite{FP95},
 \beqn
 \frac{-\beta F_{\rm teacher-student}[\{{\hat Q}(l),{\hat q}(l)\}]}{NM}
 && =\frac{\left. \partial_{s} \overline{V^{1+s}({\bf S}_{0},{\bf S}_{L}({\bf S}_{0},{\cal J}_{\rm teacher})))}^{{\bf S}_{0},{\cal J}_{\rm teacher}}\right |_{s=0}}{NM}\nonumber \\
&&  =
  \left.  \partial_{s}s_{1+s}[\{{\hat Q}(l),{\hat q}(l)\}]\right |_{s=0}.
  \label{eq-F-teacher-student}
\eeqn
where the over-line denotes the average over the imposed random inputs imposed commonly on both the teacher and student machines.
The outputs are just those of the teacher machine $a=0$, ${\bf S}_{L}({\bf S}_{0},{\cal J}_{\rm teacher})$
which  are of course functions of the inputs ${\bf S}_{0}$ and the synaptic weights of the
teacher machine ${\cal J}_{\rm teacher}=\{(J_{\bs}^{k})_{\rm teacher}\}$.
Since both the teacher and student machines are subjected to the same inputs, we set,
\beq
q_{ab}(0)=1
\eeq
for $a,b=0,1,\ldots,s$.
In addition, since the outputs of the student machine are forced to agree perfectly with that of the teacher machine we set,
\beq
q_{ab}(L)=1
\eeq
for $a,b=0,1,\ldots,s$.

\end{itemize}

\subsection{Random inputs/outputs}
\label{sec-replica-random-inputs-outputs}

Now we analyze the case of random inputs/outputs introduced in sec.~\ref{subsubsect-random-inputs-outputs}
by the replica theory using the Parisi's ansatz explained in sec.~\ref{subsubsec-parisi-ansatz-random-inputs-outputs}.

\begin{figure}[h]
  \bc
  \includegraphics[width=0.7\textwidth]{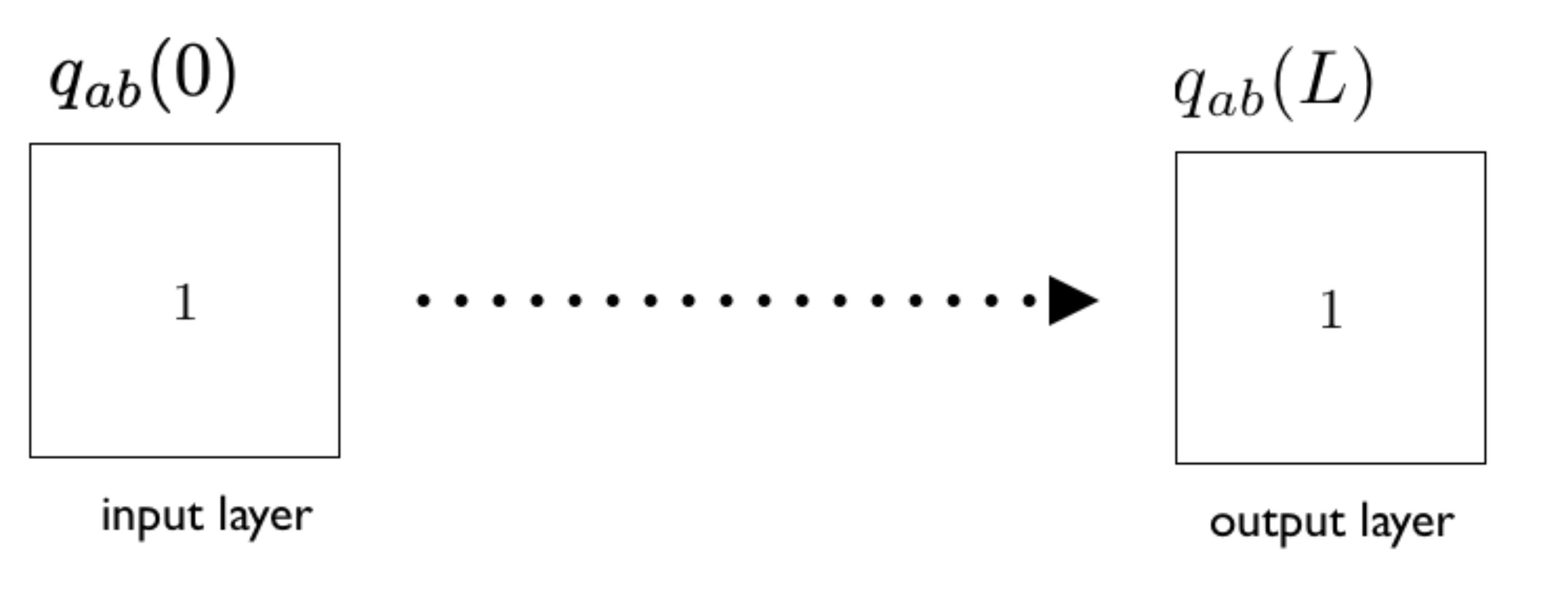}
  \ec
   \caption{``Quenched'' boundary}
   \label{fig_quenched_boundary}
\end{figure}

We assume the Pairisi's ansatz with $k$-step RSB (see sec.~\ref{subsec-parisi-ansatz}) for the order parameters
of the bonds $Q_{i}(l)$ for $l=1,2,\ldots,L$  and spins $q_{i}(l)$ for $l=1,2,\ldots,L-1$
which characterize the Parisi's matrices (see Fig.~\ref{fig-parisi-matrix}).
We solve the saddle point equations numerically to obtain the glass order parameters
as described in sec.~\ref{sec_procedure}.
For $i=0,1,2,\ldots,k$ we have parameter $m_{i}$ (see \eq{eq-m-convention-2}).
In the $k \to \infty$ limit, $Q_{i}(l)$s become continuous functions
$Q(x,l)$ which can be well approximated by $Q_{i}(l)$ plotted vs $m_{i}$ for large enough $k$
(See Fig. \ref{fig-parisi-matrix} d)).
The same holds for the order parameter of spins $q_{i}(l)$s, i.~e. we obtain
continuous functions $q(x,l)$ in $k \to \infty$ limit.
From the functions $Q(x,l)$ and $q(x,l)$, we can obtain the overlap
distribution functions $P(q,l)$ and $P(Q,l)$ (see \eq{eq-p-of-q}).
The boundary condition \eq{eq-boundary-qab} (see Fig.~\ref{fig_quenched_boundary}) amounts to set,
\beqn
&& q_{0}(0)=q_{0}(L)=1 \\
&& q_{i}(0)=q_{i}(L)=0  \qquad (i=1,2,\ldots,k).
\eeqn

In the following we present results using $k=100$ step RSB and the depth of the system $L=5-20$. Because of
the tree-approximation and the choice of the boundary condition, the system becomes symmetric
with respect to reflections at the center: we confirmed that the solutions
satisfy $q_{i}(l)=q_{i}(L-l)$ and $Q_{i}(l)=Q_{i}(L-l)$.


\subsubsection{Liquid phase}

For small $\alpha=M/N$ we find the whole system is in the liquid phase
where the glass order parameters are all zero:
for $i=1,2,\ldots,k$ $q_{i}(l)=0$ ($l=1,2,\ldots,L-1$)
and $Q_{i}(l)=0$ ($l=1,2,\ldots,L$). This means that the
parameter space is so large
that there are simply too many solutions compatible with the constraints.
Here the replica symmetry is not broken. This means that
the solution space looks like a giant continent in which all typical
solutions are continuously connected to each other.

\subsubsection{The 1st glass transition}

  \begin{figure}[h]
   \bc
   \includegraphics[width=\textwidth]{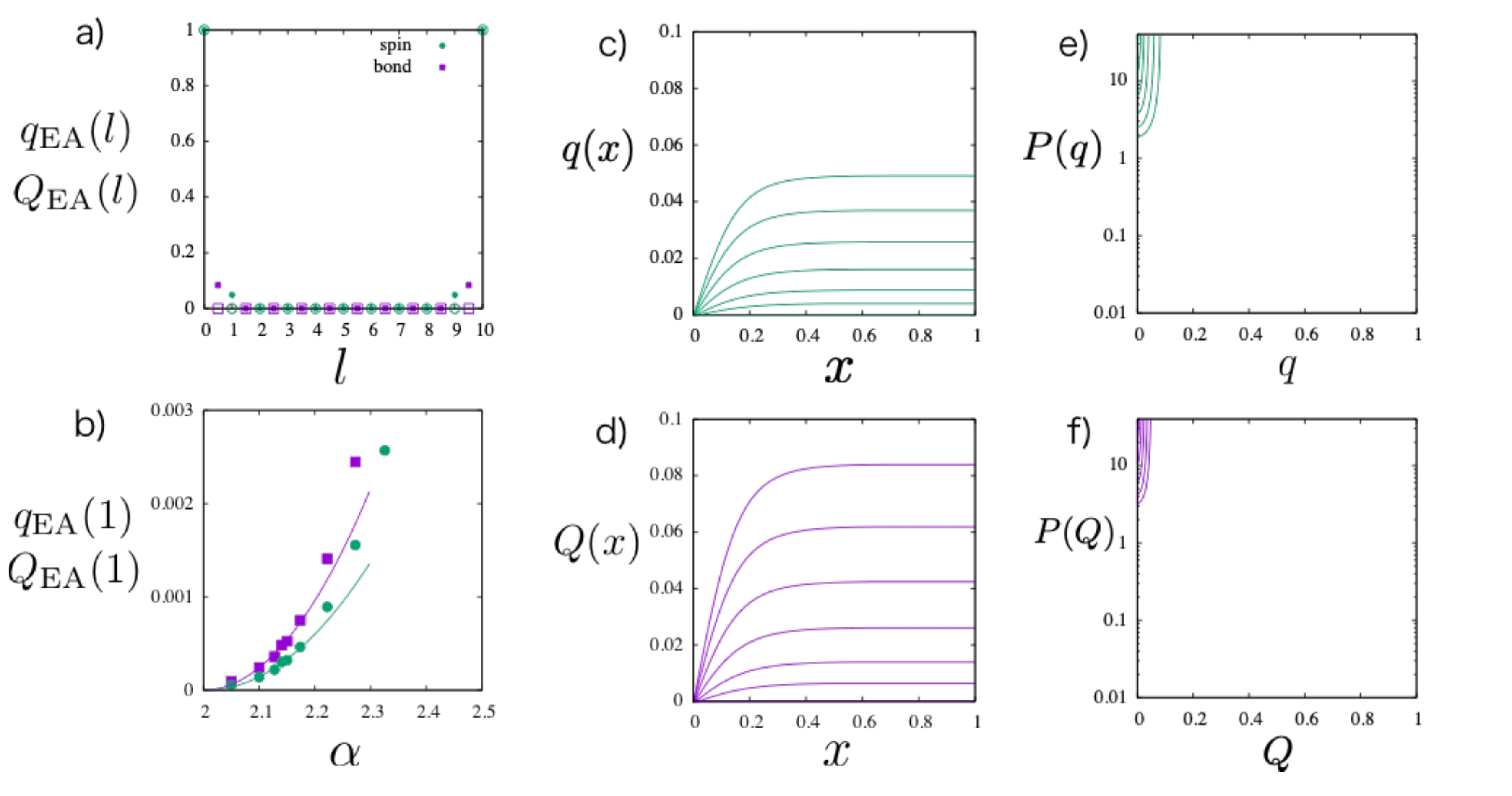}
   \ec
   \caption{The 1st glass transition:
     a) spatial profile of the Edwards-Anderson (EA) order parameter for spins
     $q_{\rm EA}(l)(=q_{k=100}(l))$ and bonds $Q_{\rm EA}(l)(=Q_{k=100}(l)$ slightly before $\alpha=2.0$
     (empty symbols)/after $\alpha=3.125$ (filled symbols)
     the 1st glass transition. The depth is $L=10$ in this example.
     b) Evolution of the EA order parameters
     $q_{\rm EA}(1)=q_{k}(1)$ and $Q_{\rm EA}(1)=Q_{k}(1)$ at the 1st layer after passing
     the critical point of the 1st glass transition $\alpha_{\rm g}(1)\simeq 2.03$.
c),d) Glass order parameter function $q(x,l)$ for spins
     and $Q(x,l)$ for bonds
     at the 1st layer $l=1$ at around the 1st glass transition.
     Here
      $\alpha=3.13, 2.94, 2.78, 2.63, 2.50,  2.38$
     from the top to the bottom.
e),f) the overlap distribution function of spins $P(q)=dx(q)/dq$
     and bonds $P(Q)=dx(Q)/dQ$  (see \eq{eq-p-of-q}).
   }
      \label{fig_1st_glasstransition}
 \end{figure}

With increasing $\alpha$, the system becomes more constrained.
We find a continuous (2nd order) glass transition at $\alpha_{\rm g}(1) \simeq 2.03$
on the 1st layers $l=1,L-1$ just beside the ``quenched'' inputs/outputs boundaries
as shown in Fig.~\ref{fig_1st_glasstransition} a).
The emergence of the finite glass order parameters signals that the solution
space is shrinking there. The rest of the system ($l=2,3,\ldots,L-2$)
remains in the liquid phase $q_{\rm EA}(l)=Q_{\rm EA}(l)=0$ at this stage.
As shown in Fig.~\ref{fig_1st_glasstransition} b)
the Edwards-Anderson (EA) order parameters
of the spins $q_{\rm EA}(l)=q_{k}(l)$ and bonds $Q_{\rm EA}(l)=Q_{k}(l)$ at the 1st layer $l=1$ 
grow continuously across the critical point $\alpha_{\rm g}(1)$.
Exactly the same happens on the other side at $l=L-1$.
The fact that the glass transition takes place in a {\it continuous} way,
is different from the random first-order transition (RFOT) in structural glass models
\cite{KT87,KW87b,KTW89,BB09,berthier2011theoretical,charbonneau2014fractal,parisi2020theory}.

Since the transition is a 2nd order transition,
the liquid sate $(Q,q)=(0,0)$ becomes unstable and a glass state can emerge smoothly at the transition.
Then what would play the role of symmetry breaking field (see sec. \ref{subsubsec-real-symmetrybreaking-field})
to pick up a particular glass state out of many candidates? 
In the learning dynamics, the random inputs/outputs data imposed
at the boundaries ($l=0$ and $l=L$) and choices of the initial condition for learning
will play the role of the symmetry breaking field.

The fact that the glassy regions emerge next to the boundaries is reasonable because the effect of constraints should be strongest there. The situation does not change even in the limit $L \to \infty$ where the two boundaries are infinitely separated. But this may appear bizarre. Why specification of the just the initial condition or finial condition for the dynamics \eq{eq-perceptron} can constrain the 1st layers ($l=1,L-1$) so much? With such a huge liquid-like region left in the bulk, any information starting from the input layer will be completely randomized before reaching the output layer. Here let us remind ourselves that we are considering statistical mechanics of the solution space which is like the statistical mechanics of hard-spheres as we noted below \eq{eq-effective-hamiltonian}. The reason for the glass transition on the 1st layers is an entirely entropic reason: a certain set of configurations of the bonds in the 1st layers ($l=1,L-1$) allow exceedingly larger fluctuation in the hidden layers compared with others so that they dominate the entropy of the solution space. In this sense it is a glass version of entropy-driven ordering like
the crystallization of hard-spheres (Alder transition) \cite{alder1957phase}
and order-by-disorder transitions oftenly observed in frustrated magnets \cite{villain1980order}.

As shown in Fig.~\ref{fig_1st_glasstransition} c),d), 
the functions $q(x,l)$ an $Q(x,l)$ at the 1st layers $l=1,L-1$
are continuous functions of $x$
with plateaus at $q_{\rm EA}$ and $Q_{\rm EA}$ for some range $x_{1}(\alpha) < x < 1$
with $x_{1}(\alpha)$ decreasing with $\alpha$. Thus the replica symmetry
is fully broken much as in the SK model for spin-glasses \cite{parisi1979infinite,MPV87}.
Correspondingly the overlap distribution functions \eq{eq-p-of-q}
$P(q)=dx(q)/dq$ and $P(Q)=dx(Q)/dQ$ shown in Fig.~\ref{fig_1st_glasstransition} e),f), exhibit delta peaks at $q=q_{\rm EA}$,
$Q=Q_{\rm EA}$ plus non-trivial continuous parts extending down to $q=0$ and $Q=0$. 

The RSB means that the solution space is now clustered, i.~e. the giant continent of the solutions is split into mutually disconnected islands.
The EA order parameters $q_{\rm EA}$ and $Q_{\rm EA}$ represent the
size of the islands, i.~e. larger EA order parameters mean smaller islands.
The probability that two solutions sampled in equilibrium
belong to the same island is given by $1-x_{1}(\alpha)$.
The continuously changing part of
the functions $Q(x)$ and $q(x)$ in the range $0 < x < x_{\rm 1}(\alpha)$
means that the islands or clusters are organized into
meta-clusters, meta-meta-clusters,... in a hierarchical way:
the mutual overlap (distance in the phase space) between the islands
is ultrametric \cite{mezard1984nature,MPV87,nemoto1987numerical,nemoto1988metastable,rammal1986ultrametricity}.
In general, the continuous RSB phase is marginally stable \cite{MPV87,nemoto1985tap,charbonneau2014exact,berthier2019gardner}.

The strong spatial heterogeneity of the glass order parameters is striking.
It means that the solution space is clustered in the
1st layers ($l=1,L-1$) but the islands of solutions merge into a big continent in the rest of the system which remains
in the liquid phase. The spatial heterogeneity is very interesting from the algorithmic point of view since this implies the learning dynamics is fast except next to the boundaries. Moreover, it is tempting to speculate that the first dynamics in the liquid region will assist the equilibration of the glassy regions close to the boundaries.

\subsubsection{The 2nd glass transition}

   \begin{figure}[h]
   \bc
   \includegraphics[width=\textwidth]{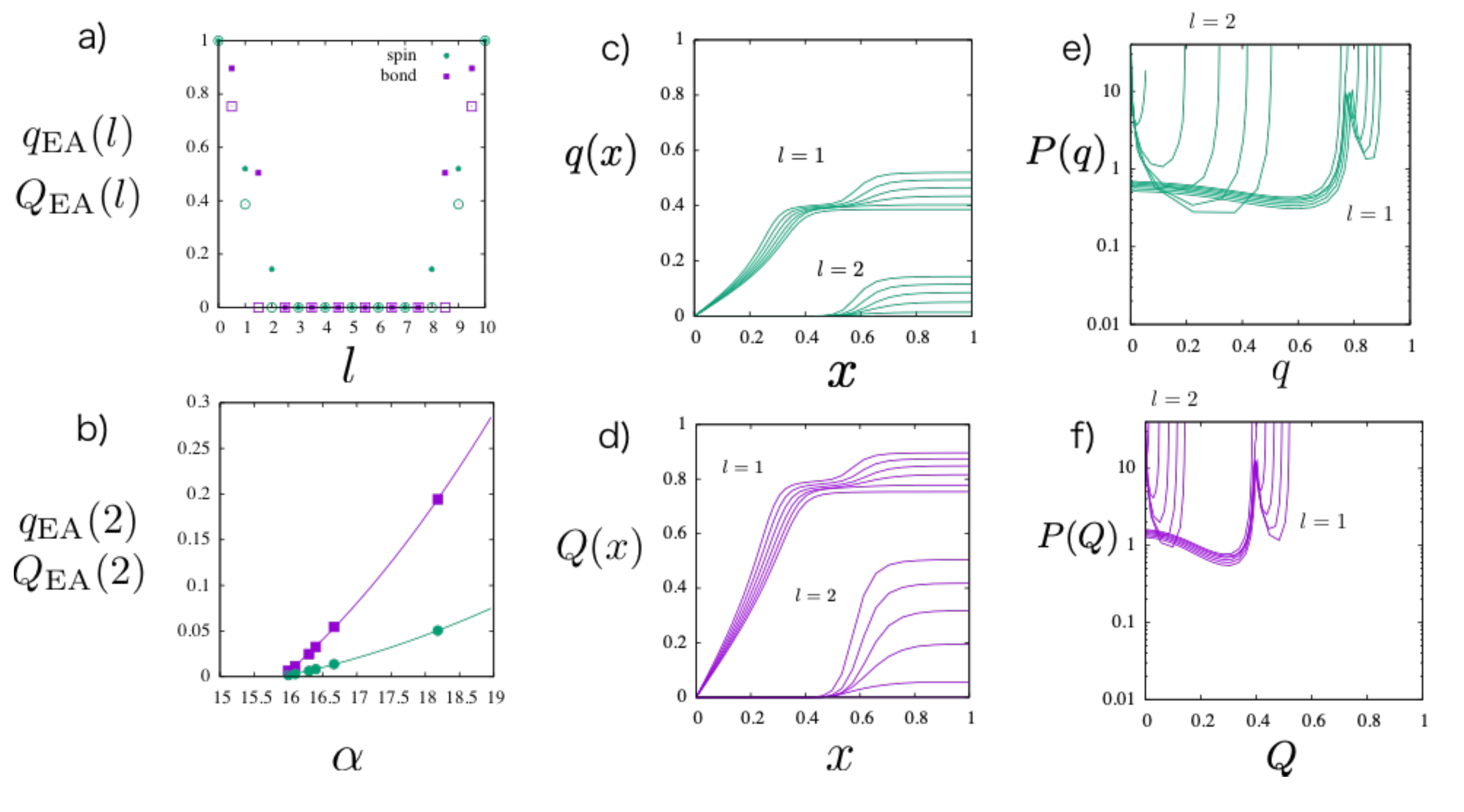}
   \ec
   \caption{The 2nd glass transition:
     a) spatial profile of the Edwards-Anderson (EA) order parameter for spins
     $q_{\rm EA}(l)$ and bonds $Q_{\rm EA}(l)$ slightly before $\alpha=15.38$
     (empty symbols)/after $\alpha=25$ (filled symbols)
     the 2nd glass transition. The depth is $L=10$ in this example.
     b) Evolution of the EA order parameters
     $q_{\rm EA}(2)$ and $Q_{\rm EA}(2)$ at the 2nd layer after passing
     the critical point of the 2nd glass transition $\alpha_{\rm g}(2)\simeq 15.9$.
     c),d) Glass order parameter function $q(x,l)$ for spins
     and $Q(x,l)$ for bonds at the 1st and 2nd layers at around the 2nd glass transition.
     Here
    $\alpha=25.0,22.2,20.0,18.2,16.7,15.4$
     from the top to the bottom.
     e),f) the overlap distribution function of spins $P(q)=dx(q)/dq$
     and bonds $P(Q)=dx(Q)/dQ$  \eq{eq-p-of-q}.}
    \label{fig_2nd_glasstransition}
  \end{figure}

  Increasing $\alpha$ further we meet another glass transition
  at $\alpha_{\rm g}(2) \simeq 15.9$ by which the 2nd layers $l=2,L-2$
  become included in the glass phase while the rest of the system $l=3,4,\ldots,L-3$ still remains
  in the liquid phase as shown in  Fig.~\ref{fig_2nd_glasstransition} a).
  The glass phase has grown one step further into the interior.
  The transition is again a continuous one as can be seen in 
  Fig.~\ref{fig_2nd_glasstransition} b) where we display the EA order parameters
  $q_{\rm EA}(l)=q_{k}(l)$ and  $Q_{\rm EA}(l)=Q_{k}(l)$ at $l=2$.  Exactly the same happens on the other side at $l=L-2$.

As shown in Fig.~\ref{fig_2nd_glasstransition} c),d), 
the functions $q(x,l)$ an $Q(x,l)$ at the 2nd layers $l=2,L-2$
are continuous functions of $x$
with plateaus at $q_{\rm EA}(2)$ and $Q_{\rm EA}(2)$
in some range $x_{2}(\alpha) < x < 1$
with $x_{2}(\alpha)$ decreasing with $\alpha$.
A marked difference to the case of the 1st glass transition which happened at
the 1st layers $l=1,L-1$ is that the order parameters become finite only in some range $x_{2}(\alpha) \lessapprox x < 1$. As a result, it looks approximately like a step function with the step located at $x_{2}(\alpha)$.
As shown in Fig.~\ref{fig_2nd_glasstransition} e),f),
this amounts to induce a delta peak not only at $q_{\rm EA}(2)$ ($Q_{\rm EA}(2)$) but also at $q=Q=0$ in the distribution of the overlaps.
In a sense the solution is approximately like one step RSB
in the random energy model \cite{derrida1980random} or
models for structural glasses
\cite{KT87,KW87b,KTW89,BB09,berthier2011theoretical,charbonneau2014fractal,parisi2020theory} 
if we neglect the smoothing part of the step like function.
This means that, roughly speaking,
the solution space in the 2nd layers are 
split into islands that are completely dissimilar from each other.
Two solutions sampled in equilibrium, in the 2nd layers,
belong to the same island
whose size is represented by $Q_{\rm EA}(2)$ and $q_{\rm EA}(2)$
with probability $1-x_{2}(\alpha)$. Otherwise, they belong to different
islands which are very far from each other.

Remarkably, the 2nd glass transition induces another continuous glass transition on the 1st layers $l=1,L-1$ 
which were already glassy. \red{Physically, this is natural because 
the 1st layers are now more constrained
than before having two glassy neighbors while they had just one glassy neighbor before.}
As can be seen in Fig.~\ref{fig_2nd_glasstransition} c),d),
an internal step-like structure emerges continuously within the region
where the glass order parameter was flat $x_{1}(\alpha) < x < 1$ before
the 2nd glass transition. As shown in Fig.~\ref{fig_2nd_glasstransition} e),f), the emergence of the internal step amounts to a continuous splitting of the delta peak at $q_{\rm EA}(1)$ ($Q_{\rm EA}(1)$) into two peaks (plus a continuous part in between) meaning that the glass phase has become more complex.
This means the smallest bundles or islands of the solutions have been split into
multiple sub-bundles.
In a sense, this is similar to the Gardner transition found originally in Ising $p$-spin spin-glass models \cite{Ga85}
and in the hard-sphere glass in large-dimensional limit \cite{kurchan2013exact,charbonneau2014fractal,parisi2020theory,berthier2019gardner}.

We could say that the situation in the 1st layers is roughly like a 2 step RSB:
if we neglect the smoothing parts, the functions $Q(x)$ and $q(x)$ look approximately like functions with two steps, one at $x_{1}(\alpha)$ and the other at $x_{2}(\alpha)$.
This means that two solutions sampled in equilibrium, in the 1st layers,
belong to the same island whose size is represented by
$1-Q_{\rm EA}(1)=1-Q(x_{2}(\alpha),1)$ and $1-q_{\rm EA}(1)=1-q(x_{2}(\alpha),1)$
with probability $1-x_{2}(\alpha)$. Otherwise they belong to different
islands. However, with a larger probability $1-x_{1}(\alpha)$,
they belong at least to the same meta-cluster of islands whose size 
is represented by $1-Q(x_{1}(\alpha),1)$ and $1-q(x_{1}(\alpha),1)$
which are larger than $1-Q_{\rm EA}$ and $1-q_{\rm EA}$.

After the 2nd glass transition, the glass order parameters have become
more heterogeneous in space. Interestingly the internal step of
the glass order parameters
on the 1st layers $l=1,L-1$ is located around $x_{2}(\alpha)$
being synchronized with the step on the 2nd layers $l=2,L-2$.
This means that two solutions sampled in equilibrium
belong to the same island in the 1st and 2nd layers
with the {\it same} probability $1-x_{2}(\alpha)$.
This implies that the same bundle of solutions continue
in the 1st and 2nd layers.
Since the EA order parameters are bigger in the 1st layers,
the bundle becomes
more spread out in the 2nd layers than in the 1st layers.
The bundles are grouped into meta-bundles in the
1st layer which becomes dissociated in the 2nd layers.
Finally, all bundles become dissociated and merge into
a gigantic liquid continent after the 3rd layers.
The two-step dissociation of the bundles of solutions
is quite interesting in the context of learning.

\subsubsection{More glass transitions}
\label{section_more_glass_transitions}

   \begin{figure}[h]
  \bc
  \includegraphics[width=0.9\textwidth]{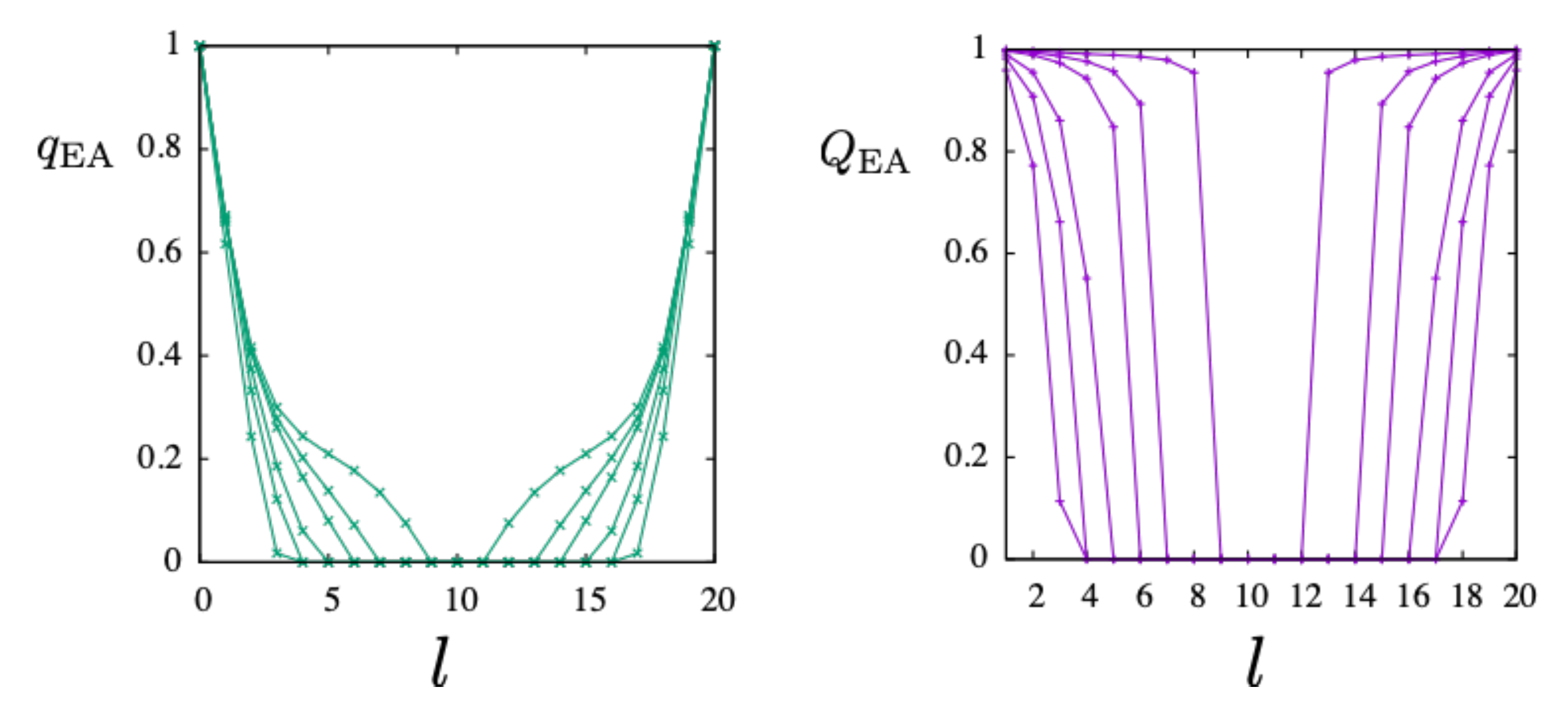}
  \ec
  \caption{The spatial profile of the EA order parameters $q_{\rm EA}(l)=q_{k}(l)$ and
    $Q_{\rm EA}(l)=Q_{k}(l)$ at
    $\alpha=50,100,200,1000,2000,4000$.     Here $L=20$.
  }
   \label{fig_more_glasstransitions}
   \end{figure}

\begin{figure}[h]
  \bc
  \includegraphics[width=0.5\textwidth]{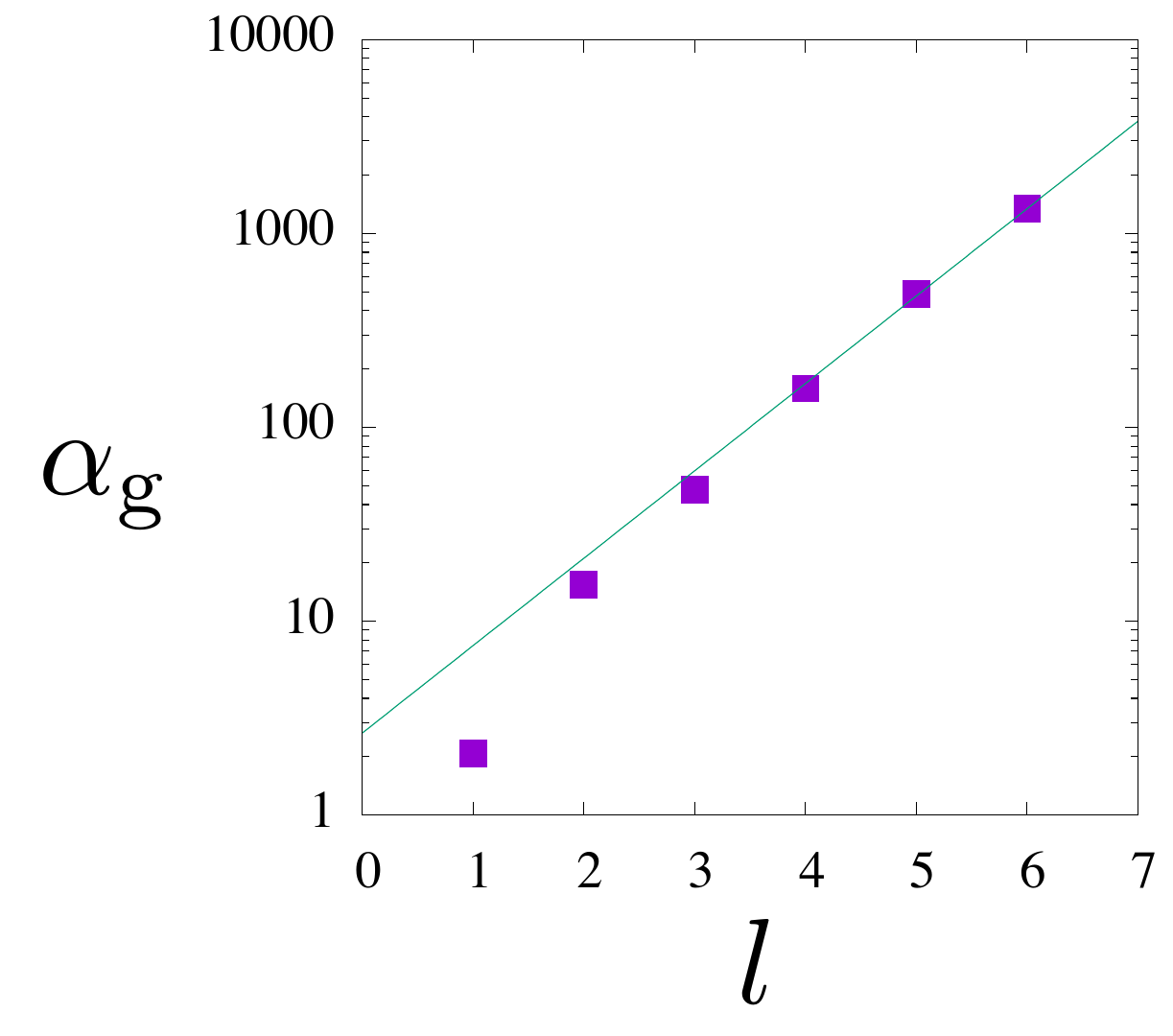}
  \ec
  \caption{The glass transition point $\alpha_{\rm g}(l)$
    of internal layers. This is obtained by numerical analysis
    of the saddle point solutions. The solid line is the exponential
    fit \eq{eq-alpha-g-l}.
  }
   \label{fig_alpha_g}
   \end{figure}

   Now it is easy to imagine that glass phase will grow further invading
   the liquid phase by increasing $\alpha$ more.
   As we show in Fig.~\ref{fig_more_glasstransitions}, this is indeed the case.
   We observe that the glass transition point $\alpha_{\rm g}(l)$ of the $l$-th
   layer (and $L-l$ the layer) grows very rapidly, exponentially fast
   with $l$ as shown in Fig.~\ref{fig_alpha_g},
   \beq
   \alpha_{\rm g}(l)\sim 2.7(3) e^{1.03(2) l}
   \label{eq-alpha-g-l}
   \eeq
   In other words, the 'penetration depth' of the glass phase $\xi_{\rm glass}$ grows very slowly with $\alpha$ as,
   \beq
   \xi_{\rm glass}(\alpha) \sim \ln \alpha
   \label{eq-xi-glass}
   \eeq

   The results shown in this section is done on systems with $L=20$
   which is still larger  than of $2\xi_{\rm glass}(\alpha)\sim 18$ of $\alpha=4000$ which is
   the largest $\alpha$ used in this section. Note that the system is under-parametrized,
   in the sense that the size of the data \eq{eq-number-of-data} is
smaller than that of the parameters \eq{eq-number-of-parameters},
only for $\alpha < 20$. However once the liquid phase is present at the center,
the solution for the glass phase does not change with larger $L$.
So that the results presented in this section
are essentially in the situation of over parametrization (for typical instances).

   The exponential growth of the glass transition point
   $\alpha_{\rm g}(l)$ with the depth $l$
   implies that the storage
   capacity $\alpha_{\rm j}(L)$, which should be greater than $\alpha_{\rm g}(L)$ by definition,
   also grow exponentially fast with the depth $L$,
\beq
\alpha_{\rm j}(L) \propto e^{{\rm const} L}
\label{eq-alpha-j-scaling}
\eeq
This is surprising because the  worst case scenarios
\cite{vapnik2015uniform} would predict linear growth with $L$.
   This means the behavior of typical instances are very different from the worst
   ones in the DNN.
   Here it is instructive to recall the case of the single perceptron.
   The storage capacity of typical instances computed by the replica method
   \cite{gardner1988space} is $\alpha_{\rm j}=2$. 
   The existence of solutions are guaranteed for all instances including
   the worst ones in the range $0 < \alpha < 1$
   while there are exponentially rare $e^{-{\rm const} N}$
   samples which lacks solutions in the range $1 < \alpha < 2$
   \cite{cover1965geometrical}.
   Our result implies the gap between the worst and typical ones become
   much more enhanced in deeper systems $L > 1$. 
   This may be related to the so-called exponential
   expressivity \cite{poole2016exponential}. The latter is due to the chaos effect
   of DNNs with non-linear activation functions like \eq{eq-perceptron}:
   trajectories starting from slightly
   different initial condition deccorrelate exponentially with the depth $l$.
   Perhaps this helps building a mapping (function) between the imposed input
   and output spin configurations, which are totally different,
   by a limited depth.

   \begin{figure}[h]
  \bc
  \includegraphics[width=0.9\textwidth]{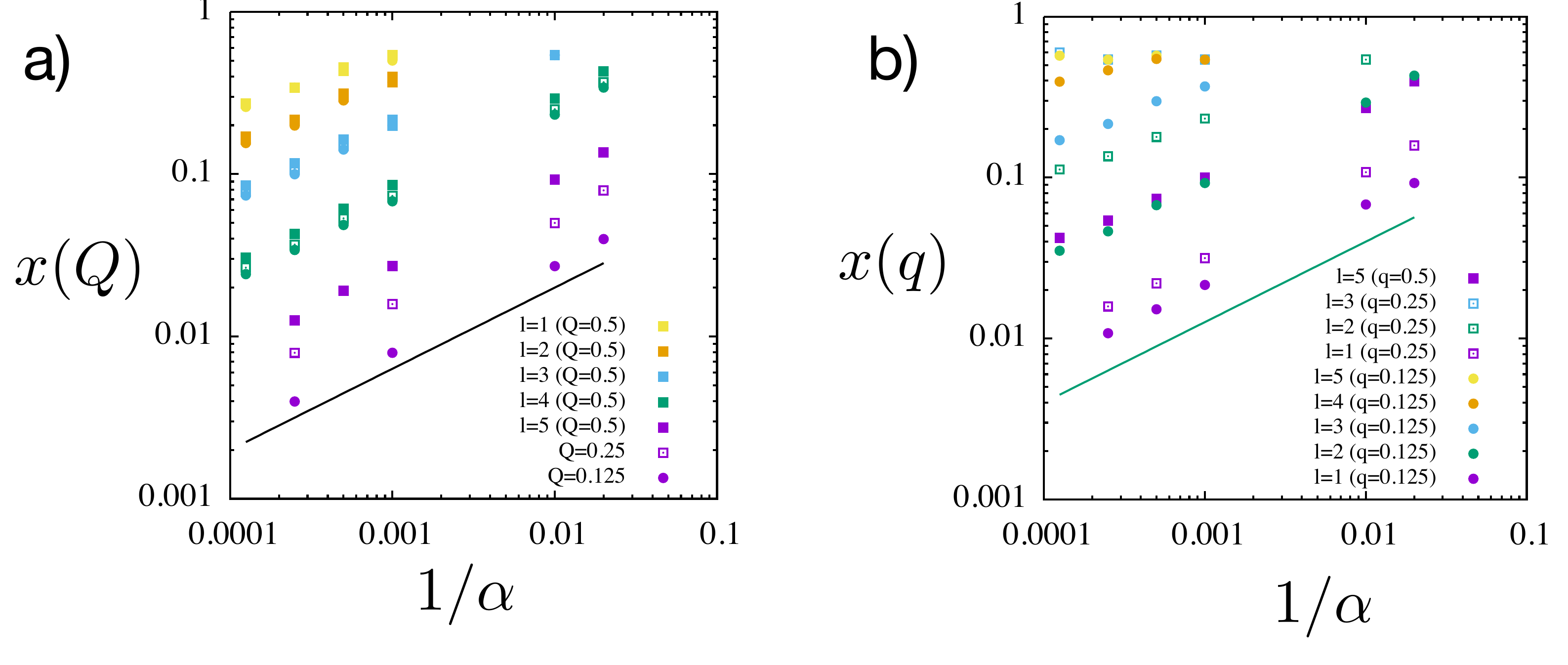}
  \ec
  \caption{Decay of $x(Q)$ and $x(q)$ with increasing $\alpha$. Here values of $x(Q,l)$,
    which is the inverse function of $Q(x,l)$ and $q(x,l)$, are shown at various $Q$, $q$ and layers $l$.
    The slope of the straight line is $1/2$.
}
   \label{fig_x_alpha}
\end{figure}

      As $\alpha$ increases, the allowed phase space volume
   becomes suppressed. In Fig.~\ref{fig_x_alpha} we display $x(Q,l)=\int_{0}^{Q}dQP(Q,l)$ and $x(q,l)=\int_{0}^{Q}dq P(q,l)$ (see \eq{eq-p-of-q}).
   The latter is the probability that two replicas (two machines learning independently) subjected to the same inputs/outputs have a mutual overlap of the bonds (spins) at $l$-th layer smaller than $Q$ ($q$).
  As can be seen in the figure, the probability appears to decay as $1/\sqrt{\alpha}$ for all $l$, $Q$ and $q$. This implies two independently
   learning machines become more
   and more similar as the number of constraints increases.

   We note however that the EA order parameter of the spins $q_{\rm EA}(l)$
   shown in Fig.~\ref{fig_more_glasstransitions} remain
   significantly smaller than that of the bonds $Q_{\rm EA}(l)$.
   Apparently, it implies that even in the jamming limit where $Q_{\rm EA}(l) \to 1^{-}$, $q_{\rm EA}(l)$
   does not reach $1$.
   A possible reason is the chaos effect. As we noted before,
   trajectories of random perceptron network with non-linear activation functions are known to show
   chaotic behavior under infinitesimal changes on the input boundary \cite{sompolinsky1988chaos,poole2016exponential}.
   We confirmed it is always the case for the present model with the 'sgn' activation function \eq{eq-perceptron} \cite{okazaki-yoshino}. Moreover the system also shows a chaotic response against infinitesimal changes made on the bonds \cite{okazaki-yoshino}. Thus even in $Q_{\rm EA}(l) \to 1^{-}$ limit, the spin configuration can fluctuate significantly.
      
   \begin{figure}[h]
  \bc
  \includegraphics[width=\textwidth]{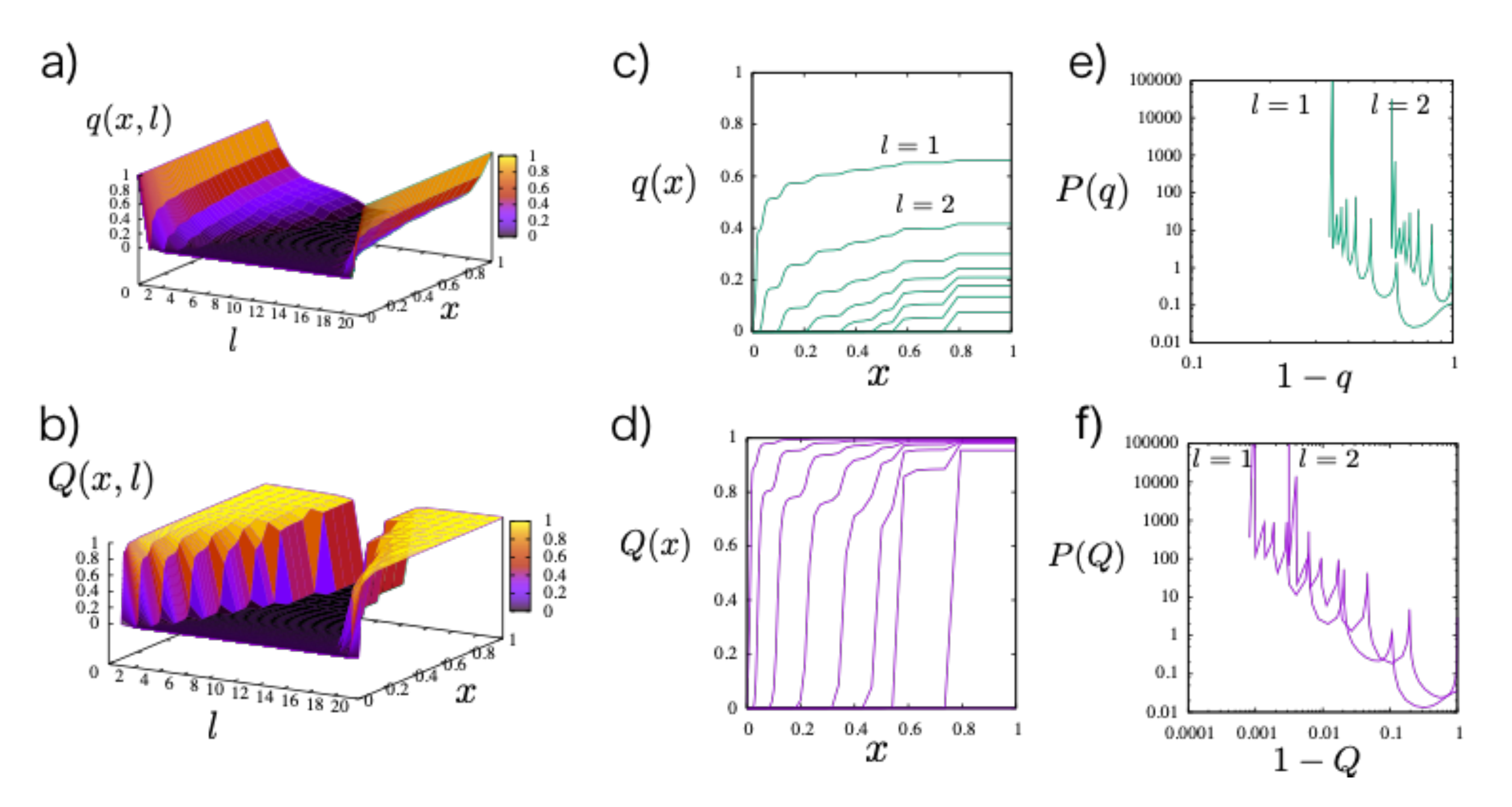}
  \ec
  \caption{Glass order parameter functions under stronger
    constraints $\alpha=4000$. In this example $L=20$
    and the central layers at $l=8,9$ still remain in the liquid phase.
    a),b) 3 dimensional plots of $q(x,l)$ and $Q(x,l)$. c),d) the same in 2 dimensional plots.
    e,f) the corresponding overlap distribution functions  \eq{eq-p-of-q}
at $l=1,2$ (for clarity others at $l=3,4,\ldots$ are not shown).
  }
   \label{fig_more_constrained_qxl_Pofq}
\end{figure}

   As shown in Fig.~\ref{fig_more_constrained_qxl_Pofq}, the glass order parameter
   functions become quite complex at large values of $\alpha$.
   Closer to the boundaries, the system has experienced larger numbers of successive glass transitions that leave behind river-terrace-like structure with many steps in the glass order parameter functions. This means distribution functions of the overlap with many delta peaks. The steps of the glass order parameter functions
   at different layers appear to be aligned with each other.

\begin{figure}[h]
  \bc
  \includegraphics[width=1.0\textwidth]{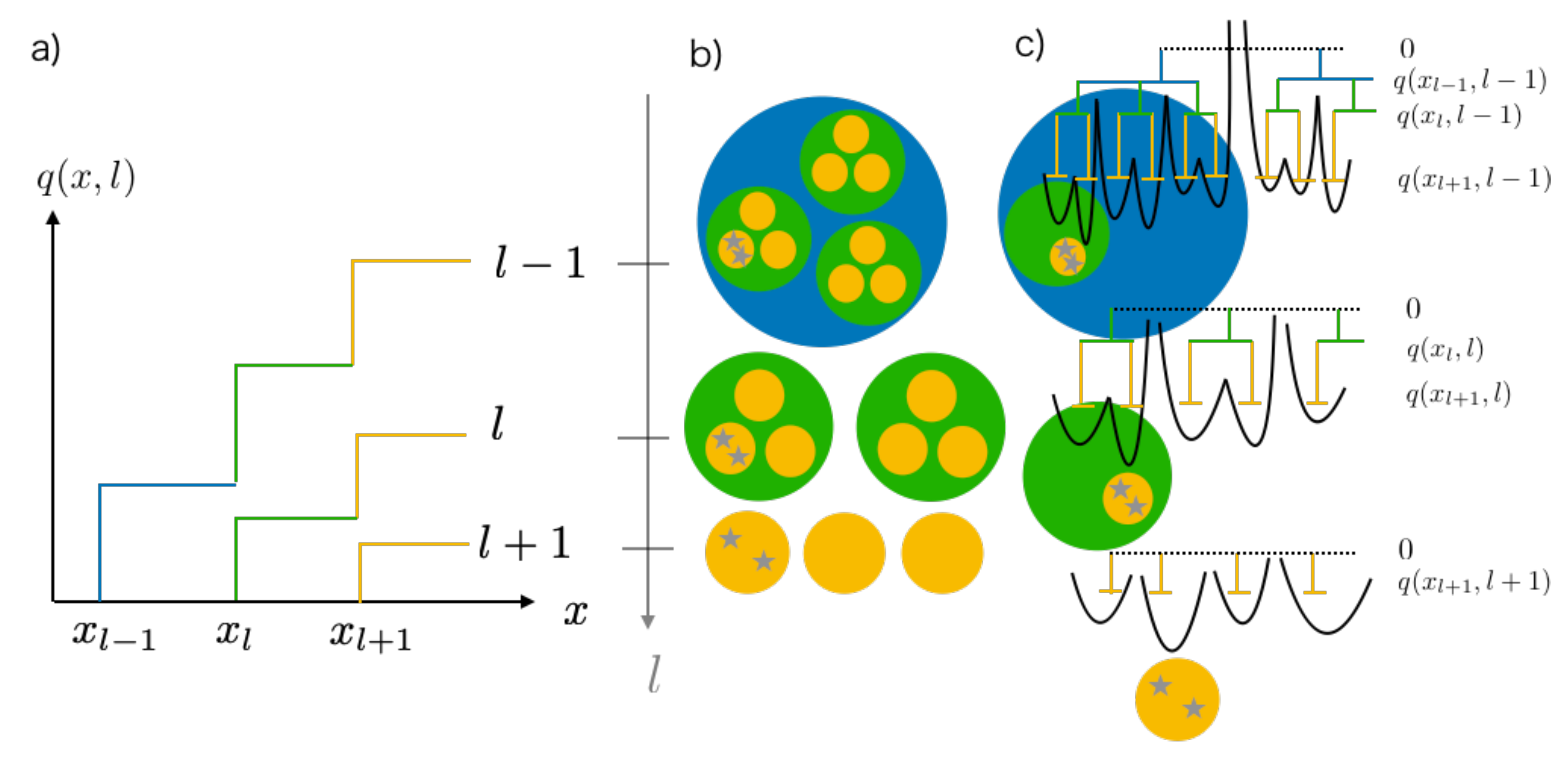}
  \ec
  \caption{River-terrace like glass order parameter and its implications:
    a) schematic picture of the river-terrace-like glass order parameter $q(x)$ (or $Q(x)$)
    b) hierarchical clustering of replicas c) schematic free-energy landscape
    and trees representing ultra-metric organization of overlaps between meta-stable states.
  }
   \label{fig_schematic}
 \end{figure}

Now let us summarize the essential features of
the  glass order parameters $q(x,l)$ and
$Q(x,l)$ shown in Fig.~\ref{fig_1st_glasstransition}, ~\ref{fig_2nd_glasstransition} and ~\ref{fig_more_constrained_qxl_Pofq}.
The essence of the river-terrace-like glass order parameter functions can be sketched schematically as shown in Fig.~\ref{fig_schematic} a).
Here we have simplified the picture representing the functions by staircases neglecting their rounding.
Comparing the river-terraces at different layers we notice an interesting feature that the steps
at different layers are synchronized: they are all located exactly at the same positions, $\ldots,x_{l-1},x_{l},x_{l+1},\ldots$.
The river-terraces reflect successive glass transitions in the following way.
At the $n$-th glass transition, a finite glass order parameter emerges continuously
in the  interval $x_{n}(\alpha) < x < 1$ at the $n$-th (and ($L-n$)-th) layer.
The glass order parameter functions at layers between the $n$-th layer and the boundary, those at $l=1,2,\ldots,n-1$
(and the corresponding layers on the other side), which are already in the glass phase, acquire additional steps in the same
interval $x_{n}(\alpha) < x < 1$.  At a given $\alpha$, the layers included in the glass phase are $l=1,2,\ldots,n$ (and the corresponding
ones on the other side) where $n$ is such that $\alpha_{\rm g}(n) < \alpha < \alpha_{\rm g}(n+1)$.
Due to the successive glass transitions $1,2,\ldots,n$, the $l$-th (and $L-l$ th) layer with $1 \leq l \leq n$
have a series of steps at $0 < x_{l}(\alpha) < x_{l+1}(\alpha) \ldots < x_{n}(\alpha) <1$.
Correspondingly the overlap distribution functions $P(Q,l)$ and $P(q,l)$ exhibit a series of
delta peaks at $q(x_{l},l) < q(x_{l+1},l) \ldots < q(x_{n},l)$ plus another delta peak at $q=0$ for $2 \leq l \leq n$.

The river-terrace-like glass order parameter function $q(x,l)$ (and $Q(x,l)$) in
Fig.~\ref{fig_schematic} a), means spatial evolution of the hierarchical clustering of
the solutions as shown schematically in Fig.~\ref{fig_schematic} b).
In panel b) clusters (and meta-clusters) of the same color represent
those associated with a common value of $x$.
Recalling the probabilistic meaning of $x$, it is natural to assume such a cluster represents
a bundle of solutions that go together through different layers.
Sampling two solutions in equilibrium, the two belong to such a common cluster with probability $1-x$.
The size of a cluster represents spreading of the solutions
$1-q(x,l)$, i.e. typical distance between the solutions belonging to the same cluster, which increases with decreasing $x$
and/or going away from the boundary $l=1,2,\ldots$.
(Meta-)clusters with smaller $x$ represent those at a higher level in the hierarchy which includes sub-clusters associated
with larger values of $x$.
Going deeper into the bulk starting from the boundary, those clusters with smaller $x$ dissociate earlier.

This, in turn, implies the hierarchical free-energy landscape with basins, meta-basins,...
which evolves in space as shown schematically in Fig.~\ref{fig_schematic} c).
The free-energy landscape
evolves in space in such a way that it progressively becomes less complex and flatter as we go deeper into the interior. For a given $\alpha$, the penetration depth
$\xi_{\rm glass}(\alpha) \sim \ln \alpha$ is finite. So that in a deep enough network $L/2 > \xi_{\rm glass}(\alpha)$,
the interior remains in the liquid phase. Moreover, the fact that the river-terraces of the glass order parameter functions
at different layers are synchronized to each other with common positions of the steps at $x_{1} < x_{2} \ldots$,
suggests that the basic backbone structure of the free-energy landscape is preserved
(but renormalized) moving away from the boundaries.
It is tempting to speculate that these features have important consequences on learning in deep neural networks.

\subsection{Fluctuating boundary}
\label{sec-replica-fluctuating-boundary}

\begin{figure}[h]
  \bc
  \includegraphics[width=0.8\textwidth]{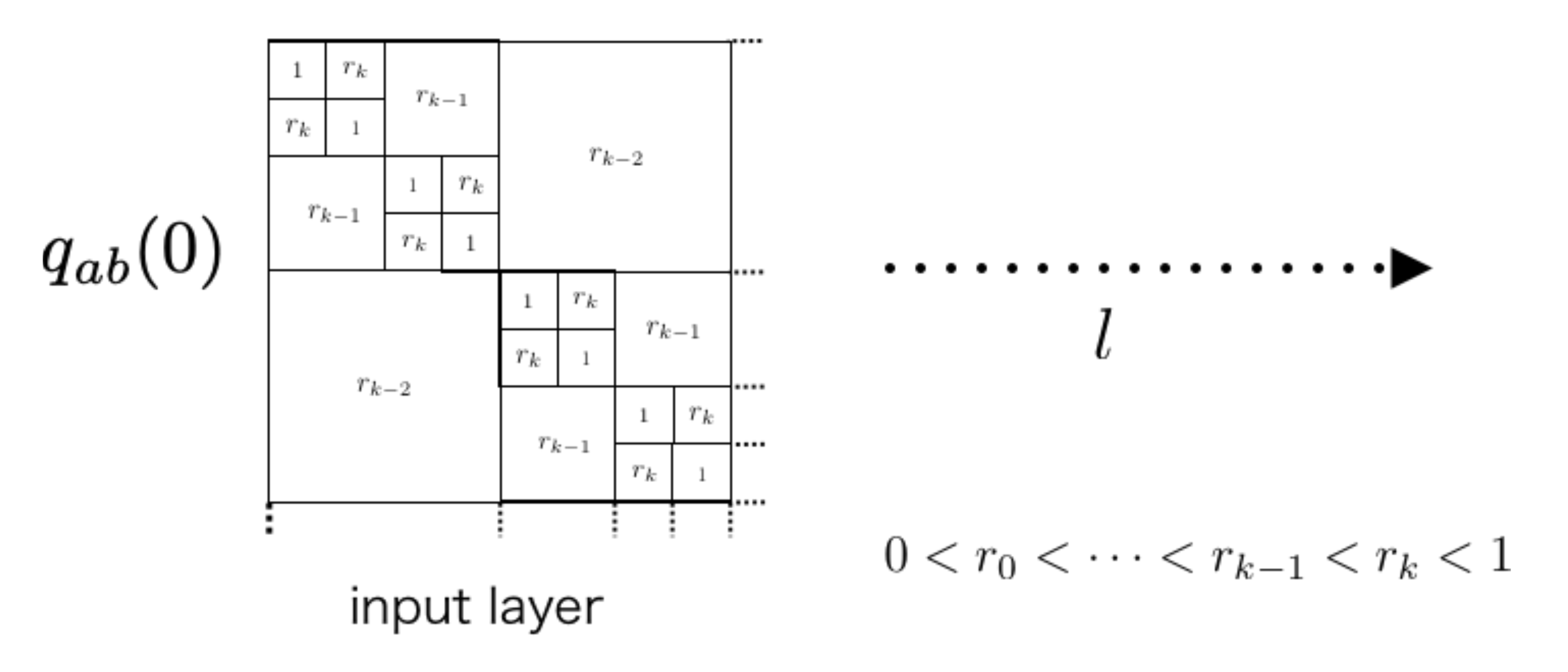}
  \ec
   \caption{Fluctuating input layer with hierarchical overlap structure}
   \label{fig_rsb_input}
\end{figure}

To obtain some further insights, we next analyze the case of fluctuating boundary:
spin configurations on the boundaries are allowed to fluctuate
during learning following certain probability distributions.
Here we consider cases such that the overlap distribution of
the spins on the input layer ($l=0$) exhibit a hierarchical structure as parametrized in the form of the  Parisi's matrix \eq{eq:parisi-matrix-q}
(Fig.~\ref{fig_rsb_input}).
There are two motivations for this analysis:
\begin{itemize}
\item The perturbation may provide some hints on the stability of the characteristic free-energy landscape of the DNN we found above.
  Given that random neural networks are typically chaotic with respect to changes made on the inputs
  \cite{poole2016exponential,okazaki-yoshino}, it is very interesting to know how training make differences.
  
\item In a typical setting of unsupervised learning, one would be interested with
  the probability distributions $P({\bf S}_{l})$ of hidden variables ${\bf S}_{l}$ ($l=1,2,\ldots$)
  when variables on the input boundary ${\bf S}_{0}$ are forced to obey some probability
  distribution $P({\bf S}_{0})$.
\end{itemize}

\subsubsection{One RSB type boundary}

Here we consider the simplest case of '1RSB'.
\beq
q_{i}(0)=
\left \{\begin{array}{cc}
r & m_{i} < x_{\rm input} \nonumber \\
1 & m_{i} > x_{\rm input}
\end{array} \right.
\label{eq-1rsb-input}
\eeq
This means the system subjected to a slightly different input data instead of the original one,
which has overlap $0 < r < 1$ with respect the original input data,
from time-to-time with some small probability $x_{\rm input}$.

\begin{figure}[h]
  \bc
  \includegraphics[width=0.9\textwidth]{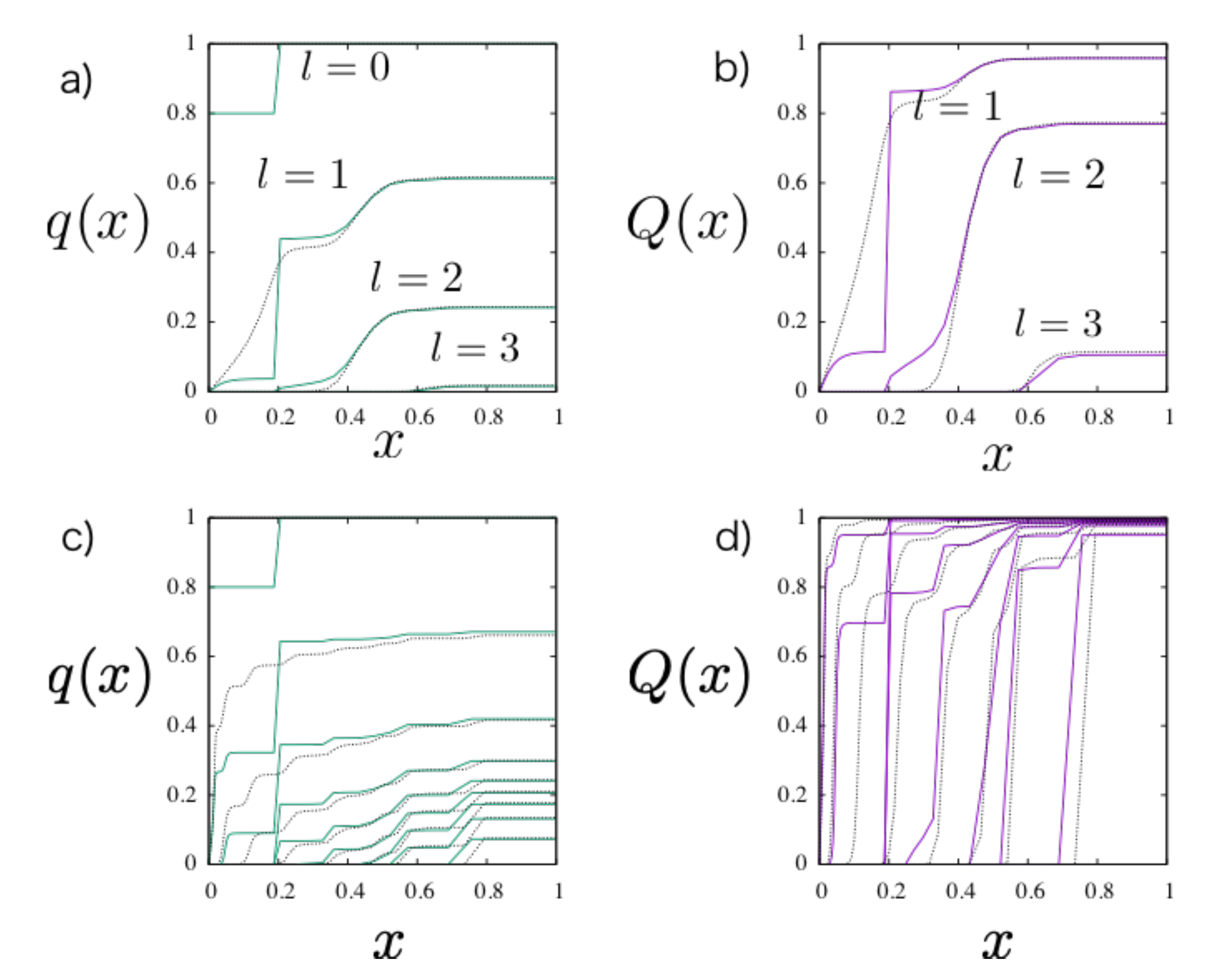}
  \ec
  \caption{Glass order parameters with 1 RSB input. Here $L=20$, $x_{\rm input}=0.2$ and $r=0.8$
    for the solid lines and $x_{\rm input}=0.2$ a),b) $\alpha=50$ c),d) $\alpha=4000$.
    Doted lines represent the glass order parameters with the frozen boundary.  }
   \label{fig_1rsb_qxl}
\end{figure}

It can be seen in Fig.~\ref{fig_1rsb_qxl} that the effect of the perturbation is strong only at $x < x_{\rm input}$. This means
that the trained system is not simply chaotic but
the hierarchical organization in the solution space has a certain degree of robustness against perturbations on the inputs
(as well as outputs).

\subsubsection{Full RSB type boundary}
\label{sec-fullRSB-boundary}

Let us next consider the 'full RSB' case.
More specifically we consider the simplest full RSB structure in the input layer,
\beq
q_{i}(0)={\rm min} (a m_{i}, 1)
\label{eq-fullrsb-input}
\eeq
with a certain constant $a>0$. Thus $q(x,0)$ function consists of
two parts: 1) 'continuous part' $q(x,0)=ax$ with slope $a$ in the interval $0 < x < 1/a$
and  2)  'plateau' $q(x,0)=q_{\rm EA}(0)=1$ in the interval $1/a < x < 1$.

\begin{figure}[h]
  \bc
  \includegraphics[width=0.9\textwidth]{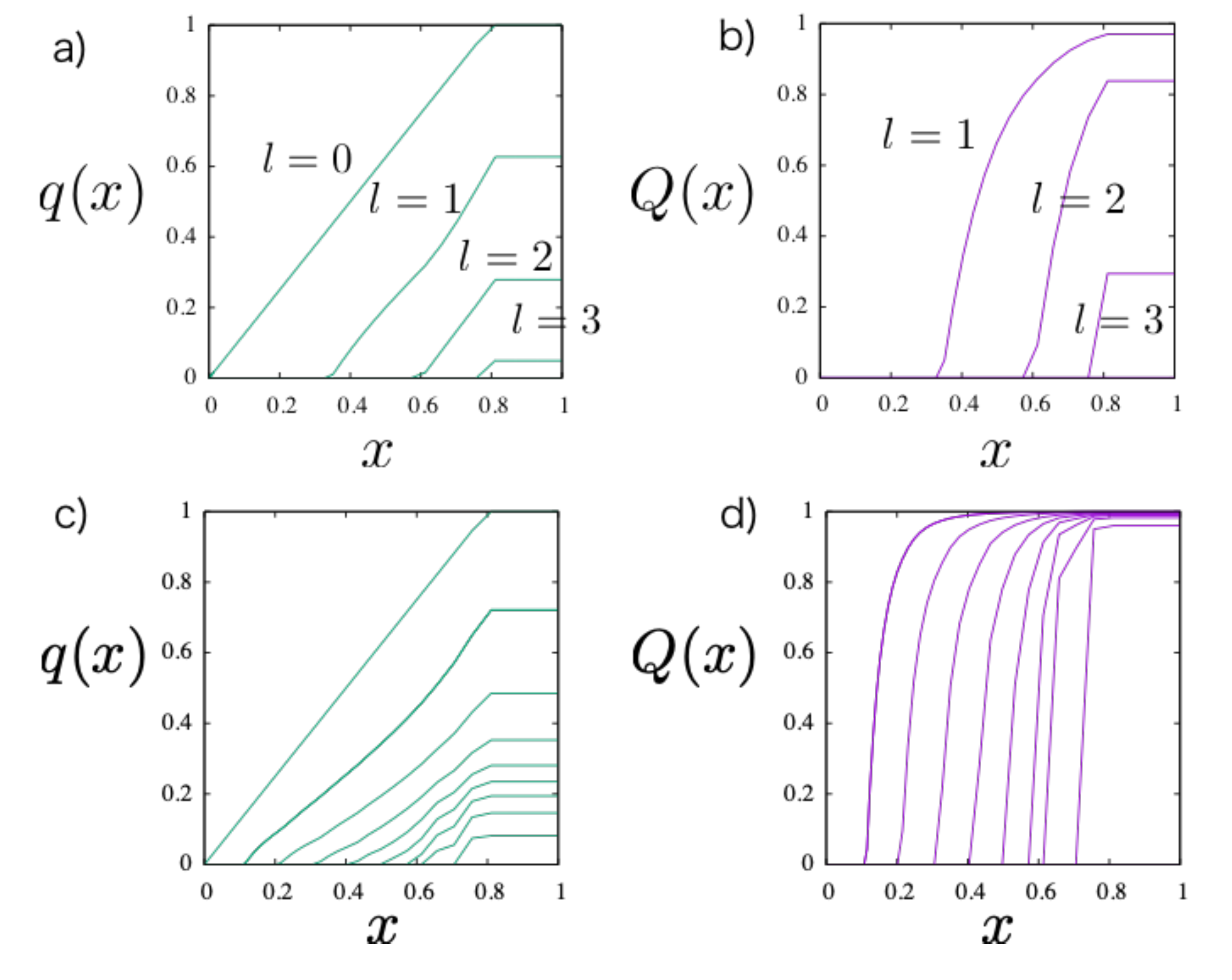}
  \ec
  \caption{Glass order parameters with full RSB input. Here $L=20$ and $1/a=0.8$.
  a),b) $\alpha=50$ c),d) $\alpha=4000$.}
   \label{fig_fullrsb_qxl}
\end{figure}

We analyze the saddle point solutions numerically as before
(see sec.~\ref{sec_procedure}). In the following, we present results
using $k=100$ step RSB and the depth of the system $L=20$.
We chose $1/a=0.8$.
As shown in Fig.~\ref{fig_fullrsb_qxl}, the glass phase grows increasing $\alpha$
much as in the case of ``quenched'' (RS) boundary condition discussed in sec.~\ref{sec-replica-random-inputs-outputs}.
We limit ourselves to $\alpha$ such that $\xi_{\rm glass}(\alpha) < L/2$
so that we have a liquid phase left at the center of the system.
In this circumstance the boundary condition on the other side $q_{ab}(L)$ is irrelevant.

A remarkable feature of the resulting glass order parameter is that the hierarchical structure put on the input propagates into the interior of the network preserving its basic hierarchical structure.
The numerical solution suggests that the $q(x,l)$ function at a given layer $l$ consists of three parts:
0) $q(x,l)=0$ for some interval $0 <  x <  x_{l}$ 1) 'continuous part'
$q(x,l)=a(x-x_{l})$
in the interval $x_{l} < x < 1/a$ with the same slope $a$ as in the input
2) 'plateau'  $q(x,l)=q_{\rm EA}(l)=1-ax_{l}$ in the last interval $1/a < x < 1$ as in the input.
Correspondingly the overlap distribution function $P(q)=dx(q)/dq$ becomes,
\beq
P(q,l)=x_{l}\delta(q)+\frac{1}{a}+\left(1-\frac{1}{a}\right)\delta(q-(1-ax_{l}))
\eeq
which consists of three parts:
0) delta peak at $q=0$
1) constant part with height $1/a$ in the interval $x_{l} < x < 1/a$ as in the input
2) delta peak at $q=q_{\rm EA}(l)$. 

Going deeper into the interior increasing $l$, we find $x_{l}$ grows and $q_{\rm EA}(l)=1-ax_{l}$ decreases.
We can regard this as a kind of 'renormalization' of the input data : the embedded overlap structure
at low overlaps in the input data become progressively renormalized into the $q=0$ sector in the hidden layers,
keeping only the important part of the hierarchical structure at higher overlaps. It will be very interesting to study further
the implication of this result in the context of data clustering where the idea of ultrametricity is very useful.

\subsection{Teacher-student setting}
\label{sec-replica-teacher-student}

   \begin{figure}[t]
  \bc
  \includegraphics[width=0.9\textwidth]{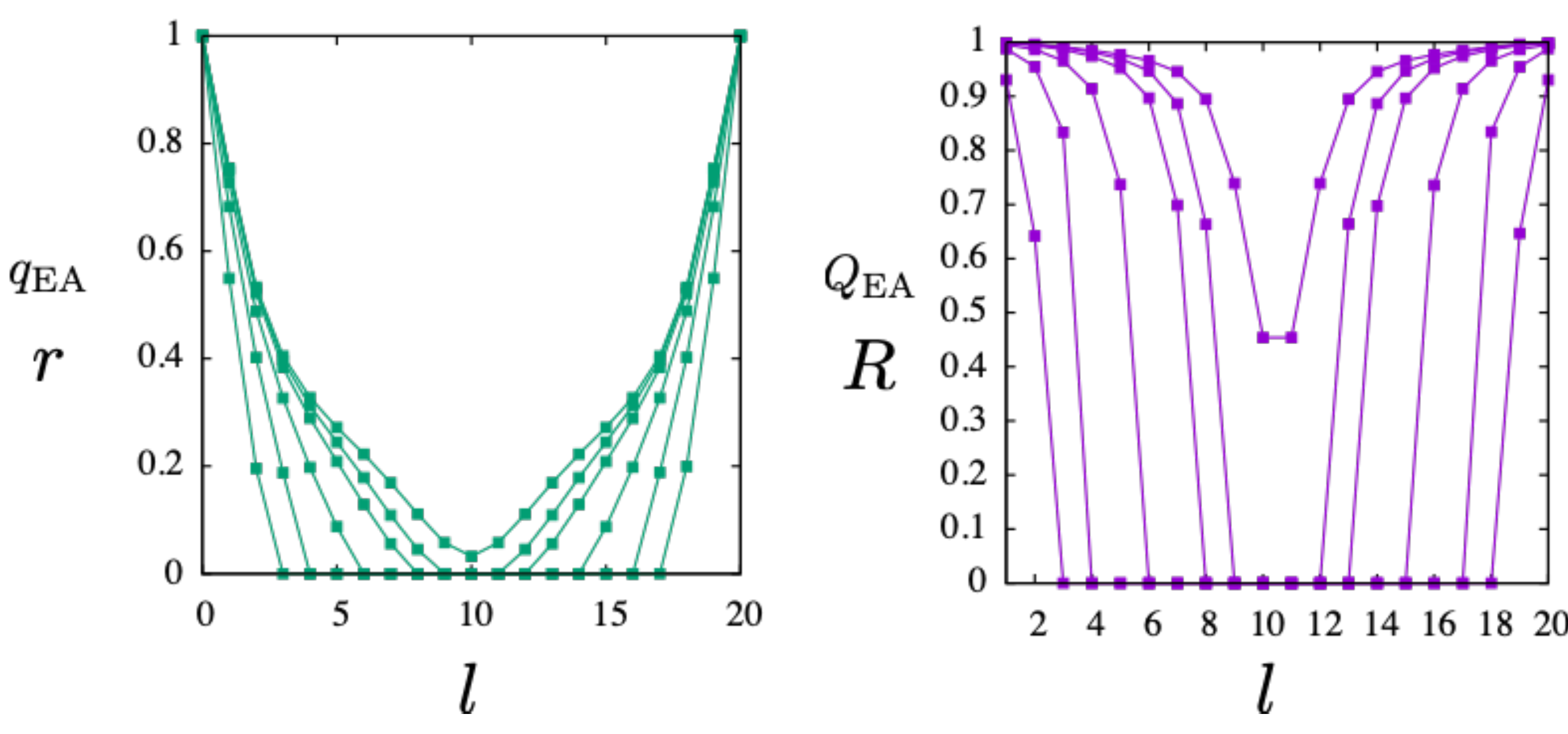}
  \ec
  \caption{The spatial profile of the EA order parameters
    $q_{\rm EA}(l)=q_{0}(l)=r(l)$,
    $Q_{\rm EA}(l)=Q_{0}(l)=R(l)$ (RS solution $k=0$)
    at
    $\alpha=25,100,250,500,625,714$.
    Here $L=20$. For the largest $\alpha$, $\xi > L$
    and the liquid phase disappear, which is a 'finite depth' effect . 
  }
   \label{fig_teacher_student}
   \end{figure}

Now let us turn to analyze the teacher-student setting introduced in sec.~\ref{subsubsect-teacher-student}
by the replica theory using the ansatz explained in sec.~\ref{subsubsec-parisi-ansatz-teacher-student}.


Since we are limiting ourselves
to the Bayes optimal case, it is sufficient to consider the replica symmetric ($k=0$) ansatz
so that the Nishimori condition holds \cite{iba1999nishimori,zdeborova2016statistical,nishimori2001statistical}, which reads in the present system as,
\beq
r(l)=q_{0}(l) \qquad R(l)=Q_{0}(l).
\eeq
The saddle point equations in sec. \ref{sec-sp-teacher-student} admit such solutions.

In Fig.~\ref{fig_teacher_student} we show the profile of the solutions obtained at various $\alpha=M/N$.
Remarkably the spatial profile of the order parameters are very similar to those of random inputs/outputs
(See Fig.~\ref{fig_more_glasstransitions}).
This is again due to successive layer-by-layer, 2nd order 'crystalline' phase transitions which start from the boundaries.
The overlap of the student machine to the teacher machine
grows from the boundary and the penetration depth grows again as
\beq
\xi_{\rm teacher-student} \propto \ln \alpha.
\eeq
Remarkably the central part of the student machine remains de-correlated from the teacher machine if the system is deep enough, i. e. $L > \xi$.
The solution (for the case $L > \xi$) in the crystalline region does not change even in $L \to \infty$ limit.
The reason for the crystalline transition starting from the 1st layers ($l=0,L-1$) is again the entropic effect:
some set of configurations of the bonds in the 1st layers ($l=0,L-1$) allow exceedingly larger fluctuation in the hidden layers compared with others
so that they dominate the entropy of the solution space.

Now let us discuss what the above theoretical results mean for practice.
The fact that the transitions are 2nd order transitions is a very good news. This is because
it implies that inference will not be too difficult
 \cite{zdeborova2016statistical}: we do not need to worry about the possibility to be trapped
in the solution of $R=0$ (failure of inference) because it becomes unstable at the transition. 
However, very importantly, we have to ask what would play the role of symmetry
breaking field by which the student machine can detect the teacher's configuration during learning.
In our theory, we had the convenient 'fictitious' symmetry breaking
field (see sec. ~\ref{subsubsec-explicit-rsb}) but it must be realized by some 'real' field (in the computer!).
Actually, if the central part remains really random, how can the student machine 
ever develop some finite overlap to the teacher machine
at the opposite ends disconnected by the liquid phase in between?
Our analysis for the case of the random inputs/outputs would suggest otherwise: the student machine should
not be able to pick up the minima planted by the teacher machine correctly hidden in the ocean of many (wrong) minima, all of which correctly satisfy the constraints on the input and output boundaries.

\red{At the moment we do not have a
  proposal for the real symmetry breaking field which works during learning.
  Instead, we can think of the following {\it unlearning}.
 Suppose that we give the student machine the complete
 configuration of the teacher machine at the beginning
 then let the student machine relax under the constraint
 by the training data of size $M=\alpha N$.
 Our theory implies that the student machine will keep the configuration of the teacher machine close the boundaries over the region of size $\xi \propto \ln \alpha$ but forget the teacher machine in the center.
}

\red{However, during unlearning, some weak correlation between the teacher and student machines of order, say
  $O(\log(N)/N)$ which does not contribute the order parameter \eq{eq-glass-order-parameters} in the limits $N \to \infty$ (with fixed $\alpha=M/N$), can remain in the central part of the system.  Once established this would play the role of the symmetry-breaking field: the free-energy of the selected state (teacher's configuration) will be lowered by an amount of order $O(\log(N))$ to other low lying (wrong) states. In this way the teacher's configuration may survive close to the boundaries. Such logarithmic correction naturally arises by integrating out the fluctuation of the order parameters around the saddle point.
We leave the detailed analysis of the correction for future studies.
}

\red{The next question is how the performance of the
  student-machine compares with the output of the teacher machine
  with respect to unseen test data. Increasing $\alpha$,
  i.~.e. the size of training data,
  not only the thickness of the crystal phase $\xi$ grows but also the
  the amplitude of the bias field, that is the polarization
  of the student machine with respect to the
  teacher machine in the liquid-like region will become larger.
  Because of these two aspects, we expect the output of the student machine
  against the unseen test data
  is not totally decorrelated from that of the teacher machine even in the
  over-parametrized situation but the similarity
  of their outputs (generalization ability of the student machine)
  increases with the size of the training data $\alpha$.}
  
\red{The above scenario based on unlearning is obviously artificial
 (we are not interested in unlearning but learning!)
but may help us to understand better generalization.}

   \subsection{Summary}
   \label{sec-summary-theory}

             \begin{figure}[h]
  \bc
  \includegraphics[width=0.9\textwidth]{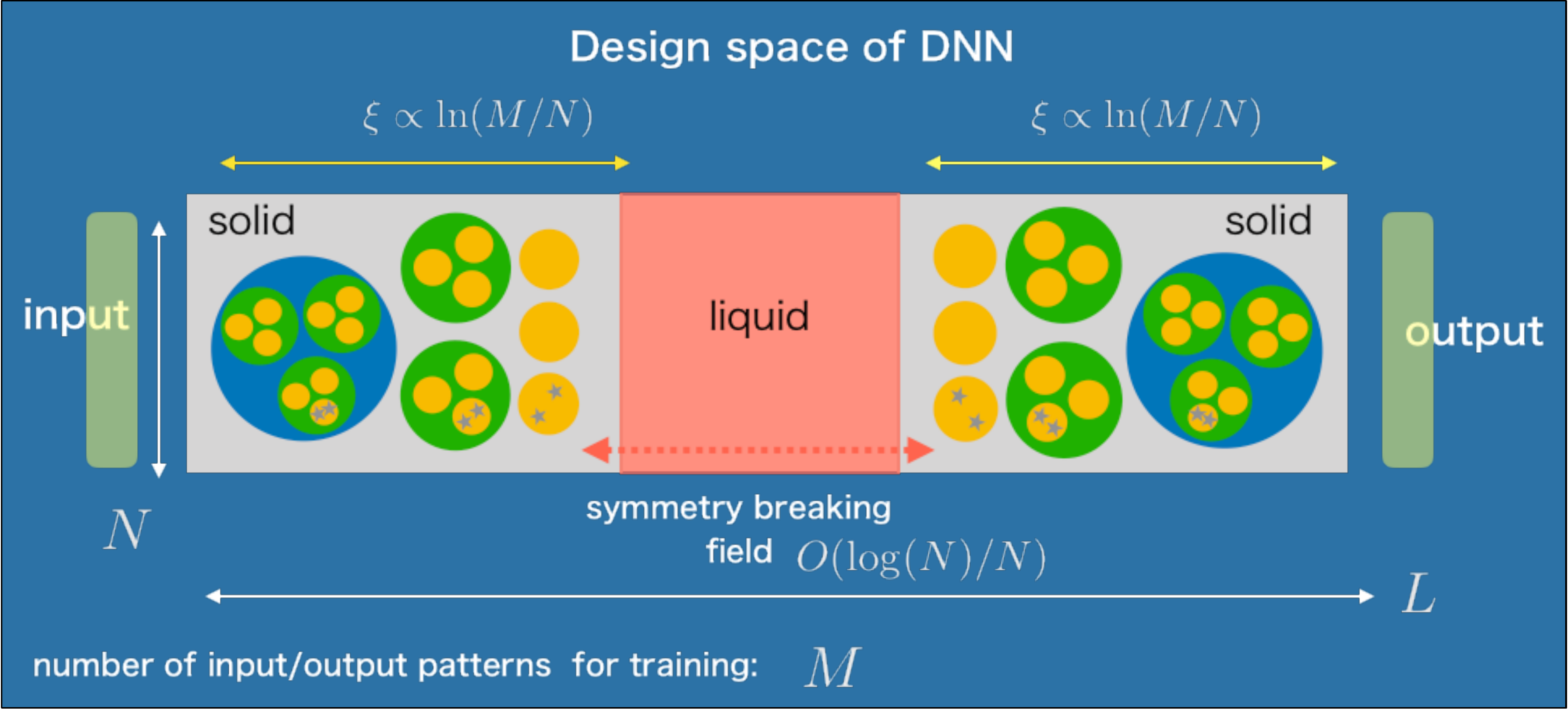}
  \ec
  \caption{Schematic picture of the design space of deep neural network based on the present theory}
   \label{fig_DNN_sandwich}
             \end{figure}

             The non-linear dynamics \eq{eq-perceptron} in random perceptron networks
             is known to be  highly chaotic \cite{sompolinsky1988chaos,poole2016exponential}.
             Among all such random perceptrons, which typically give chaotic dynamics, 
             we considered statistical mechanics on the ensemble of extremely {\it rare} samples
             which happen to meet a large number of externally imposed inputs/outputs boundary conditions.
             Our theory predicts that such a  {\it selection} (learning)
             on the ensemble of chaotic trajectories give rise to 
             a hierarchical clustering of the trajectories (solutions) which evolves in space
             as shown in Fig.~\ref{fig_DNN_sandwich}.
             The presence of the liquid phase in the center is consistent with the chaos.
             The spatial evolution of RSB smoothly connects the free-liquid like center
             and strongly constrained boundaries.

             Imagine that we are monitoring the behavior of multiple machines that are subjected to the same inputs/outputs boundary condition but evolving (learning) independently from each other. 
             Their configurations are represented by 'stars' in Fig.~\ref{fig_DNN_sandwich}.
             Starting from the input layer, we notice that they progressively become more separated
             going deeper into the bulk but they become closer again approaching the output layer.
             The initial part, where different clusters of solutions (machines)
             merge into bigger cluster look like forgetting the detailed differences
             (renormalization)
             and the latter is the reverse: it amplifies mutual differences to produce the desired
             output (label). This picture appears to be consistent with some intuitions gained
             in some studies of machine learning in DNNs \cite{shwartz2017opening}.

             Usually (by definition) chaotic systems are extremely weak against perturbations.
             It is very interesting to ask what happens if selections (learning) come into play.
             Our theory implies that, during learning, a machine can diffuse chaotically
             within the huge continent of solutions (liquid) at the center
             without violating the imposed boundary conditions.
             Larger fluctuation means entropic stability.
             Thus it is not inconceivable that
             the combination of the strong internal chaos and the selection made
             at the boundaries can create a machine whose output is strong against perturbations.
             Our theoretical results suggest this is the case.

             One can view Fig.~\ref{fig_DNN_sandwich} as a picture of the phase space of hard-spheres
             bounded by two walls made by frozen particles. The frozen boundaries
             act like pinning field for the particles and successive layer-by-layer
             glass transitions start from the boundaries as the pressure is increased.
             This is similar to the layer-by-layer growth by
             physical adsorption on substrates \cite{de1978lattice}.   
             As the glass region grows in space,
             the interior of the glass region experiences further glass transitions
             (like the Gardner transition \cite{Ga85}) by which their phase space become split further.
             In the teacher-student setting, one of the glass corresponding to that of the  teacher and the student tries to find it.

   \section{Simulations of learning}
   \label{sec-simulation}

   Now we turn to discuss some numerical simulations
   to examine our theoretical predictions regarding
   the setting of a random constraint satisfaction
   problem with random inputs/outputs at boundaries.


In sec. \ref{sec-replica-random-inputs-outputs} we found theoretically that
the free-energy landscape of the perceptron network
subjected to random constraints on the boundaries
exhibit spatially heterogeneous structure: it is
very complicated close to the boundaries but very simple in the central part.
This naturally implies that learning dynamics is also heterogeneous in space. 

To examine the nature of the learning
dynamics we perform Monte Carlo simulations of
the multi-layer neural network 
with depth $L$, width $N$ and randomly quenched inputs/outputs spins.
The effective Hamiltonian of the system
\eq{eq-effective-hamiltonian}
reads as,
\beq
H=\sum_{\mu}\sum_{\bs}V(r^{\mu}_{\bs})
\qquad 
r^{\mu}_{\bs} \equiv
\sum_{i=1}^{N}\frac{J_{\bs}^{i}}{\sqrt{N}}S^{\mu}_{\bs(i)}S^{\mu}_{\bs}
\label{eq-hamiltonian-soft}
\eeq
For convenience for the simulation, we replace
the hard-core potential \eq{eq-hardcore} by a soft-core potential,
\beq
V(h)=\epsilon h^{2}\theta(-h)
\label{eq-soft-potential}
\eeq
where $\epsilon$ is the unit of energy.
Note that the statistical mechanics of a system
with the soft-core potential becomes the same as the hardcore potential in the zero-temperature limit $k_{\rm B}T/\epsilon \to 0$, where $k_{\rm B}$ is the Boltzmann's constant
and $T$ is the temperature, in the region where all the constraints are satisfied (SAT).

The dynamical variables are the $M$-component vector spins and bonds,
\beqn
S_{\bs}^{\mu}
\qquad &  (\mu=1,2,\ldots,M)  \qquad & (\bs=1,2,\ldots,N(L-1)) \\
J_{\bs}^{i} \qquad& (i=1,2,\ldots,N) \qquad & (\bs=1,2,\ldots,NL) 
\eeqn
(Here we excluded the spins on the boundaries which are fixed.)
Each component of the spins only takes Ising values $\pm 1$
while each of the bonds takes continuous values.
In order to satisfy the normalization condition \eq{eq-J-normalization}
$\sum_{i=1}^{N}(J_{\bs}^{i})^{2}=N$, we assume that
$J_{\bs}$ follows a Gaussian distribution with $0$ mean and variance $1$.
We performed simple Metropolis updates of the dynamical variables
at very low temperatures $T$ to simulate the relaxational dynamics.
In a sense the finite temperature Monte Carlo dynamics
mimic the 'stochastic' nature of the standard Stochastic Gradient Descent (SGD)
algorithms used for training of DNNs \cite{lecun2015deep}.
To propose a new spin configuration for
the Metropolis algorithm, first we 
select a spin component $S_{\bs}^{\mu}$
randomly out of the $N\times L \times M$ possibilities
and then flip it as $S_{\bs}^{\mu} \to -S_{\bs}^{\mu}$.
To update the bond configuration, 
first we 
select a bond $J_{\bs}^{i}$
randomly out of the $N\times L \times N$ possibilities
and then shift its value as,
\beq
J_{\bs}^{i}  \to  \frac{J_{\bs}^{i} + r x}{\sqrt{1+r^{2}}}
\eeq
where $x$ is a random number following the Gaussian distribution
with zero mean and variance $1$. We set $r=0.1$ in our simulations.
Within  $1$ MCS (Monte Carlo Step), the unit step of the Monte Carlo simulation,
we try updates of the spins $N \times L \times M$ times and
updates of the bonds $N \times L \times N$ times.

At first the configurations of the frozen spins on the input 
$l=0$ and the output $l=L$ layers are generated randomly.
The initial configurations of the mobile spins are
bonds are also generated randomly.
Then spins and bonds are updated using the Metropolis algorithm
at a low temperature $T$.
In our simulations we set $k_{\rm B}T/\epsilon=0.015$.
Here we prepare two machines $a$ and $b$ which evolves from the same initial configurations,
common boundary configuratinos for the spins on the boundaries. The two machines are updated
independently by the Monte Calro method using independent series of random numbers. 

We measure the following overlaps between the two machines (replicas),
\beq
Q(t,l)=\frac{1}{N^{2}}\sum_{\bs \in l}\sum_{i=1}^{N}
\overline{(J_{\bs}^{i})^{a}(t)(J_{\bs}^{i})^{b}(t)}
\qquad
q(t,l)=\frac{1}{MN}\sum_{\bs \in l}\sum_{\mu=1}^{M}
\overline{(S_{\bs}^{\mu})^{a}(t)(S_{\bs}^{\mu})^{b}(t)}
\label{eq-c-bond-spin-simulation}
\eeq
where $\bs \in l$ stands for summation over the perceptrons
within the $l$-th layer. \red{Since the two replicas start from the same initial conditions
$q(0,l)=Q(0,l)=1$ and the overlaps decay with time $t$.}
As we noted in sec.~\ref{sec-order-parameters} the system is symmetric under permutations of the perceptrons $\bs$
within the same layer. This does
not matter here as long as we limit ourselves on the time scales where the correlation functions defined above remain positive.
The overline $\overline{\cdots}$ represents the average over different samples:
the realization of the inputs/outputs spins are chosen randomly for each sample.
In the following, we used $60-240$ samples. 

 \begin{figure}[h]
   \bc
   \includegraphics[width=\textwidth]{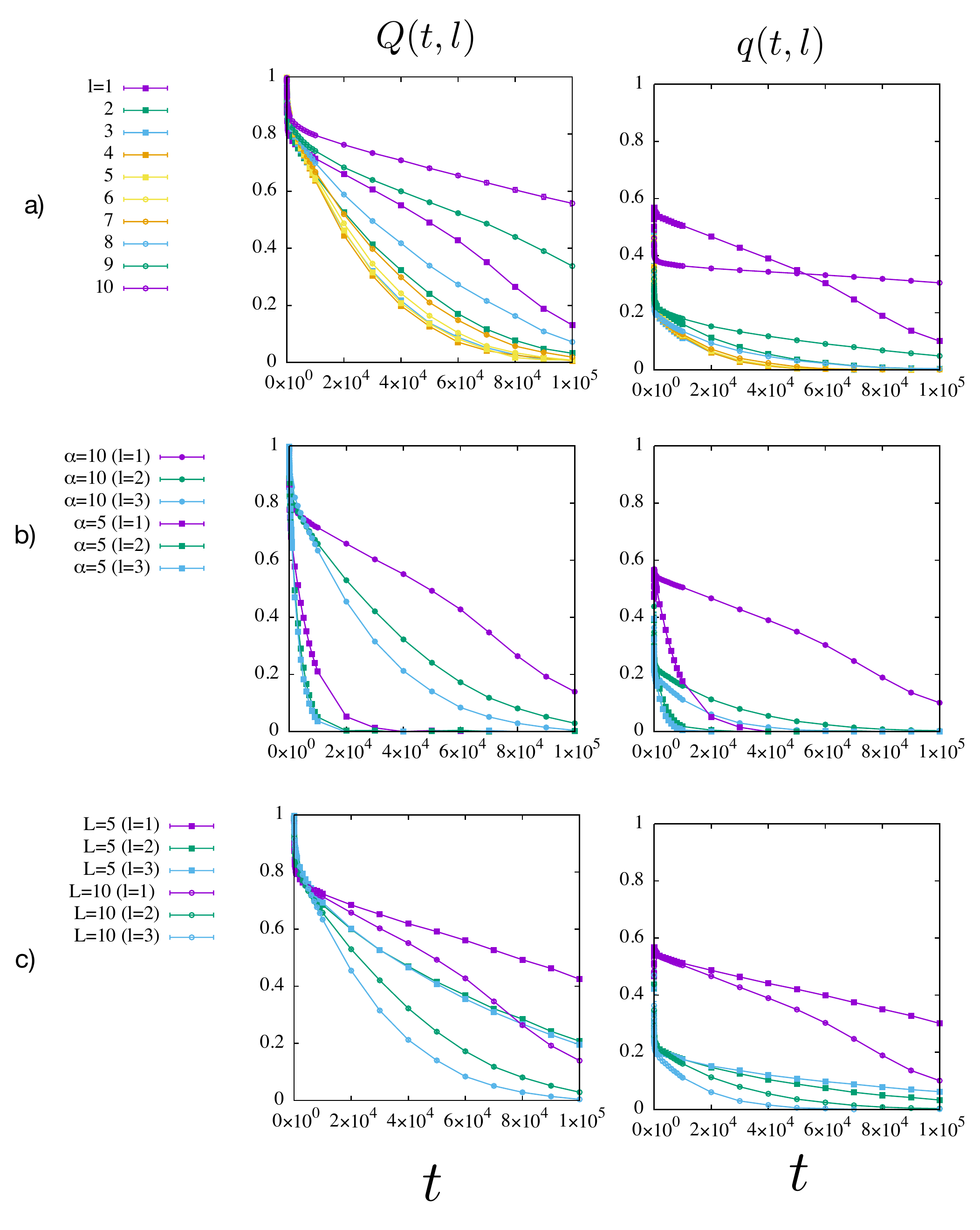}
   \ec
   \caption{
     Relaxation of the replica overlaps of the bonds $Q(t,l)$
     and spins $q(t,l)$. The unit of time is $1$ MCS.
     In the 1st row a) data at different layers $l=1,2,\ldots,10$ ($L=10$,$N=20$,$\alpha=10$) are shown. In the 2nd row b) data for $\alpha=5$ and $10$ ($L=10$,$N=20$) are shown.
     In the 3rd row c) data obtained with different depth $L=5$ and $10$ ($N=20$,$\alpha=10$) are shown. Error bars are smaller than the size of the symbols.
   }
   \label{fig_relaxation}
 \end{figure}

  \begin{figure}[h]
   \bc
   \includegraphics[width=0.9\textwidth]{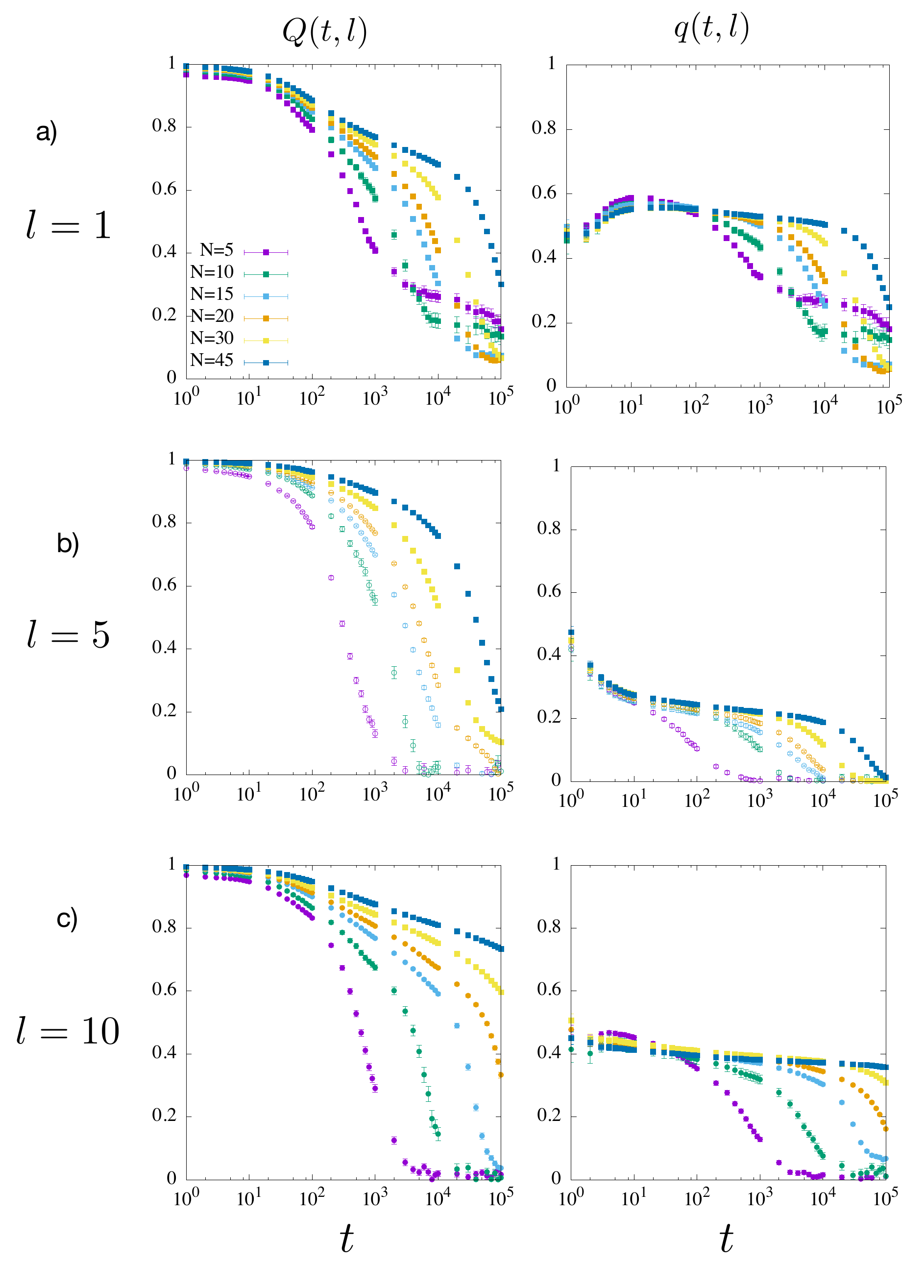}
   \ec
   \caption{
     Relaxation of the replica overlaps of the bonds $Q(t,l)$
     and spins $q(t,l)$ at various width $N$ plotted against logarithmic time.
     The unit of time is $1$ MCS.
     Here $L=10$, $T=0.015$ and $\alpha=5$.
     Data at different layers (a) close to the input $l=1$, (b) at the center $l=5$ and (c) close to the output $l=10$ are shown.
   }
   \label{fig_relaxation_overlap_finiteN_alpha=5}
  \end{figure}

    \begin{figure}[h]
   \bc
   \includegraphics[width=0.5\textwidth]{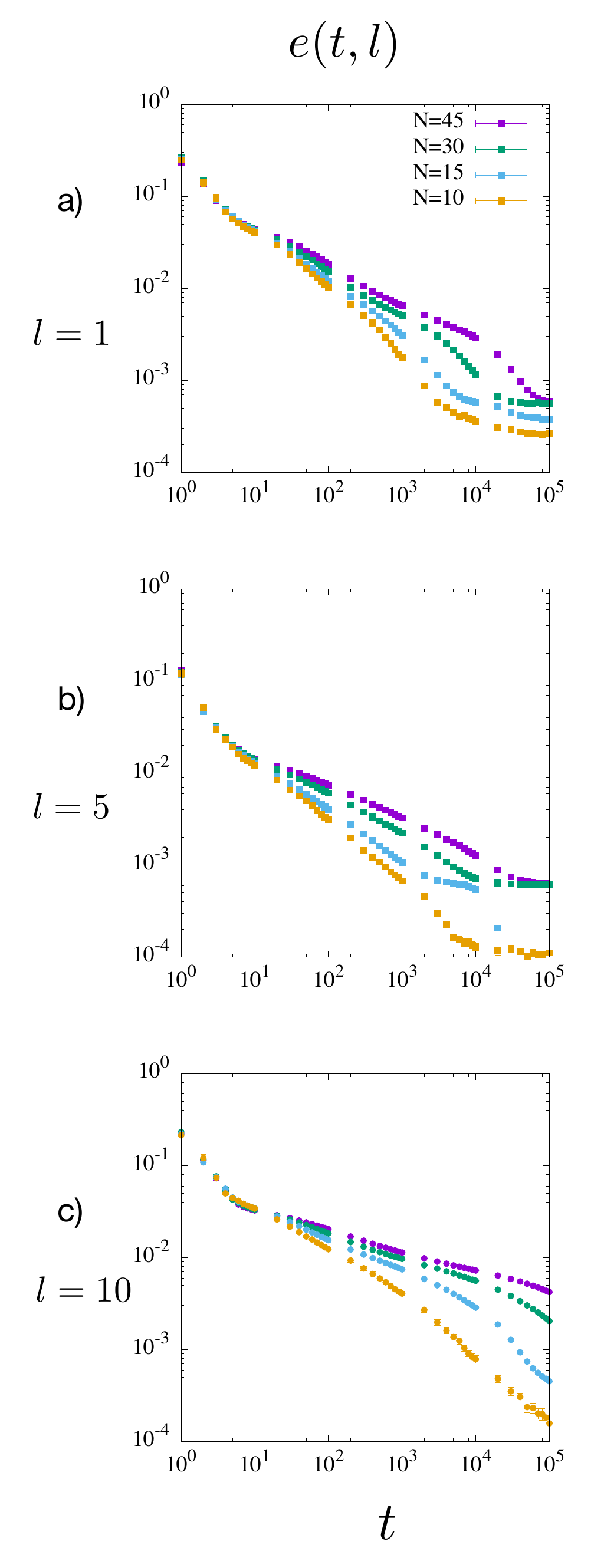}
   \ec
   \caption{
     Relaxation of energy $e(t,l)$ of various width $N$. Here $L=10$, $T=0.015$ and $\alpha=10$.
     The unit of time is $1$ MCS.
     Data at different layers (a) close to the input $l=1$, (b) at the center $l=5$ and (c) close to the output $l=10$ are shown.
     Error bars are smaller than the size of the symbols.
   }
   \label{fig_relaxation_energy}
 \end{figure}

\red{In Fig.~\ref{fig_relaxation} we show data of replica overlap of the spins and bonds
plotted against linear time $t$. Here the width is $N=20$. 
From the panels in the row a), it can be seen that the dynamics is
actually faster in the center and slower close to the boundaries as we expected.
From the panels in row b), it can be seen that the dynamics
become slower as the strength of the constraints $\alpha$ increases. 
This is consistent with the theoretical expectation
that the system becomes more glassy with larger $\alpha$.
From the panels in row c), it can be seen that 
relaxation is apparently faster in the deeper system.
Interestingly this happens even in the layer just next to the boundaries.
Presumably, this implies that the deeper system is more flexible in the center
and the fast dynamics at the center assists the relaxation
of the whole system.}

\red{In the data shown above, the overlap of the spins and both tend to decay down to $0$
  at the long times. This implies the finite $N$ system is in the liquid phase everywhere
  at long enough time scales, as it should be.
  In Fig.~\ref{fig_relaxation_overlap_finiteN_alpha=5}
we show data of the overlaps of various width $N$ against {\it logarithmic} time so that
we can also see dynamics at shorter time scales.
Apparently system with smaller width $N$ decay faster suggesting that there is a
finite relaxation time $\tau(N,\alpha,l)$ which increases with the width $N$
(increases also with $\alpha$ and becomes smallest at the center $l \sim L/2$).
It can bee seen that at short enough time scales $t \ll \tau(N,\alpha,l)$, the data do not
depend on width $N$ suggesting there is a limiting curve $N \to \infty$ with fixed $\alpha$.
This is consistent with our theory in which the parameter $\alpha$ is the essential control
parameter. Some of such limitting curves suggest complex dynamics with plateaus which are signatures of glassy dynamics \cite{berthier2011theoretical}.
Overlap of the bonds appear to be larger than those of spins.
These features are consistent with our theory.
We consider that the slow dynamics at the shorter time scales
$t \ll \tau(N,\alpha,L)$ reflect the complex free-landscape and the
truncation of the slow dynamics at longer time scales
is a finite width $N$ effect.
It is interesting to note that somewhat similar truncation of the slow
dynamics has been observed in a study of SGD dynamics in DNNs \cite{baity2019comparing}.
}

\red{Finally, in Fig.~\ref{fig_relaxation_energy}, we display the relaxation of energy
$e(t,l)=E(t,l)/N$ at each layer $l$. Here $E(t,l)$ is the contribution of $l$ th layer
to the energy (see \eq{eq-hamiltonian-soft}-\eq{eq-soft-potential}) at time $t$ (MCS).
It can be seen again that relaxation is slower closer to the boundaries.
The data also suggest that there is a $N \to \infty$ limit curve with fixed $\alpha$.
Note also that the behavior of the system is not symmetric to the
exchange of input and output sides. The asymmetry becomes
stronger closer to the boundaries as one naturally anticipates.
}

To summarize the numerical observation of the relaxations
is qualitatively consistent with the theoretical prediction
which implies spatially heterogeneous dynamics.

\clearpage    
    \section{Conclusion and outlook}
    \label{sec-conclusions}

    \subsection{Conclusion}

In the present paper, we constructed a statistical mechanical theory for the solution space of a deep perceptron network of depth $L$ and width $N$ subjected to $M=\alpha N$ patterns of data
  using the internal representation,  based on the replica method
  in the limit $N,M \to \infty$ with fixed $\alpha$.
  We studied two scenarios :
  \begin{itemize}
  \item[1)] Random inputs/outputs, which 
  is a random constraint satisfaction problem
\item[2)] Teacher-student setting, which is a statistical inference problem.
  \end{itemize}
In addition, we performed simulations to examine the theoretical predictions
on 1).

  The main outcome of the theory is the prediction of
  the strongly heterogeneous spatial profile of states inside the layered network :
  1)  'glass-liquid-glass' in the case of the random inputs/outputs
  2) 'crystal-liquid-crystal' in the case of the teacher-student setting.
\red{We find $\alpha=M/N$ is the key parameter which plays the role similar to the
  the inverse temperature in condensed matters.}
  The thickness of the glass/crystal phase grows as $\xi \propto \ln \alpha$.
  This implies exponential growth of the storage capacity of DNN with the depth $\alpha_{\rm j}(L) \propto e^{{\rm const} L}$
  for typical instances. 
Moreover, in the case of setting 1) random inputs/outputs,
even the pattern of the replica symmetry breaking
(RSB) varies in space: it is most complex close to the boundaries
with $k$(+continuous)-RSB, which becomes $k-1$(+continuous)-RSB
in the next layer, ... down to a replica symmetric ($0$ RSB) state in the central part. The hierarchical structures
in different layers are synchronized.
We argued that in the 2)  teacher-student setting,
the small positive overlap can remain in the liquid phase
as a finite $N$ correction and plays the role of
symmetry-breaking field.

There are some weak points in our theory which must be clarified by further works,
\begin{itemize}
\item Our theory assumes the wide limit
$N \to \infty$ (with $\alpha=M/N$ fixed)
while real networks have {\it finite} width $N$ so that
'phase transitions' we found here become at most just crossovers.
Nonetheless, we believe our results still provide useful guidelines to understand real DNNs.
\item \red{Technically our theory relies on the tree-approximation
  which does not take into account the 1-dimensional inter-layer
  fluctuations along the $z$-axis (See Fig.~\ref{fig_multilayer_network})
  faithfully.
  We believe that the qualitative nature of the system in the limit $N \to \infty$  do not change by loop-corrections much as the Ginzburg-Landau theory in $1+\infty$ dimension
  (here $1$ is for the $z$-axis  and $\infty$ dimension for the transverse directions) do not change the mean-field
  nature. 
  However consideration of loop-corrections would improve the theory
  at small scales
  especially close to the boundaries where spatial heterogeneity is strongest.
  It is also important in systems with smaller width $N$ where we 
  expects the effects of inter-layer fluctuations become larger.
  Indeed if we take $L \to \infty$ limit with finite width $N$,
  the system becomes truly an one dimensional system.
}

\item Our theory is based on order parameters.
  This is a strong point but the theory itself does not
  answer what plays the role of symmetry breaking field
conjugated to the order parameters in practice.
This is a particularly important open question in the inference problem (teacher-student setting).
\end{itemize}

For 1), we also performed some simulations of
the relaxational dynamics.
 \red{We found crossover from the complex glassy slow dynamics to rapid decay at longer time scales.
  For fixed $\alpha=N/M$, the crossover time increases with the width $N$.
  As the theory suggests, the glassy dynamics is controlled by the strength of $\alpha$
  and spatially heterogeneous.
It is faster and simpler in the center and slower and more complex closer to the boundaries
in agreement with the theory.
}
\red{We leave numerical investigation for the teacher-student setting 2) for future works.}

To summarize, both the theory and the simulation suggest spatially heterogeneous free-energy landscape in DNNs
which is controlled by the parameter $\alpha=M/N$ (See Fig.~\ref{fig_DNN_sandwich}).
We speculate that this is responsible for the efficiency of DNNs in three respects:
\begin{itemize}
\item[a)] The presence of the liquid phase in the center 
  may facilitate the equilibration (learning) of the whole system.
  This is an interesting point which deserves further studies.
  In this respect, it is interesting to note that the so-called ultra-stable glasses
  \cite{swallen2007organic,leonard2010macroscopic,perez2014suppression} are created by vapor deposition
  which allows rapid equilibration at layers close to the surface.
  As we noted in sec.~\ref{sec-summary-theory}, the analogy to the hard-spheres is suggestive.
  One could also think about the analogy with the replica-exchange Monte Carlo method \cite{hukushima1996exchange} which dramatically accelerates the equilibration of complex systems.
\item[ b)] Despite the over-parametrization the system
  may still generalize because of the crystalline phase close to the boundaries and weak bias field in the liquid phase. However, how the bias field can be prepared in practice is an important open question.

\item[c)] Hierarchical free-energy landscapes with ultra-metric structure has been known in glass physics since the discovery of replica symmetry breaking \cite{mezard1984nature,MPV87,rammal1986ultrametricity}.  Our theory suggests that it evolves in space in the DNN, during learning, progressively from the complex to simple ones going deeper into the
  interior from both the input and output boundaries.
  \red{Probably the spatial 'renormalization' of the hierarchical clustering and the presence of the
    liquid phase at the center stabilizes the system against external perturbations or incompleteness of equilibration and contributes positively to the generalization ability of DNNs. As discussed in sec. \ref{sec-fullRSB-boundary}, 
   it would also be very interesting to study  implications
    for unsupervised learning including in particular hierarchical data clustering\cite{johnson1967hierarchical}.}
  \end{itemize}

\subsection{Outlook}

There are many possibilities for further investigations including the following.
First of all, it will be very interesting to perform extensive numerical
simulations with realistic algorithms and realistic large-scale data to examine our predictions.
Second, a more detailed theoretical/numerical analysis of the remnant bias field in the liquid phase
is needed.
\red{Third, the storage capacity and critical properties at jamming (SAT/UNSAT transition)
  \cite{franz2016simplest,franz2017universality,yoshino2018,baity2019comparing,franz2019jamming,geiger2019jamming} can be studied in detail by analyzing the regime $\xi \gg L$.
}
Lastly, it will be interesting to extend the present work to
other cases regarding the activation function, architecture,
and learning methods.

Our system may be view as a $1+\infty$ dimensional glass which is an interesting playground to analyze spatial heterogeneity
with mean-field theoretical approaches \cite{de1978lattice,franz1994interfaces,franz2007analytic,ikeda2015one}.
 The present work may also have implications on various complex systems with spatial
 heterogeneity, including ultra-stable glass \cite{swallen2007organic,leonard2010macroscopic,perez2014suppression} mentioned above, gene regulatory networks \cite{wagner1994evolution,wagner1996does,nagata2019emergence} which are often viewed like perceptrons  and allosteric systems \cite{monod1965nature,rocks2017designing,yan2017architecture}. The central liquid region may be related to the idea of  {\it neutral space} \cite{wagner1994evolution,wagner1996does} which is considered as responsible for robustness of biological systems. 

  \section*{Acknowledgments}
  We thank
  Giulio Biroli,
  Silvio Franz,
  Koji Hashimoto,
  Sungmin  Hwang,
  Kyogo Kawaguchi,
  Macoto Kikuchi,
  Kota Mitsumoto,
  Tomoyuki Obuchi,
  Haruki Okazaki,
  Akinori Tanaka,
  Pierfrancesco Urbani,
  Takaki Yamamoto,
  Lenka Zdeborov{\'a}
  and Francesco Zamponi
  for useful discussions.

  Numerical analysis in this work has been done using the supercomputer systems OCTOPUS and SX-ACE at the Cybermedia Center, Osaka University. The author thanks the Simons collobration on ``cracking the glass problem'' for opportunities of stimulating discussions. The author thanks the Yukawa Institute for Theoretical Physics at Nyoto University for discussions during the YITP workshop YITP-W-19-18 "Deep Learning and Physics 2019".

{\bf Founding information} 
This work was supported by KANENHI (No. 19H01812) from MEXT, Japan.  

\clearpage

\bibliography{paper_v5.1_scipost.bbl}

 \clearpage

 \begin{appendix}

   \section{Replicated free-energy}
   \label{appendix-replicated-free-energy}

The replicated phase space volume (the Gardner volume) can be written as,
\beqn
 V^{n}\left({\bf S}_{0},{\bf S}_{L}\right)
&=&e^{N M {\cal S}_{n}\left({\bf S}_{0},{\bf S}_{l}\right)} \nonumber \\
& = &\prod_{a=1}^{n}
\left(\prod_{\bs} {\rm Tr}_{{\bf J}^{a}_{\bs}}\right)
\left( \prod_{\bs\backslash {\rm output}}{\rm Tr}_{{\bf S}^{a}_{\bs}}  \right)
\left\{ \prod_{\mu,\bs,a}
\int \frac{d\eta_{\mu,\bs,a}}{\sqrt{2\pi}}
W_{\eta_{\mu,\bs,a}}e^{i \eta_{\mu,\bs,a} (r^{\mu}_{\bs})^{a}}\right\} \nonumber \\
&=& \left ( \prod_{\mu,\bs,a}
\int \frac{d\eta_{\mu,\bs,a}}{\sqrt{2\pi}}
W_{\eta_{\mu,\bs,a}}\right)
\left(\prod_{\bs,a} {\rm Tr}_{{\bf J}^{a}_{\bs}}\right)
\left( \prod_{\bs\backslash {\rm output},a}{\rm Tr}_{{\bf S}^{a}_{\bs}}  \right)
\nonumber \\
&& \hspace*{2cm}\prod_{\mu,\bs,a}
e^{i \eta_{\mu,\bs,a}
(S^{\nu}_{\bs})^{a}
  \sum_{i=1}^{N}\frac{(J_{\bs}^{i})^{a}}{\sqrt{N}}
  (S^{\nu}_{\bs(i)})^{a}} \qquad
\label{eq-replicated-gardner-volume-appendix}
\eeqn
where we introduced a Fourier representation,
\beq
e^{-\beta V(r)}=
\int \frac{d\eta}{\sqrt{2\pi}}
W_{\eta}e^{-i \eta r}
\eeq

In the following we derive the free-energy  functional
of the replicated system
starting from \eq{eq-replicated-gardner-volume-appendix},
following similar steps as
in \cite{yoshino2018}.

\subsection{Basic strategy}

Before going to the details of the computations, let us sketch the basic strategy to extract properties of glassy phases using the replica approach
in the present work as well as \cite{yoshino2018}. 
Very importantly, this applies to systems without the quenched disorder.
Actually, this strategy lies behind the replica approach to
structural glasses \cite{FP95,Mo95,MP99,parisi2020theory,yoshino2018translational}. This is an important point for our present problem which is essentially disorder-free except for the boundaries.

\subsubsection{Explicit replica symmetry breaking}
\label{subsubsec-explicit-rsb}

For simplicity,
suppose that we have a generic system which consists of $N$ degrees of freedom $\{x\}=(x_{1},x_{2},\ldots,x_{N})$
whose Hamiltonian is $H[\{x\}]$, which can be with/without the quenched disorder. 
Let us introduce $n$ replicas $a=1,2,\ldots,n$ and a replicated Hamiltonian,
\beq
H_{n}[\hat{\epsilon}]=\sum_{a=1}^{n}H[\{x_{i}^{a}\}]-\sum_{a < b}\epsilon_{ab}\sum_{i=1}^{N} x^{a}_{i}x^{b}_{i}.
\label{eq-replica-under-symmetry-breaking-field}
\eeq
Here we introduced the 2nd term on the r.h.s. which represts an artificial
attractive coupling $\epsilon_{ab} >0$ between replicas. The field $\epsilon_{ab}$
explicitely breaks
the replica symmetry, i.e. the permutation symmetry of replica index.
The free-energy of the replicated system can be defined as,
\beq
-\beta G[\hat{\epsilon}]=\ln \prod_{a=1}^{n} \prod_{i=1}^{N} {\rm Tr}_{x_{i}^{a}}
e^{-\beta H_{n}[\hat{\epsilon}]}.
\eeq
This allows us to evaluate the overlap between the replicas,
\beq
q_{ab}=\frac{1}{N} \sum_{i=1}^{N}\langle x^{a}_{i}x^{b}_{i} \rangle_{\epsilon}
=-\frac{1}{N} \frac{\partial G_{\rm \epsilon}}{\epsilon_{ab}}
\eeq
We are interested with,
\beq
\lim_{\hat{\epsilon} \to 0}\lim_{N \to \infty} q_{ab}
\label{eq-spontaneous-q}
\eeq
and consider that this is the glass order parameter of the system. This is the idea of explicit
replica symmetry breaking (RSB)
by Parisi and Virasoro \cite{parisi1989mechanism}. Similarly to the magnetic field $h$ for magnetization in ferromagnetic systems, the field $\epsilon_{ab}$ is conjugated to the glass order parameter $q_{ab}$ and plays the role of symmetry breaking field. One can define a 'glass' susceptibility,
\beq
\chi_{ab,cd} \equiv -\frac{1}{N} \frac{\partial}{\partial \epsilon_{ab}
  \partial \epsilon_{cd}} \beta G[\hat{\epsilon}]=
\frac{\partial q_{ab}}{\partial \epsilon_{cd}}= 
\frac{1}{N}\sum_{i,j=1}^{N}
\left[
  \langle x_i^{a}x_i^{b}x_j^{c}x_j^{d} \rangle_{\epsilon}
  - \langle x_i^{a}x_i^{b}\rangle_{\epsilon}\langle x_j^{c}x_j^{d} \rangle_{\epsilon}
  \right]
\label{eq-glass-susceptibility}
\eeq
Instability toward spontaneous replica symmetry breaking 
may accompany divergence of the glass susceptibility at $\epsilon=0$.

Let us then consider the Legendre transform of the free-energy,
\beq
-\beta F[\hat{q}]=-\beta G[\hat{\epsilon}^{*}]-N \sum_{a< b} \epsilon^{*}_{ab}q_{ab}
\label{eq-G-to-F}
\eeq
where $\hat{\epsilon}^{*}=\hat{\epsilon}^{*}[\hat{q}]$ is defined such that,
\beq
\frac{1}{N}\left. \frac{\partial}{\partial \epsilon_{ab}} (-\beta G[\hat{\epsilon}])\right |_{\hat\epsilon=\hat\epsilon^{*}[\hat{q}]}=q_{ab}.
\label{eq-epsilon-star}
\eeq
The inverse of the Legendre transform is,
\beq
-\beta G[\hat{\epsilon}]=-\beta F[\hat{q}^{*}]+N \sum_{a< b} \epsilon_{ab}q^{*}_{ab}
\eeq
where $\hat{q}^{*}=\hat{q}^{*}[\hat{\epsilon}]$ is defined such that,
\beq
\frac{1}{N}\left. \frac{\partial}{\partial q_{ab}} (-\beta F[\hat{q}])\right |_{\hat{q}=\hat{q}^{*}[\hat{\epsilon}]}=-\epsilon_{ab}.
\label{eq-q-star}
\eeq
The last expression tells us that the order parameter of our intest, which detects
the spontaneous RSB \eq{eq-spontaneous-q},
can be obtained by minimizing the free-energy $F[\hat{q}]$ which yields $\epsilon_{ab}=0$.
Related to \eq{eq-glass-susceptibility} is the Hessian matrix,
\beq
H_{ab,cd}\equiv
\frac{1}{N}\left. \frac{\partial^{2}}{\partial q_{ab}\partial q_{cd}} (\beta F[\hat{q}])\right |_{\hat{q}} =(\chi^{-1})_{ab,cd}
\label{eq-hessian}
\eeq
The divergence of the glass susceptibility \eq{eq-glass-susceptibility}
imply vanishig of the eigen value(s) of the Hessian matrix \cite{de1978stability}. Thermodynamic stability implies $\chi_{ab,cd} < \infty$ or positive
(semi-)definiteness of the eigenvalues of the Hessian matrix.

\subsubsection{Ergodicity breaking}
\label{subsubsec-ergodicity-breaking}

If the field does not depend on the replica indexes $\epsilon_{ab}=\epsilon$, we are not breaking
the replica symmetry. But the replica symmetric (RS) field
can be used at least to detect the {\it ergodicity breaking} where
the Edwards-Anderson (EA) order parameter \cite{edwards1975theory},
\beq
q_{\rm EA}=\frac{1}{N}\sum_{i=1}^{N} \langle x_{i} \rangle^{2}
\eeq
becomes non-zero. Here $\langle \ldots \rangle$ represents an appropriate thermal average.
For example, in a model with $3$-body interaction $H=-J/N^{2} \sum_{i,j,k=1}^{N}x_{i}x_{j}x_{k}$,
the liquid phase where $q_{\rm EA}=0$ is realized at high enough
temperatures while crystalline or glassy phases where $q_{\rm EA}>0$ emerge at lower temperatures
\cite{franz2001ferromagnet,yoshino2018}.

In the context of relaxational dynamics, the EA order parameter can be considered as the long-time limit
of the time autocorrelation function \cite{edwards1975theory},
\beq
q_{\rm EA}=\lim_{t \to \infty}C(t)\qquad C(t) \equiv \frac{1}{N} \sum_{i=1}^{N}\langle x_{i}(0)x_{i}(t)\rangle.
\eeq
In the liquid phase the auto-correlation function decays down to $0$ after finite relaxation time.
The latter diverges at the transition leading to $q_{\rm EA} >0$. Thus the EA order parameter
detects the Ergodicity breaking (either due to crystalline or glass transitions).
Note that the RSB discussed previously in sec. \ref{subsubsec-explicit-rsb}
automatically also means an ergodicity breaking,
but of a more complicated version involving the hierarchical organization of
relaxations \cite{sompolinsky1982relaxational,cugliandolo1993analytical,franz1994off,cugliandolo1994out,franz1998measuring}.

In some cases, like a model with $2$-body interaction $H=-J/N \sum_{i,j,k=1}^{N}x_{i}x_{j}$,
the system is symmetric under global 'spin flip'
$x_{i} \to -x_{i}$ for $\forall i$, the perturbation \eq{eq-replica-under-symmetry-breaking-field}
with the RS field $\epsilon_{ab}=\epsilon$
then breaks this symmetry. In our present problem \eq{eq-effective-hamiltonian}, the 'spins' have this symmetry.
One observes that the system is no longer invariant under global flip 
in one replica, say $a$, $x^{a}_{i} \to -x^{a}_{i}$  for $\forall i$. Thus the role played by
the RS field $\epsilon_{ab}=\epsilon$, in this case,
is like the magnetic field conjugated to the magnetization
which is the order parameter for usual ferromagnets.

\subsubsection{What is the symmetry breaking field?}
\label{subsubsec-real-symmetrybreaking-field}

In the theoretical formulation, we introduced conveniently the symmetry-breaking fields
$\epsilon_{ab}$ as \eq{eq-replica-under-symmetry-breaking-field}.
But we have to ask ourselves what plays the role of the somewhat fictitious field in reality.
The perturbations like \eq{eq-replica-under-symmetry-breaking-field}
can be introduced by considering 'random pinning fields' \cite{parisi1989mechanism,monasson1995structural}.
Some sorts of weak random pinning fields may exist in nature. But what about computers?
In the context of machine learning, the role of symmetry breaking field may be played by
1) choices of boundary conditions (inputs/outputs data) 2) choices of initial condition for learning.

\subsubsection{Plefka expansion}
\label{sec-plefka}

Now our task is to compute the free-energy $F[\hat{q}]$ defined in \eq{eq-G-to-F}.
To this end, we will follow the idea of
Plefka expansion \cite{plefka1982convergence}. The computations presented in the following sections follow this strategy.

Suppose that the effect of the interactions between the dynamical variables $x_{i}$ $(i=1,2,\ldots,N)$
can be treated perturbatively which enable the following decompositions,
\beq
F=F_{0}+\lambda F_{1}+\ldots \qquad G=G_{0}+\lambda G_{1} +\ldots
\qquad \epsilon_{ab}=(\epsilon_{0})_{ab}+\lambda (\epsilon_{1})_{ab}+\ldots
\label{eq-plefka}
\eeq
Here the quantities with suffix $0$ repsent those which are present in the absence of interactions
(like the ideal gass free-energy) and those with suffix $1$ repsent those due to interactions.
Here we omitted the higher-order terms.
The parameter $\lambda$, which is introduced to organize a perturbation theory, is put back to $\lambda=1$ in the end.

The Legendre transform \eq{eq-G-to-F} becomes, at $O(\lambda^{0})$,
\beq
-\beta F_{0}[\hat{q}]=-\beta G_{0}[\hat{\epsilon}_{0}^{*}]-N \sum_{a< b} (\epsilon^{*}_{0})_{ab}q_{ab}
\eeq
where $(\epsilon^{*}_{0})_{ab}$ is defined such that,
\beq
\frac{1}{N}\left. \frac{\partial}{\partial \epsilon_{ab}} (-\beta G_{0}[\hat{\epsilon}])\right
|_{\hat\epsilon=\hat\epsilon_{0}^{*}[\hat{q}]}=q_{ab}.
\label{eq-epsilon-star-0}
\eeq
Then at $O(\lambda)$ we find,
\beqn
 -\beta F_{1}[\hat{q}]&&=-\beta G_{1}[\hat\epsilon_{0}^{*}[\hat{q}]]
+\sum_{a< b} \left. \frac{\partial}{\partial \epsilon_{ab}}
(-\beta  G_{0}[\hat\epsilon])\right|_{\hat\epsilon=\hat\epsilon^{*}_{0}[\hat{q}]}(\epsilon^{*}_{1})_{ab}
-N \sum_{a< b} (\epsilon^{*}_{1})_{ab}q_{ab} \nonumber \\
&& =-\beta G_{1}[{\hat\epsilon_{0}^{*}[\hat{q}]]}
\label{eq-G-1}
\eeqn
In the 2nd equation we used \eq{eq-epsilon-star-0}. Minimization of the
free-energy $F[\hat{q}]$ (see \eq{eq-q-star}) implies $(\epsilon^{*}_{0})_{ab}=-\lambda (\epsilon^{*}_{1})_{ab}$ up to this order.

\red{If higher order terms $O(\lambda^{2})$ in the expansion
  \eq{eq-plefka} vanish in $N \to \infty$ limit, the treatment described above is
  sufficient. This happens in the derivation of the
the exact free-energy functional of a family of glassy spin models
\cite{yoshino2018} which include the family of p-spin (Ising/spherical) mean-field spin-glass models (with or without the quenched disorder) (see section 6 of \cite{yoshino2018}),
glassy hard-spheres \cite{parisi2020theory} in large-dimensional limit
and aspherical particles \cite{yoshino2018translational} in large-dimensional limit.
Unfortunately, in the present system with the layered geometry,
we will find that $O(\lambda^{3})$ do not vanish because of
the loop effects across different layers (see Fig.~\ref{fig_loop}).
To make an analytical progress we
invoke a tree-approximation which neglects contributions of such loops.
}

\subsection{Evaluation of the entropic part of the free-energy}

We introduce 'local' order parameters \cite{yoshino2018}, for each perceptron $\bs$,
\beq
Q_{ab,\bs}=\frac{1}{N}\sum_{i=1}^{N}(J_{\bs}^{i})^{a}(J_{\bs}^{i})^{b}
\qquad q_{ab,\bs}=\frac{1}{M}\sum_{\mu=1}^{M}(S_{\bs}^{\mu})^{a}(S_{\bs}^{\mu})^{b}
\label{eq-qab-Qab-def}
\eeq
through the identities,
\beqn
1&=&
\int_{-\infty}^{\infty}\int_{-i \infty}^{i \infty}
    \prod_{a <  b} \left(\frac{N}{2\pi}\right) dQ_{ab,\bs}d\epsilon_{ab,\bs}
e^{N \sum_{a < b} \epsilon_{ab,\bs}\left(Q_{ab,\bs}-N^{-1}\sum_{i=1}^{N}(J_{\bs}^{i})^{a}(J_{\bs}^{i})^{b}\right)}\nonumber \\
1&=&
\int_{-\infty}^{\infty}\int_{-i\infty}^{i\infty}
\prod_{a < b} \left(\frac{M}{2\pi}\right)dq_{ab,\bs}d\varepsilon_{ab,\bs}
e^{ M\sum_{a < b} \varepsilon_{ab,\bs}\left(q_{ab,\bs}-M^{-1}\sum_{\mu=1}^{M}(S_{\bs}^{\mu})^{a}(S_{\bs}^{\mu})^{b}\right)}
\eeqn
by which we can write the summation over the configurations
of the bonds and spins of each replica
which appear in 
\eq{eq-replicated-gardner-volume-appendix}
as,
\beqn
\prod_{a}{\rm Tr}_{{\bf J}^{a}_{\bs}} \ldots &=&
\left(   \prod_{a < b}  \inti d(Q_{ab,\bs}) \right) e^{N s_{\rm ent,bond}[\hat{Q}_{\bs}]}
\prod_{i=1}^{N} \langle \cdots \rangle_{J_{\bs}^{i}}
\label{eq-trace-bond}
\\
\prod_{a}    {\rm Tr}_{{\bf S}^{a}_{\bs}} \ldots &=&
\left(   \prod_{a < b}  \inti d(q_{ab,\bs}) \right) e^{M s_{\rm ent,spin}[\hat{q}_{\bs}]}
\prod_{\mu=1}^{M} \langle \cdots \rangle_{S_{\bs}^{\mu}}
\label{eq-trace-spin}
\eeqn
where we have performed integrations over $\epsilon_{ab}$ and $\varepsilon_{ab}$ by the saddle point method assuming $N \gg 1$ and $M \gg 1$. We dropped irrelevant prefactors.
\red{In \eq{eq-trace-bond} and  \eq{eq-trace-spin}, in the products $\prod_{i=1}^{N} \langle \cdots \rangle_{J^{i}}$ 
  and $\prod_{\mu=1}^{M} \langle \cdots \rangle_{S^{\mu}}$, the symbol $\cdots$ refere to quantities factorized in terms of
  $i$ and $\mu$.
Note that \eq{eq-qab-Qab-def}-\eq{eq-trace-spin} are defined for each perceptron $\bs$, which are assumed to be
independent from each other following the prescription formulated in sec. \ref{sec-plefka}.
Thus in the following, we dropp the subscript $\bs$ for simplicity.}

For the trace over the configurations of bonds, we find using \eq{eq-bond-trace}, 
     \beqn
&&  s_{\rm ent,bond}[\hat{Q}]=\frac{1}{2}\sum_{a,b}
     \epsilon^{*}_{ab}Q_{ab}+\ln \prod_{c=1}^{n}
     \int_{-\infty}^{\infty}dJ^{c}
  e^{-\frac{1}{2}\sum_{a,b}\epsilon^{*}_{ab}J^{a}J^{b}}
\\
  &&  \langle \cdots \rangle_{J^{i}}= 
\frac{\prod_{c=1}^{n}
     \int_{-\infty}^{\infty}d(J^{i})^{c}
  e^{-\frac{1}{2}\sum_{a,b}\epsilon_{ab}^{*} (J^{i})^{a} (J^{i})^{b}} \cdots
  }{
  \prod_{c=1}^{n}
       \int_{-\infty}^{\infty}d(J^{i})^{c}
       e^{-\frac{1}{2}\sum_{a,b}\epsilon_{ab}^{*} (J^{i})^{a} (J^{i})^{b}} }
\label{eq-s-ent-bond-replica}
\eeqn
where we introduced $Q_{aa}=1$ and $\epsilon_{aa}=\lambda_{a}$ to
include the integral \eq{eq-bond-trace} which 
enforces the spherical constraint \eq{eq-J-normalization}.
Simiarly, for the trace over the spin configuration we find using
\eq{eq-spin-trace},
\beqn
      &&  s_{\rm ent,spin}[\hat{q}]=\frac{1}{2}\sum_{a,b}
\varepsilon^{*}_{ab}q_{ab}+\ln \prod_{c=1}^{n}
\sum_{S^{c}=\pm 1}
  e^{-\frac{1}{2}\sum_{a,b}\varepsilon^{*}_{ab}S^{a}S^{b}}
  \\
  &&
    \langle \cdots \rangle_{S^{\mu}}= 
    \frac{\prod_{c=1}^{n}
\sum_{(S^{\mu})^{c}=\pm 1}
  e^{-\frac{1}{2}\sum_{a,b}\varepsilon_{ab}^{*} (S^{\mu})^{a} (S^{\mu})^{b}} \cdots
  }{
      \prod_{c=1}^{n}
      \sum_{(S^{\mu})^{c}=\pm 1}
      e^{-\frac{1}{2}\sum_{a,b}\varepsilon_{ab}^{*} (S^{\mu})^{a} (S^{\mu})^{b}} 
    }
    \label{eq-s-ent-spin-replica}
    \eeqn
    \red{where we introduced $\varepsilon_{aa}=0$.}
The saddle points $\epsilon^{*}_{ab}=\epsilon^{*}_{ab}(\hat{Q})$
and $\varepsilon^{*}_{ab}=\varepsilon^{*}_{ab}(\hat{q})$
are determined by
\beqn
   Q_{ab}=\left. 
   \frac{\prod_{c}\Tr_{J^{c}} e^{-\sum_{a < b}\epsilon_{ab} J^{a}J^{b}} J^{a}J^{b}}{\prod_{c}\Tr_{J^{c}} e^{-\sum_{a<b}\epsilon_{ab} J^{a}J^{b}}}
   \right |_{\epsilon_{ab}=\epsilon^{*}_{ab}(\hat{Q})} 
   \qquad
   q_{ab}=\left. 
   \frac{\prod_{c}\Tr_{S^{c}} e^{-\sum_{a < b}\varepsilon_{ab} S^{a}S^{b}} S^{a}S^{b}}{\prod_{c}\Tr_{S^{c}} e^{-\sum_{a<b}\varepsilon_{ab} S^{a}S^{b}}}
   \right |_{\varepsilon_{ab}=\varepsilon^{*}_{ab}(\hat{q})}
   \label{eq-sp-q-Q}
   \eeqn
      Let us note that \eq{eq-sp-q-Q} correspond
   to \eq{eq-epsilon-star-0} in the program outlined in sec.~\ref{sec-plefka}.
   \red{The $\varepsilon^{*}$ and $\varepsilon^{*}$ determined here are
     $\varepsilon^{*}_{0}[\hat{q}]$ and $\epsilon^{*}_{0}[\hat{Q}]$ which are evaluated in the absence of the interaction term.}
   
   The above equations imply in particular,
   \beqn
   && \langle (J_{\bs}^{i})^{a} \rangle_{J^{i}}=0 \qquad
   \langle (J_{\bs}^{i})^{a}(J_{\bs'}^{j})^{b} \rangle_{J^{i}}=Q_{ab}\delta_{ij} \delta_{\bs,\bs'} \\
      && \langle (S_{\bs}^{\mu})^{a} \rangle_{S^{\mu}}=0 \qquad
   \langle (S_{\bs}^{\mu})^{a}(S_{\bs'}^{\nu})^{b} \rangle_{S^{\mu}}=q_{ab}\delta_{\mu\nu} \delta_{\bs,\bs'}
   \eeqn

   \subsubsection{Entropic part of the 'bonds'}

   The entropic contribution of the bonds \eq{eq-s-ent-bond-replica}
   can be readily evaluated as (see (77) of \cite{yoshino2018}),
\beq
s_{\rm ent,bond}[\hat{Q}]=\frac{n}{2}+\frac{n}{2}\ln(2\pi)
+\frac{1}{2}\ln {\rm det}\hat{Q}
\label{eq-S-ent-Q}
\eeq

\subsubsection{Entropic part of the 'spins'}

The spins are 'Ising spins'. 
The entropic part of the free-energy of the spins
are (see (79) of \cite{yoshino2018}),
\beqn
s_{\rm ent,spin}[\hat{q}]
&=&\frac{1}{2}\sum_{a,b} 
 \varepsilon^{*}_{ab}q_{ab}+
  \ln  
e^{-\frac{1}{2}\sum_{a,b}\varepsilon^{*}_{ab}\frac{\partial^{2}}{\partial h_{a}\partial h_{b}}}
\left. \prod_{a} 2\cosh (h_{a})  \right |_{\{h_{a}=0\}}
\label{eq-S-ent-q}
\eeqn
with $\epsilon_{aa}=0$. Here we performed the spin trace formally as
\beqn
&& \Tr_{\vS^{c}}e^{-\frac{1}{2}\sum_{a,b}\varepsilon_{ab}S^{a}S^{b}}
=
\Tr_{\vS^{c}}e^{-\frac{1}{2}\sum_{a,b}\varepsilon_{ab}S^{a}S^{b}}
=
\Tr_{\vS^{c}}e^{-\frac{1}{2}\sum_{a,b}\varepsilon_{ab}\frac{\partial^{2}}{\partial h_{a}\partial h_{b}}}
\left. e^{\sum_{a}h_{a}S^{a}}  \right |_{\{h_{a}=0\}} \nonumber \\
&& =
e^{-\frac{1}{2}\sum_{a,b}\varepsilon_{ab}\frac{\partial^{2}}{\partial h_{a}\partial h_{b}}}
\left. \prod_{a} 2\cosh (h_{a})  \right |_{\{h_{a}=0\}}
\nonumber
\eeqn
For the integration over $\varepsilon_{ab}$, the saddle point $\varepsilon_{ab}^{*}=\varepsilon^{*}_{ab}[\hat{q}]$ is obtained formally as,
\beqn
&& q_{ab}=-\left. \frac{\delta}{\delta \varepsilon_{ab}}\ln e^{-\frac{1}{2}\sum_{a,b}\varepsilon_{ab}\frac{\partial^{2}}{\partial h_{a}\partial h_{b}}} 
\left. \prod_{a} 2\cosh (h_{a})  \right |_{\{h_{a}=0\}}
\right |_{\varepsilon_{ab}=\varepsilon^{*}_{ab}[\hat{q}]} (a \neq b)
\label{eq-saddle-epsilon}
\eeqn

\subsection{Evaluation of interaction part of the free-energy}
\label{subsec-eval-interaction-part}

Now we wish to evaluate the partition function \eq{eq-replicated-gardner-volume-appendix}
using the tools developed above. 
What we are doing below is the evaluation of the interaction part of the free-energy
\eq{eq-G-1} in the program outlined in sec.~\ref{sec-plefka}.

In \eq{eq-trace-spin}
we notice that different spin components $\mu$ are decoupled
in the average $\prod_{\mu}\langle \ldots \rangle_{_{S^{\mu}}}$.
Then we obtain the following cumulant expansion which will become very useful.
For any observable $A=A(S^{\mu})$
and writing $\langle \ldots \rangle_{S^{\mu}}=\langle \ldots \rangle$ for simplicity
we find,
\beqn
\ln \langle e^{\frac{1}{\sqrt{M}}\sum_{\mu=1}^{M}A_{\mu}} \rangle
&=&
\sqrt{M}\langle A_{\mu} \rangle+ \frac{1}{2!} (\langle A_{\mu}^{2} \rangle
-\langle A_{\mu} \rangle^{2})\nonumber \\
&& +\frac{1}{3!\sqrt{M}}(\langle A^{3}_{\mu}\rangle-3\langle A^{2}_{\mu}\rangle\langle A_{\mu}\rangle+2\langle A_{\mu}\rangle^{3} )+ \ldots
\label{eq-cumulant-expansion}
\eeqn
Here we just used the fact that $\langle A^{\mu}A^{\nu} \rangle = \langle A^{\mu} \rangle \langle A^{\nu} \rangle$ holds for $\mu \neq \nu$. Thus in the $M \to \infty$,
the lowest non-vanishing cumulant dominates the r.h.s..
For instance if $\langle A_{\mu} \rangle=0$
and $\langle A^{2}_{\mu} \rangle \neq 0$, then
$\lim_{M \to \infty} \ln \langle e^{\frac{1}{\sqrt{M}}\sum_{\mu=1}^{M}A_{\mu}} \rangle
=\frac{1}{2!} \langle A_{\mu}^{2} \rangle$.
Note that the same property also holds for the averaging in the 'bond space'
$\langle \ldots \rangle_{J^{i}}$ in \eq{eq-trace-bond}.


Now we are ready to evaluate Gardner's volume \eq{eq-replicated-gardner-volume}.
We find, introducing a small parameter $\lambda$,
   \begin{eqnarray}
     && \ln   \left\langle   \prod_{\mu,\bs,a} \exp  \left[
     i \eta_{\mu,\bs,a}
     \sum_{i=1}^{N}\sqrt{\frac{\lambda}{N}} (J_{\bs}^{i})^{a}
(S^{\mu}_{\bs})^{a}(S^{\mu}_{\bs(i)})^{a}
\right]     \right \rangle_{J^{i},S^{\mu}} 
     \nonumber  \\
         &&
     =\ln \left \langle 1+
     \sum_{a}
     \sum_{\bs}
                 \sqrt{\frac{\lambda}{N}}
      \sum_{i,\mu}
      i \eta_{\mu,\bs,a}
                  (J_{\bs}^{i})^{a}
(S^{\mu}_{\bs})^{a}
     (S^{\mu}_{\bs(i)})^{a}
     \right. \nonumber \\
     && \hspace*{1cm}  \left.
     + \frac{1}{2!}
     \sum_{a,b}
     \sum_{\bs,\bs'}
                 \frac{\lambda}{N}
      \sum_{i,j,\mu,\nu}
      i \eta_{\mu,\bs,a}
      i \eta_{\nu,\bs',b}
      (J_{\bs}^{i})^{a}
                        (J_{\bs'}^{j})^{b}
      (S^{\mu}_{\bs})^{a}
      (S^{\nu}_{\bs'})^{b}
     (S^{\mu}_{\bs(i)})^{a}
       (S^{\nu}_{\bs'(j)})^{b}
  + \ldots \right \rangle_{J^{i},S^{\mu}} \nonumber \\
  && =
\frac{1}{2!}
     \sum_{a,b}
     \sum_{\bs,\bs'}
                 \frac{\lambda}{N}
      \sum_{i,j,\mu,\nu}
      i \eta_{\mu,\bs,a}
      i \eta_{\nu,\bs',b}
      \delta_{\bs,\bs'}\delta_{ij}\delta_{\mu,\nu}
      Q_{ab,\bs}q_{ab,\bs}q_{ab,\bs(i)}\nonumber \\
&&      +      \frac{1}{4!}
     \sum_{a,b,c,d}
     \sum_{\bs_{1},\bs_{2},\bs_{3},\bs_{4}}
                 \left(\frac{\lambda}{N} \right)^{2}
      \sum_{i,j,k,l,\mu_{1},\mu_{2},\mu_{3},\mu_{4}}
      i \eta_{\mu_{1},\bs_{1},a}
      i \eta_{\mu_{2},\bs_{2},b}
      i \eta_{\mu_{4},\bs_{3},c}
      i \eta_{\mu_{4},\bs_{4},d}\nonumber \\
&&  \times     \delta_{\bs_{1},\bs_{2}}\delta_{\bs_{1},\bs_{3}}\delta_{\bs_{1},\bs_{4}}
\delta_{ij}     \delta_{ik}   \delta_{il} 
      \delta_{\mu_{1},\mu_{2}}   \delta_{\mu_{1},\mu_{3}}  \delta_{\mu_{1},\mu_{4}} \times \nonumber \\
&&       \left[
                \langle       (J_{\bs}^{i})^{a}(J_{\bs}^{i})^{b}(J_{\bs}^{i})^{c}(J_{\bs}^{i})^{d}  \rangle_{J^{i}}
        \langle (S_{\bs}^{\mu})^{a}(S_{\bs}^{\mu})^{b}(S_{\bs}^{\mu})^{c}(S_{\bs}^{\mu})^{d}\rangle_{S^{\mu}}
        \langle (S_{\bs(i)}^{\mu})^{a}(S_{\bs(i)}^{\mu})^{b}(S_{\bs(i)}^{\mu})^{c}(S_{\bs(i)}^{\mu})^{d}\rangle_{S^{\mu}} \right.
        \nonumber \\
&&        \left.        - Q_{ab,\bs}Q_{cd,\bs}q_{ab,\bs}q_{cd,\bs}q_{ab,\bs(i)}q_{cd,\bs(i)}
        - Q_{ac,\bs}Q_{bd,\bs}q_{ac,\bs}q_{bd,\bs}q_{ac,\bs(i)}q_{bd,\bs(i)} \right.
                \nonumber \\
&&  \left.           - Q_{ad,\bs}Q_{bc,\bs}q_{ad,\bs}q_{bc,\bs}q_{ad,\bs(i)}q_{bc,\bs(i)}
                \right]  
        +\ldots + \mbox{``loop correction''}
  \nonumber \\
  && = 
\frac{1}{2}  \sum_{a,b} \sum_{\bs}\sum_{\mu}
i \eta_{\mu,\bs,a}  i \eta_{\mu,\bs,b}
  q_{ab,\bs}
    Q_{ab,\bs}
  \frac{\lambda}{N} \sum_{i=1}^{N}
  q_{ab,\bs(i)} + \mbox{``loop correction''}
  \label{eq-F-int-original}
   \end{eqnarray}
   In the last equation we assumed $N \gg 1$ by which only the 2nd order term in the cumulant
   survives. The contribution of 4th order term of order $O(\lambda^{2})$  writtten explicitely 
   above is smaller than the 2nd order term by a factor $O(\lambda/N)$ so that it can be neglected
   in the $N \to \infty$ limit.

   \red{   However, it is easy to realize that
   the correction due to the interaction loop shown in Fig.~\ref{fig_loop}
   make a contribution of order
   $O(\lambda^{3})$
   and it does {\it not} vanish in the $N \to \infty$ limit.
   In the present paper, we invoke a tree-approximation discarding
   correction terms due to such loops and more extended ones.
   Within this approximation, the expansion of the free-energy
     \eq{eq-plefka} stops at $O(\lambda)$ 
     so that \eq{eq-G-1} gives the free-energy.
   }

\

\subsection{Total free-energy}
\label{sec-total-free-energy}

To sum up, we find,
   \beqn
&& V^{n}\left({\bf S}_{0},{\bf S}_{L}\right)
   =
 \int \prod_{a< b}
\left(
 \prod_{\bs} dQ_{ab,\bs} e^{N S_{\rm ent,bond}[\hat{Q}_{\bs}]}
\prod_{\bs\backslash{\rm output}} dq_{ab,\bs} e^{MS_{\rm ent,spin}[\hat{q}_{\bs}]}
\right)  \nonumber \\
&&  \hspace*{2cm}\prod_{\bs}
\left \{
\left (\prod_{a}
\int \frac{d\eta_{a}}{\sqrt{2\pi}}
W_{\eta_{a}} \right)
e^{-\frac{1}{2}\sum_{a,b}\eta_{a}\eta_{b}Q_{ab,\bs}q_{ab,\bs}\frac{\sum_{i=1}^{N}q_{ab,\bs(i)}}{N}}
\right\}^{M} \qquad \nonumber \\
\hspace*{-1cm}&& = \int \prod_{a< b}\left(
\prod_{\bs} dQ_{ab,\bs} e^{N S_{\rm ent,bond}[\hat{Q}_{\bs}]}
\prod_{\bs\backslash{\rm output}} dq_{ab,\bs} e^{MS_{\rm ent,spin}[\hat{q}_{\bs}]}\right)
\nonumber\\
&& \hspace*{2cm}\prod_{\bs}
\left\{
e^{-\frac{1}{2}\sum_{a,b}\frac{\partial^{2}}{\partial_{h_{\bs,a}}\partial_{h_{\bs,b}}}Q_{ab,\bs}q_{ab,\bs}\frac{\sum_{i=1}^{N}q_{ab,\bs(i)}}{N}}
\left. \prod_{a} e^{-\beta V(h_{\bs,a})}\right|_{h_{\bs,a}=0}
\right\}^{M} \qquad
\eeqn

Given the structure of the network, it is natural to assume that the saddle point values only depend on the
layer $l=0,1,2,\ldots,L$, as \eq{eq-layerwise-orderparameter},
\beq
Q^{*}_{ab,\bs}=Q_{ab}(l) \qquad q^{*}_{ab,\bs}=q_{ab}(l)
\eeq
Then we find,
\beq
s_{n}[\{{\hat Q}(l),{\hat q}(l)\}]
=\frac{1}{\alpha} \sum_{l=1}^{L} s_{\rm ent,bond}[\hat{Q}(l)]
+\sum_{l=1}^{L-1} s_{\rm ent,spin}[\hat{q}(l)]
-\sum_{l=1}^{L} {\cal F}_{\rm int}[\hat{q}(l-1),\hat{Q}(l),\hat{q}(l)]
\label{eq-S-total}
\eeq
with
\beq
    -{\cal F}_{\rm int}[\hat{q}(l-1),\hat{Q}(l),\hat{q}(l)]= \ln 
    e^{\frac{1}{2}\sum_{ab}q_{ab}(l-1)Q_{ab}(l)q_{ab}(l)\partial_{h_{l,a}}\partial_{h_{l,b}}}\prod_{a=1}^{n}
    \left. e^{-\beta V(h_{l,a})}  \right |_{h_{l,a}=0}
    \label{eq-F-int}
\eeq

The order parameters must verify saddle point equations,
\beqn
 \frac{\partial}{\partial Q_{ab}(l)}\left. \partial_{n} s_{n}[\{{\hat Q}(l),{\hat q}(l)\}] \right |_{n=0}&=&0
\label{eq-saddle-Q}
\qquad l=1,2,\ldots,L \\
 \frac{\partial}{\partial q_{ab}(l)}\left. \partial_{n}s_{n}[\{{\hat Q}(l),{\hat q}(l)\}]\right|_{n=0}&=&0
\label{eq-saddle-q}
\qquad l=1,2,\ldots,L-1
\eeqn

\subsection{Parisi's ansatz}
\label{subsec-parisi-ansatz}

\subsubsection{Random inputs/outputs}
\label{subsubsec-parisi-ansatz-random-inputs-outputs}

\begin{figure}[h]
 \includegraphics[width=\textwidth]{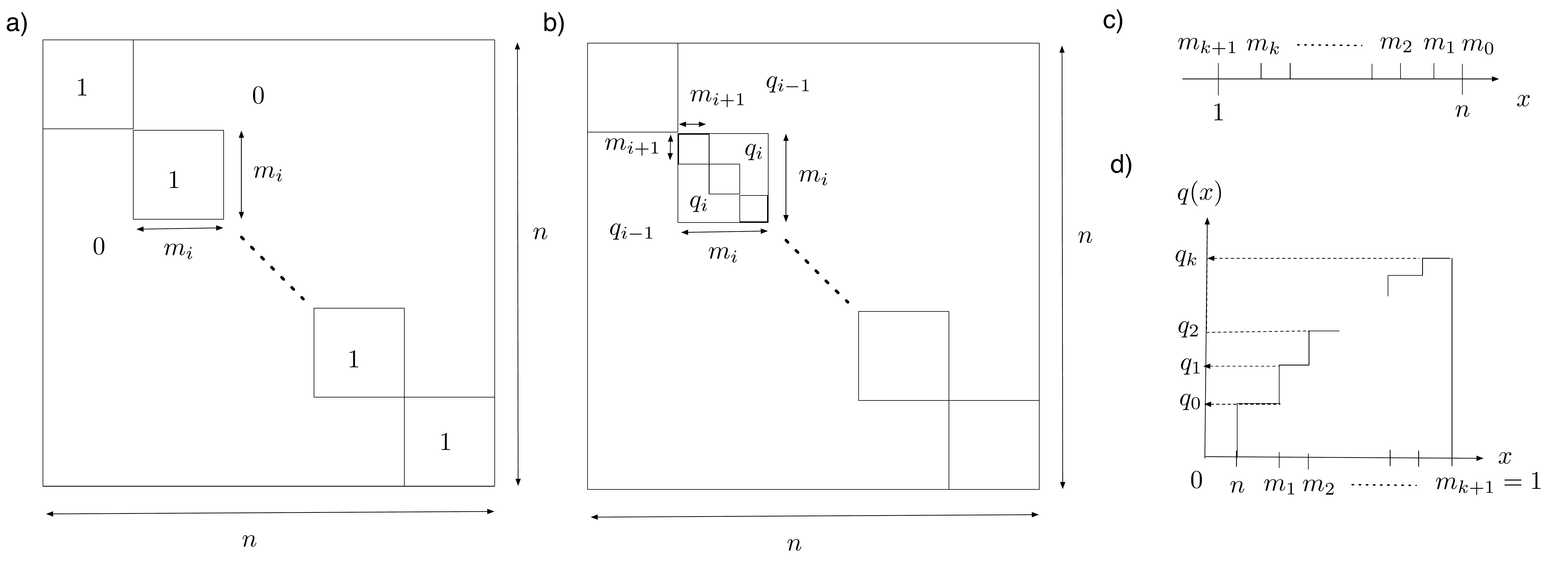}
 \caption{Parametrization of the Parisi's matrix  a) the 'fat' identity matrix
   $I_{ab}^{m_{i}}$ b)  Parisi's matrix given by \eq{eq:parisi-matrix-q} (\eq{eq:parisi-matrix-Q} and \eq{eq:parisi-matrix-epsilon} have the same structure with $q_{i}$'s replaced by
   $Q_{i}$'s and $\epsilon_{i}$'s.)
             c) the hierarchy of the sizes $m_{i}$ of the sub-matrices d) the $q(x)$ function with $0 < n < 1$ ($Q(x)$,$\epsilon(x)$ functions have the same structure).}
           \label{fig-parisi-matrix}
\end{figure}

In the case of random inputs/outputs  (sec. \ref{subsubsect-random-inputs-outputs})
we have $n$ replicas $a=1,2,\ldots,n$. Then it is natural to 
consider the standard Parisi's ansatz with
$k$-step RSB (including RS as $k=0$ and continuous RSB as $k=\infty$) \cite{parisi1979infinite,MPV87}
(See Fig.~\ref{fig-parisi-matrix}),
\beqn
Q_{ab}(l)&=&\sum_{i=0}^{k+1}Q_{i}(l)(I_{ab}^{m_{i}}-I_{ab}^{m_{i+1}})
\qquad l=1,2,\ldots,L
\label{eq:parisi-matrix-Q}
\\
q_{ab}(l)&=&\sum_{i=0}^{k+1}q_{i}(l)(I_{ab}^{m_{i}}-I_{ab}^{m_{i+1}})
\qquad l=1,2,\ldots,L-1
\label{eq:parisi-matrix-q}
\\
\varepsilon_{ab}(l)&=&\sum_{i=0}^{k}\varepsilon_{i}(l)(I_{ab}^{m_{i}}-I_{ab}^{m_{i+1}})
\qquad l=1,2,\ldots,L-1
\label{eq:parisi-matrix-epsilon}
\eeqn
where $I_{ab}^{m}$ is a generalized ('fat') Identity matrix of size $n \times n$ composed of blocks of size $m \times m$. Here we aassumed $q_{k+1}(l)=Q_{k+1}(l)=1$ and $I_{ab}^{m_{k+2}}=0$.
In the Parisi's ansatz one considers 
\beq
1=m_{k+1} <m_{k} <...<m_{1} <m_{0} =n
       \label{eq-m-convention}
\eeq 
which becomes
              \beq
              0= m_{0} < m_{1} < \ldots <  m_{k} < m_{k+1}=1
                 \label{eq-m-convention-2}
              \eeq
              in the $n \to 0$ limit.
In the $k \to \infty$ limit, the matrix elements can be parametrized by
functions $q(x,l)$, $Q(x,l)$ and $\epsilon(x,l)$ defined in the range $0 \leq x \leq 1$  (See Fig. \ref{fig-parisi-matrix} d)).

The order parameter functions encode characteristics of the complex free-energy landscape  \cite{MPV87}. For example the distribution functions of the overlaps between two replicas (two independent machines) can be related to the
order parameter functions as,
\beqn
P(q,l)&=&\frac{dx(q,l)}{dq} \qquad x(q,l)=\int_{0}^{q}dq'P(q',l) \nonumber\\
P(Q,l)&=&\frac{dx(Q,l)}{dQ} \qquad x(Q,l)=\int_{0}^{Q}dQ'P(Q',l)
\label{eq-p-of-q}
\eeqn
Thus $x(q,l)$ ($x(Q,l)$) is the probability that the mutual overlap of the spin (bond) patterns at $l$-th layer
between two machine are smaller than $q$ ($Q$).
Equivalently $1-x(q,l)$ ($1-x(Q,l)$) is the probability that the mutual overlap of the spin (bond) patterns at $l$-th layer
between two machine are {\it larger} than $q$ ($Q$).

The functions $q(x,l)$ and $Q(x,l)$ are expected to have a 'plateau' close to $x=1$, which gives rise to a delta function in
the overlap distribution functions $P(q,l)$ and $P(Q,l)$. As usual we regard the plateau values as
the self-overlaps of the meta-stable states or the Edwards-Anderson order parameters  $q_{\rm EA}(l)$ and $Q_{\rm EA}(l)$.
In practice we will use the values of $q_{k}$ and $Q_{k}$ in the $k$-RSB ansatz as the Edwards-Anderson order parameters.

Analysis of the free-energy functional $-\beta F[\{{\hat Q}(l),{\hat q}(l)\}]/MN$ \eq{eq-F-random-inputs-outputs}
can be done using these matrices in \eq{eq-S-total}.
In appendix~\ref{appendix-rsb-random-inputs-outputs}, we present details of the RSB solution.

\subsubsection{Teacher-student setting}
\label{subsubsec-parisi-ansatz-teacher-student}

\begin{figure}[h]
  \begin{center}
    \includegraphics[width=\textwidth]{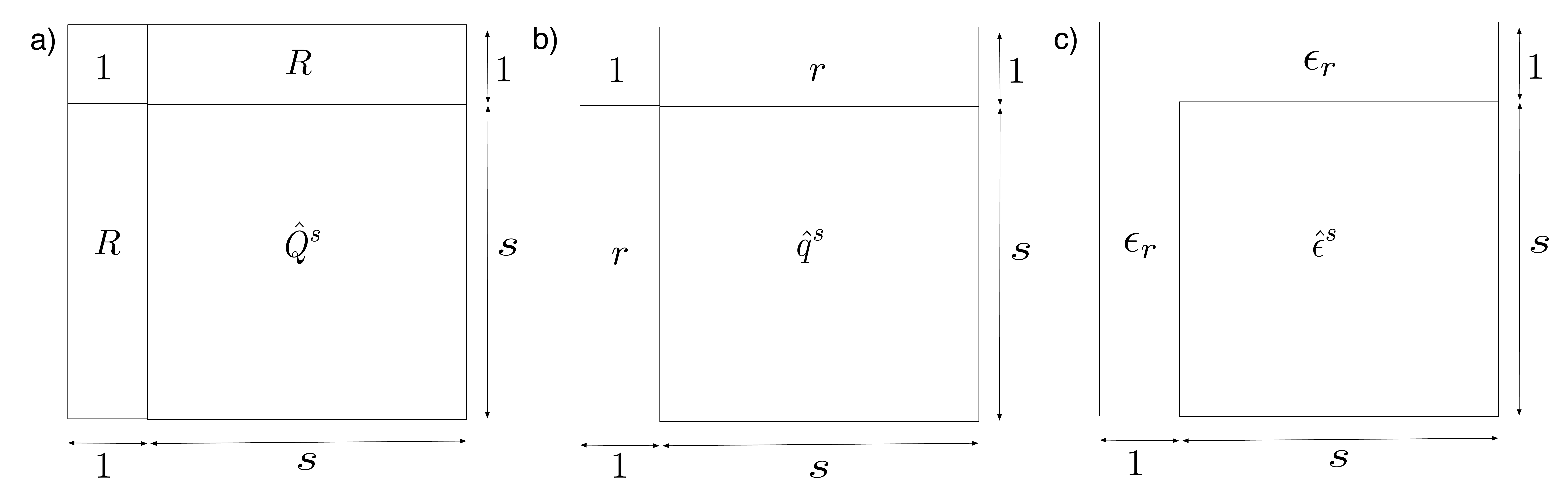}
      \end{center}
  \caption{Parametrization of the Parisi's matrices for the teacher-student setting :
    a) $\hat{Q}^{1+s}$ b) $\hat{q}^{1+s}$ and c) $\hat{\epsilon}^{1+s}$.
For the sub-matrices $\hat{q}^{s}$, $\hat{Q}^{s}$ and  $\hat{\epsilon}^{s}$
    of size $s\times s$ we assume the same hierarchical structure as those in
    \eq{eq:parisi-matrix-Q}-\eq{eq:parisi-matrix-epsilon} (see
    Fig.~\ref{fig-parisi-matrix}) but with $n$ replaced by $s$.
  }
           \label{fig-parisi-matrix-teacher-student}
\end{figure}

For the teacher-student setting (sec.~\ref{subsubsect-teacher-student}) we have to modify the matrices
$\hat{Q}$,$\hat{q}$ and $\hat{\epsilon}$ slightly to
include $a=0$ for the teacher machine in addition to  $a=1,2,\ldots,s$ for the student 
as shown in Fig.~\ref{fig-parisi-matrix-teacher-student}.
We denote the modified matrices as $\hat{Q}^{1+s}$,$\hat{q}^{1+s}$ and $\hat{\epsilon}^{1+s}$.
The sub-matrices $\hat{q}^{s}$, $\hat{Q}^{s}$ and  $\hat{\epsilon}^{s}$ are for the student for which we assume
the same hierarchical structure as before \eq{eq:parisi-matrix-Q}-\eq{eq:parisi-matrix-epsilon}.

Analysis of the Franz-Parisi potential $-\beta F_{\rm teacher-student}[\{{\hat Q}(l),{\hat q}(l)\}]/MN$ \eq{eq-F-teacher-student}
can be done using these matrices in \eq{eq-S-total}.
In appendix~\ref{appendix-rsb-teacher-student}, we present details of the RSB solution.

\section{RSB solution for the random inputs/outputs}
\label{appendix-rsb-random-inputs-outputs}.

Here we derive the RSB solution using the Parisi's ansatz explained in sec.~\ref{subsubsec-parisi-ansatz-random-inputs-outputs}.

\subsection{Entropic part of the free-energy}

\subsubsection{Entropic part of the free-energy due to 'bonds'}

In the $k$-RSB ansatz, the entropic part of the free-energy
\eq{eq-S-ent-Q} due to the 'bonds' can be evaluated as follows.
We find \cite{mezard1991replica,yoshino2018},
\beqn
\ln {\rm det} \hat{Q}&=&
\ln \left(1+\sum_{j=0}^{k}(m_{j}-m_{j+1})Q_{j} \right) \\
&+& n \sum_{i=0}^{k}\left (\frac{1}{m_{i+1}}-\frac{1}{m_{i}} \right)
\ln \left(1+\sum_{j=i}^{k}(m_{j}-m_{j+1})Q_{j}-m_{i}Q_{i}\right)
\label{eq-det-Q-k-RSB}
\eeqn
Remembering that $m_{0}=n$ we find,
\beqn
\left. \partial_{n} S_{\rm ent,bond}[\hat{Q}] \right|_{n=0}=
\frac{1}{2}
\left. \partial_{n}\ln {\rm det} \hat{Q} \right |_{n=0}&=&
\frac{1}{2}
\frac{Q_{0}}{G_{0}}
+ \frac{1}{2}\frac{1}{m_{1}}\ln G_{0}
\nonumber \\
&+&
\frac{1}{2}\sum_{i=1}^{k}\left (\frac{1}{m_{i+1}}-\frac{1}{m_{i}} \right)\ln G_{i}
\label{eq-ln-det-Q}
\eeqn
with
\beqn
G_{i}&=&1+\sum_{j=i}^{k}(m_{j}-m_{j+1})Q_{j}-m_{i}Q_{i} \qquad i=0,1,\ldots,k
\label{eq-G-Q}
\eeqn
which implies
\beqn
Q_{i}&=& 1-G_{k}+\sum_{j=i+1}^{k}\frac{1}{m_{j}}(G_{j}-G_{j-1}) \qquad i=0,1,\ldots,k
\label{eq-Q-G}
\eeqn

\subsubsection{Entropic part of the free-energy due to 'spins'}
\label{subsubsec-ent-spin-kRSB}

In the $k$-RSB ansatz, the entropic part of the free-energy
\eq{eq-S-ent-q} due to the 'spins' can be evaluated as follows
\beqn
S_{\rm ent,spin}[\hat{\epsilon},\hat{q}]
&=&\frac{n}{2}\sum_{i=0}^{k}\epsilon_{i}q_{i}(m_{i}-m_{i+1})
+\frac{n}{2}\epsilon_{k} \nonumber \\
&& +\ln \prod_{i=0}^{k}
 \left. \exp \left[
  \frac{\Lambda^{\rm Ising}_{i}}{2}\sum_{a,b=1}^{n}I_{ab}^{m_{i}}\frac{\partial^{2}}{\partial h_{a}\partial h_{b}}\right]
 \prod_{a=1}^{n} (2\cosh(h_{a})) \right |_{\{h_{a}=0\}} 
\eeqn
which implies
\beqn
\left. \partial_{n} S_{\rm ent,spin}[\hat{\epsilon},\hat{q}]\right |_{n=0}
&=& \frac{1}{2}\sum_{i=0}^{k}\epsilon_{i}q_{i}(m_{i}-m_{i+1})
+\frac{\epsilon_{k}}{2}
-f_{\rm Ising}(m_{0}=0,0) \nonumber \\
&=& \frac{1}{2}\sum_{i=0}^{k}\epsilon_{i}q_{i}(m_{i}-m_{i+1})
+\frac{\epsilon_{k}}{2}
-
\int D z_{0}f_{\rm Ising}(m_{1},\sqrt{\Lambda^{\rm Ising}_{0}}z_{0})
\label{eq-S-ent-q-k-RSB}
\eeqn
where $\epsilon_{i}$'s must be fixed through saddle point equations with respect to variations of them (see below).

In the last two-equation of \eq{eq-S-ent-q-k-RSB}
we used a family of functions which can be obtained recursively
as follows \cite{duplantier1981comment}.
Using
\beqn
\Lambda^{\rm Ising}_{i}=\left \{
\begin{array}{cc}
  -\epsilon_{0} & (i=0)\\
  -\epsilon_{i}+\epsilon_{i-1} & (i=1,2,\ldots,k)
\end{array}
\right.
\label{eq-def-Lambda-ising}
\eeqn
we introduce a family of functions defined recursively
for $i=0,1,2,\ldots,k$,
\beqn
e^{-m_{i}f_{\rm Ising}(m_{i},h)}&=&e^{\frac{\Lambda^{\rm Ising}_{i}}{2}\frac{\partial^{2}}{\partial h^{2}}}
e^{-m_{i}f_{\rm Ising}(m_{i+1},h)}
\nonumber \\
&=&\gamma_{\Lambda^{\rm Ising}_{i}} \otimes
e^{-m_{i}f(m_{i+1},h)} \nonumber \\
&=&\int Dz_{i}
e^{-m_{i}f_{\rm Ising}(m_{i+1},h-\sqrt{\Lambda^{\rm Ising}_{i}}z_{i})}
\label{eq-recursion-f-ising}
\eeqn
with the initial condition
\beq
f_{\rm Ising}(m_{k+1},h)=-\ln 2\cosh(h).
\label{eq-recursion-f-ising-initial-condition}
\eeq

Here we used an identity
\beq
\exp\left(\frac{a}{2}\frac{\partial^{2}}{\partial h^{2}}\right)A(h)=\gamma_{a} \otimes A(h)
\label{eq-formula1}
\eeq
and the following short hand notations: $\gamma_{a}(x)$ is a Gaussian

\beq
\gamma_{a}(x)=\frac{1}{\sqrt{2\pi a}}e^{-\frac{x^{2}}{2a}},
\eeq
by which we write a convolution of a function $A(x)$ with the Gaussian as,
\beq
\gamma_{a}\otimes A(x) \equiv \int dy \frac{e^{-\frac{y^{2}}{2a}}}{\sqrt{2\pi a}}A(x-y)=\int {\cal D}zA(x-\sqrt{a}z)
\label{eq-formula2}
\eeq
where
\beq
\int \cD z \ldots \equiv \int
dz \frac{e^{-\frac{z^{2}}{2}}}{\sqrt{2\pi}} \cdots
\eeq

The saddle point equation with respect to variations of  $\hat{\epsilon}$
\eq{eq-saddle-epsilon} becomes in the $k$-RSB ansatz, for $i=0,1,2,\ldots,k$,
\beqn
q_{i}&=&\frac{2}{m_{i}-m_{i+1}}
\left[\left(-\frac{\partial}{\partial \epsilon_{i}} \right)
  (-f_{\rm Ising}(m_{0}=0,0))
  -\frac{1}{2}\delta_{ik}
  \right]
  \nonumber \\
&=&\int dh P_{\rm Ising}(m_{i},h)
(-f'_{\rm Ising}(m_{i}+1,h))^{2}
\label{eq-saddle-epsilon-kRSB}
\eeqn
where $f'_{\rm Ising}(m,h)=\partial_h f_{\rm Ising}(m,h)$ and
we used (see  \cite{yoshino2018} appendix C)
\beq
\left(-\frac{\partial}{\partial \epsilon_{i}} \right)
(-f_{\rm Ising}(m_{0}=0,0))=\frac{1}{2}(m_{i}-m_{i+1})
\int dh P_{\rm Ising}(m_{i},h)(-f'_{\rm Ising}(m_{i+1},h))^{2}
+\frac{1}{2}\delta_{i,k}
\eeq
with
    \beq
    P_{\rm Ising}(m_{i},h) \equiv \frac{\delta f_{\rm Ising}(m_{0},0)}{\delta f_{\rm Ising}(m_{i+1},h)}
    \eeq
        which follows a recursion formula  (see \cite{yoshino2018} sec. 8.3.1),
\beq
P_{\rm Ising}(m_{j},h)=e^{-m_{j}f_{\rm Ising}(m_{j+1},h)} \gamma_{\Lambda^{\rm Ising}_{j}}\otimes_{h}
\frac{P_{\rm Ising}(m_{j-1},h)}{e^{-m_{j}f_{\rm Ising}(m_{j},h)}} \qquad j=1,2,\ldots,k
\label{eq-recursion-p-ising}
\eeq
 with the 'boundary condition'
\beq
P_{\rm Ising}(m_{0},h)=\frac{1}{\sqrt{2\pi \Lambda^{\rm Ising}_{0}}}e^{-\frac{h^{2}}{2\Lambda_{0}}}
\label{eq-boundary-P-ising}
\eeq
In \eq{eq-recursion-p-ising} 
$\otimes_{h}$ stands for a convolution with respect to the variable $h$.

The derivatives $f'_{\rm Ising}(m,h)=\partial_h f_{\rm Ising}(m,h)$ can also be
obtained recursively. From \eq{eq-recursion-f-ising} and
\eq{eq-recursion-f-ising-initial-condition} we find,
\beq
f'_{\rm Ising}(m_{i},h)
=e^{m_{i}f_{\rm Ising}(m_{i},h)}\gamma_{\Lambda^{\rm Ising}_{i}}\otimes f'_{\rm Ising}(m_{i+1},h)e^{-m_{i}f_{\rm Ising}(m_{i+1},h)}
\label{eq-recursion-f-dash-ising}
\eeq
for $i=1,2,\ldots,k$ with the 'boundary condition',
\beq
f_{\rm Ising}'(m_{k+1},h)=-\tanh(h)
\label{eq-boundary-f-dash-ising}
\eeq

    \subsection{Interaction part of the free-energy}

    The interaction part of the free-energy \eq{eq-F-int} becomes in the $k$-RSB ansatz,
  \beqn
-\left. \partial_{n}     {\cal F}_{\rm int}[\hat{q}(l-1),\hat{Q}(l),\hat{q}(l)]  \right |_{n=0}
&=&
\left. \partial_{n} \ln 
\prod_{i=0}^{k+1}
 \left. \exp \left[
  \frac{\Lambda_{i}(l)}{2}\sum_{a,b=1}^{n}I_{ab}^{m_{i}}\frac{\partial^{2}}{\partial h_{a}\partial h_{b}}\right]
 \prod_{a=1}^{n} e^{-\beta V(r(h_{a}))} \right |_{\{h_{a}=0\}}  \right |_{n=0}
\nonumber \\
&=& f(m_{0}=0,0,l) \nonumber \\
&=& \left. \int D z_{0}f(m_{1},h-\sqrt{\Lambda_{0}},l) \right |_{h=0}
\label{eq-F-int-k-RSB}
\eeqn
with, for $l=1,2,\ldots,L$,
\beq
\Lambda(l)=\left \{\begin{array}{cc}
\lambda_{0}(l) &  (i=0)\\
\lambda_{i}(l)-\lambda_{i-1}(l) & (i=1,2,\ldots,k+1)
\end{array}\right.
\label{eq-def-Lambda}
\eeq
and
\beq
\lambda_{i}(l)=q_{i}(l-1)Q_{i}(l)q_{i}(l)
\label{eq-def-lambda}
\eeq

We introduced a family of functions defined recursively
for $i=0,1,2,\ldots,k$ and $l=1,2,\ldots,L$,
\beqn
e^{-m_{i}f(m_{i},h,l)}&=&e^{\frac{\Lambda_{i}(l)}{2}\frac{\partial^{2}}{\partial h^{2}}}
e^{-m_{i}f(m_{i+1},h,l)}
\nonumber \\
&=&\int Dz_{i}
e^{-m_{i}f(m_{i+1},h-\sqrt{\Lambda_{i}(l)}z_{i},l)}
\label{eq-recursion-f}
\eeqn
with the initial condition
\beq
f(m_{k+1},l)=-\ln \gamma_{\Lambda_{k+1}(l)}\otimes e^{-\beta V(h)}
=-\ln \int Dz_{k+1} e^{-\beta V(h-\sqrt{\Lambda_{k+1}(l)}z_{k+1})}
\label{eq-recursion-f-initial-condition}
\eeq

For the hard-core potential \eq{eq-hardcore} we find,
\beq
f(m_{k+1},h,l)=-\ln \Theta \left( \frac{h}{\sqrt{2\Lambda_{k+1}(l)}} \right)
\label{eq-recursion-f-initial-condition-hardcore}
\eeq
where
\beq
\Theta(x)=\int_{-\infty}^{x}\frac{dy}{\sqrt{\pi}}e^{-y^{2}}
\label{eq-function-Theta}
\eeq

\subsection{Saddle point equations}

\subsubsection{Variation of $q_{i}(l)$'s}
The saddle point equations \eq{eq-saddle-q} becomes, for $i=0,1,2,\ldots,k$ and $l=1,2,\ldots,L-1$,
\beqn
0&=&\frac{\partial}{\partial q_{i}(l)} \left. \partial_{n} S^{\rm ent,spin}_{n}[\hat{q},\hat{Q}] \right|_{n=0} \nonumber \\
&=& \frac{1}{2}\epsilon_{i}(l)(m_{i}-m_{i+1})-\frac{\partial}{\partial q_{i}(l)}
\sum_{l'=1}^{L}{\cal F}_{\rm int}[\hat{q}(l'-1),\hat{Q}(l'),\hat{q}(l')] \nonumber \\
&=& \frac{1}{2}\epsilon_{i}(l)(m_{i}-m_{i+1})
+\sum_{l'=1}^{L}\frac{\partial \lambda_{i}(l')}{\partial q_{i}(l)}
\left(- \frac{\partial f(m_{0}=0,0,l')}{\partial \lambda_{i}(l')}
\right)
\label{eq-saddle-q-2}
\eeqn
from which we find,
\beqn
\epsilon_{i}(l)&=&-\sum_{l'=1}^{L}\frac{\partial \lambda_{i}(l')}{\partial q_{i}(l)}\kappa_{i}(l') \nonumber \\
&=&-q_{i}(l-1)Q_{i}(l)\kappa_{i}(l)-Q_{i}(l+1)q_{i}(l+1)\kappa_{i}(l+1)
\label{eq-saddle-q-k-RSB}
\eeqn
where we introduced, for $i=0,1,2,\ldots,k$ and $l=1,2,\ldots,L$,
\beqn
\kappa_{i}(l) \equiv \int dh P(m_{i},h,l)(-f'(m_{i+1},h,l))^{2}
\label{eq-def-kappa}
\eeqn
with
    \beq
    P(m_{i},h,l) \equiv \frac{\delta f(m_{0},0,l)}{\delta f(m_{i+1},h,l)}
    \eeq
    which follows a recursion formula  (see \cite{yoshino2018} sec. 8.3.1),
\beq
P(m_{j},h,l)=e^{-m_{j}f(m_{j+1},h,l)} \gamma_{\Lambda_{j}}\otimes_{h}
\frac{P(m_{j-1},h,l)}{e^{-m_{j}f(m_{j},h,l)}} \qquad j=1,2,\ldots,k+1
\label{eq-recursion-p}
\eeq
 with the 'boundary condition'
\beq
P(m_{0},h,l)=\frac{1}{\sqrt{2\pi \Lambda_{0}(l)}}e^{-\frac{h^{2}}{2\Lambda_{0}(l)}}
\label{eq-boundary-P}
\eeq
The last equation of \eq{eq-saddle-q-2} is obtained  using the following (see \cite{yoshino2018} appendix C)
\beq
\frac{\partial}{\partial \lambda_{i}(l)} 
(-f(m_{0}=0,0,l))=\frac{1}{2}(m_{i}-m_{i+1})
\int dh P(m_{i},h,l)(-f'(m_{i+1},h,l))^{2}
\eeq

The derivatives $f'(m,h,l)=\partial_h f(m,h,l)$ can also be
obtained recursively. From \eq{eq-recursion-f} and
\eq{eq-recursion-f-initial-condition} we find,
\beq
f'(m_{i},h,l)
=e^{m_{i}f(m_{i},h,l)}\gamma_{\Lambda_{i}(l)}\otimes f'(m_{i+1},h,l)e^{-m_{i}f(m_{i+1},h,l)}
\label{eq-recursion-f-dash}
\eeq
for $i=1,2,\ldots,k$ with the 'boundary condition',
\beq
f'(m_{k+1},h,l)=-\frac{\int Dz_{k+1} (d/dh)(e^{-\beta V(h-\sqrt{\Lambda_{k+1}(l)})})
}{
  \int Dz_{k+1} e^{-\beta V(h-\sqrt{\Lambda_{k+1}(l)})}
  }
\label{eq-boundary-f-dash}
\eeq
which becomes for the hardcore potential
(using \eq{eq-recursion-f-initial-condition-hardcore} and \eq{eq-function-Theta}),
\beq
f'(m_{k+1},h,l)=-\frac{
1
}{\Theta \left( \frac{h}{\sqrt{2\Lambda_{k+1}(l)}} \right)}
  \frac{1}{\sqrt{2\pi\Lambda_{k+1}(l)}}\exp\left(-\frac{h^{2}}{2\Lambda_{k+1}(l)}\right)
\label{eq-recursion-f-dash-initial-condition-hardcore}
\eeq

\subsubsection{Variation of $G_{i}(l)$'s}

For the saddle point equations \eq{eq-saddle-Q}
it is convenient to consider instead, for $i=0,1,2,\ldots,k$ and $l=1,2,\ldots,L$,
\beqn
0= \left. \frac{\partial}{\partial G_{i}(l)}\partial_{n} S[\hat{q},\hat{Q}] \right|_{n=0}
\eeqn
where $G_{i}(l)$'s are defined in \eq{eq-G-Q}. We obtain (see \cite{yoshino2018} sec. 8.4),
\beqn
\frac{Q_{0}(l)}{G^{2}_{0}(l)}&=& \alpha q_{0}(l)q_{0}(l-1)\kappa_{0}(l)\nonumber\\
\frac{1}{G_{i}(l)}-\frac{1}{G_{0}(l)}&=&\alpha \left(\sum_{j=0}^{i-1}(m_{j}-m_{j+1})q_{j}(l)q_{j}(l-1)\kappa_{j}(l)
+m_{i}q_{i}(l)q_{i}(l-1)\kappa_{i}(l)\right) \qquad
\label{eq-saddle-point-G}
\eeqn
for $i=1,2,\ldots,k$.

\subsubsection{Procedure to solve the saddle point equations}
\label{sec_procedure}

The saddle point equations for a generic finite $k$-RSB ansatz
with some fixed values of $0 < m_{1} < m_{2} < \ldots < m_{k} <1$
can be solved numerically as follows.

\begin{itemize}
\item [0.] Choose a boundary condition by fixing
  $q_{i}(0)$ and $q_{i}(L)$ for $i=0,1,2,\ldots,k$.
\item[1.] Make some guess for the initial values of
  $q_{i}(l)$ ($l=1,2,\ldots,L-1$) and $Q_{i}(l)$ ($l=1,2,\ldots,L$)
  for $i=0,1,2,\ldots,k$. 
  Then compute $G_{i}(l)$ for $i=0,1,\ldots,k$
  and $l=1,2,\ldots,L-1$ using \eq{eq-G-Q}.
\item[2.] Do the following (1)-(8) for $l=1,2,\ldots,L$.
  (1) Compute $\lambda_{i}(l)$ for $i=0,1,2,\ldots,k$ and 
  using \eq{eq-def-lambda}.  
  (2) Compute $\Lambda_{i}(l)$ for $i=0,1,2,\ldots,k+1$
    using \eq{eq-def-Lambda}.  
  (3) Compute functions $f(m_{i},h,l)$ recursively
    for $i=k,k-1,\ldots,0$ using \eq{eq-recursion-f} with the boundary condition given by \eq{eq-recursion-f-initial-condition} (which is
    \eq{eq-recursion-f-initial-condition-hardcore} for the hardcore potential).
    (4) Compute also the derivatives  $f'(m_{i},h,l)$
    recursively  for $i=k,k-1,\ldots,2,1$ 
  using  \eq{eq-recursion-f-dash} with the boundary condition given by \eq{eq-boundary-f-dash} (which is \eq{eq-recursion-f-dash-initial-condition-hardcore} for the hardcore potential). (5) Compute functions $P(m_{i},h,l)$
  recursively for $i=1,\ldots,k$
using \eq{eq-recursion-p} with the boundary condition given by \eq{eq-boundary-P}.
(6) Compute $\kappa_{i}(l)$
for $i=0,1,\ldots,k$
using \eq{eq-def-kappa}.
(7) Compute
  $G_{i}(l)$ for $i=0,1,\ldots,k$ using \eq{eq-saddle-point-G}.
  (8) Compute  $Q_{i}(l)$ for $i=0,1,\ldots,k$ using \eq{eq-Q-G}.
\item[3.] Do the following (1)-(6) for $l=1,2,\ldots,L-1$.
  (1) Compute $\epsilon_{i}(l)$ for $i=0,1,2,\ldots,k$
  using \eq{eq-saddle-q-k-RSB}. (2) Compute
  $\Lambda^{\rm Ising}_{i}(l)$ for $i=0,1,2,\ldots,k$
  using \eq{eq-def-Lambda-ising}.
  (3) Compute functions $f_{\rm Ising}(m_{i},h,l)$ recursively
  for $i=k,k-1,\ldots,0$ using \eq{eq-recursion-f-ising} with the boundary condition given by \eq{eq-recursion-f-ising-initial-condition}.
  (4) Compute also the derivatives  $f'_{\rm Ising}(m_{i},h,l)$
    recursively  for $i=k,k-1,\ldots,2,1$ 
    using  \eq{eq-recursion-f-dash-ising}
    with the boundary condition given by \eq{eq-boundary-f-dash-ising}.
   (5) Compute functions $P_{\rm Ising}(m_{i},h,l)$
  recursively for $i=1,\ldots,k$
  using \eq{eq-recursion-p-ising} with the boundary condition given by \eq{eq-boundary-P-ising}.
  (6) Compute $q_{i}(l)$ for $i=0,1,\ldots,k$ using
  \eq{eq-saddle-epsilon-kRSB}.
\item[4.] Return to 2.
\end{itemize}
The above procedure 1.-4. must be repeated until the solution converges.
The values of $m_{i}$s ($0 < m_{1} < m_{2} \ldots < m_{k}< 1$ (see Fig.~\ref{fig-parisi-matrix} c)))
are chosen such that $\log m_{i}$s are equally spaced between $\log m_{1}$ and $\log m_{k+1}=0$.
We chose $m_{1}=0.0001$ in the numerical analysis shown in this paper. Numerical integrations
are done by the simple rectangle rule with an integration step $0.01$.

\section{RSB solution for the teacher-student setting}
\label{appendix-rsb-teacher-student}

Here we derive the RSB solution using the Parisi's ansatz explained in sec.~\ref{subsubsec-parisi-ansatz-teacher-student}.

\subsection{Entropic part of the free-energy}

\subsubsection{Entropic part of the free-energy due to 'bonds'}

Within the ansatz for the teacher-student setting, the entropic part of the free-energy
\eq{eq-S-ent-Q} due to the 'bonds' can be evaluated as follows. First we find,
\beqn
{\rm det} \hat{Q}^{1+s}&=& {\rm det} (\hat{Q}^{s}-R^{2}).
\eeqn
Thus we find 
\beqn
\left. \partial_{s} S_{\rm ent,bond}[\hat{Q}^{1+s}] \right |_{s=0} &=&\left. \frac{1}{2}\partial_{s}\ln {\rm det} (\hat{Q}^{s}-R^{2}) \right |_{s=0} \nonumber \\
&=&
\frac{1}{2}
\frac{Q_{0}-R^{2}}{G_{0}}
+ \frac{1}{2}\frac{1}{m_{1}}\ln G_{0}
\nonumber \\
&+&
\frac{1}{2}\sum_{i=1}^{k}\left (\frac{1}{m_{i+1}}-\frac{1}{m_{i}} \right)\ln G_{i}
\label{eq-S-ent-bond-teacher-student}
\eeqn
with 
\beqn
G_{i}&=&1+\sum_{j=i}^{k}(m_{j}-m_{j+1})Q_{j}-m_{i}Q_{i} \qquad i=0,1,\ldots,k
\label{eq-G-Q-teacher-student}
\eeqn
which implies
\beqn
Q_{i}&=& 1-G_{k}+\sum_{j=i+1}^{k}\frac{1}{m_{j}}(G_{j}-G_{j-1}) \qquad i=0,1,\ldots,k
\label{eq-Q-G-teacher-student}
\eeqn
Note that above equations slightly modify \eq{eq-G-Q} and \eq{eq-Q-G}.

\subsubsection{Entropic part of the free-energy due to 'spins'}

Within the same ansatz, the entropic part of the free-energy 
\eq{eq-S-ent-q} due to the 'spins' can be evaluated as follows,

\beqn
\hspace*{-1cm} &&
S_{\rm ent,spin}[\hat{\epsilon}^{1+s},\hat{q}^{1+s}]
=s\epsilon_{r}r+\frac{1}{2}\epsilon_{r}+\frac{s}{2}\sum_{i=0}^{k}\epsilon_{i}q_{i}(m_{i}-m_{i+1})
+\frac{s}{2} \epsilon_{k} \nonumber \\
\hspace*{-2cm} && \hspace*{1cm}+
\ln 
\exp \left[\frac{\Lambda^{\rm Ising}_{\rm com}}{2}\sum_{a,b=0}^{s}\frac{\partial^{2}}{\partial h_{a}\partial h_{b}}\right]
\prod_{i=0}^{k}
 \left. \exp \left[
  \frac{\Lambda^{\rm Ising}_{i}}{2}\sum_{a,b=1}^{s}I_{ab}^{m_{i}}\frac{\partial^{2}}{\partial h_{a}\partial h_{b}}\right]
 \prod_{a=0}^{s} (2\cosh(h_{a})) \right |_{\{h_{a}=0\}}
 \nonumber \\
\hspace*{-2cm} &&  \hspace*{1cm}=s\epsilon_{r}r+\frac{1}{2}\epsilon_{r}+\frac{s}{2}\sum_{i=0}^{k}\epsilon_{i}q_{i}(m_{i}-m_{i+1}) 
+\frac{s}{2} \epsilon_{k} \nonumber \\
\hspace*{-2cm} && \hspace*{2cm}+\left. \ln \gamma_{\Lambda_{\rm com}} \otimes (2\cosh(h) \gamma_{\Lambda_{0}^{\rm Ising}} \otimes e^{-s f^{\rm Ising}(m_{1},h)}
 \right |_{h=0} \qquad
 \label{eq-S-ent-q-k-RSB-teacher-student}
\eeqn
where $\epsilon_{r}$ and $\epsilon_{i}$'s must be fixed through saddle point equations with respect to variations of them (see below).

In  \eq{eq-S-ent-q-k-RSB-teacher-student} $I_{ab}^{m_{i}}$ is defined similarly as those used in
\eq{eq:parisi-matrix-Q}-\eq{eq:parisi-matrix-epsilon} (see Fig.~\ref{fig-parisi-matrix}) but with size $s \times s$ instead of $n \times n$.
We have also introduced,
\beqn
&& \Lambda^{\rm Ising}_{\rm com}=-\epsilon_{r} \nonumber \\
&& \Lambda^{\rm Ising}_{i}=\left \{
\begin{array}{cc}
  -\epsilon_{0}+\epsilon_{r} & (i=0)\\
  -\epsilon_{i}+\epsilon_{i-1} & (i=1,2,\ldots,k)
\end{array}
\right.
\label{eq-def-Lambda-ising-teacher-student}
\eeqn
and used the family of functions defined recursively for $i=0,1,2,\ldots,k$ using \eq{eq-recursion-f-ising}
and the initial condition \eq{eq-recursion-f-ising-initial-condition}. One must keep in mind
that $\Lambda^{\rm Ising}_{0}$ in \eq{eq-def-Lambda-ising-teacher-student} is shifted with respect
to that in \eq{eq-def-Lambda-ising} due to $\epsilon_{r}$.

The saddle point equations with respect to variations  of  $\hat{\epsilon}^{1+s}$ \eq{eq-saddle-epsilon}
yield $q_{i}$'s and $r$. 
Variation with respect to $\epsilon_{i}$  yields the equation for the $q_{i}$'s, which
is formally the same as \eq{eq-saddle-epsilon-kRSB},
\beqn
q_{i}&=&
\int dh P_{\rm Ising}(m_{i},h)
(-f'_{\rm Ising}(m_{i},h))^{2}
\label{eq-saddle-epsilon-kRSB-teacher-student}
\eeqn
for $i=0,1,2,\ldots,k$. Here   $P_{\rm Ising}(m_{i},h)$  can be obtained from the equation \eq{eq-recursion-p-ising}.
However the initial condition is modified from \eq{eq-boundary-P-ising} to
\beq
P_{\rm Ising}(m_{0},h)=\frac{
  \int Dz_{\rm com} 2\cosh (\sqrt{\Lambda^{\rm Ising}_{\rm com}}z_{\rm com})
  \frac{1}{\sqrt{2\pi \Lambda^{\rm Ising}_{0}}}e^{-\frac{(h-\sqrt{\Lambda_{\rm com}}z_{\rm com})^{2}}{2\Lambda^{\rm Ising}_{0}}}
}{
\int Dz_{\rm com} 2\cosh (\sqrt{\Lambda^{\rm Ising}_{\rm com}}z_{\rm com})}
\eeq

Variation with respect to $\epsilon_{r}$ yields the equation for the $r$ as the following. Using 
\beqn
&& 0=\frac{\partial}{\partial \epsilon_{r}} S_{\rm ent,spin}[\hat{\epsilon},\hat{q}^{1+s}]
= sr+\frac{1}{2} -\frac{1}{2}
 \nonumber \\
&&  -s\frac{\int Dz_{\rm com} 2\sinh(\sqrt{\Lambda_{\rm com}}z_{\rm com})
         \int Dz_{0}(-f'_{\rm Ising}(m_{1},\sqrt{\Lambda_{\rm com}}z_{\rm com}+\sqrt{\Lambda^{\rm Ising}_{0}}z_{0}))
   }{\int Dz_{\rm com} 2\cosh(\sqrt{\Lambda_{\rm com}}z_{\rm com})}
 +O(s^{2}) \hspace*{2cm}
 \eeqn
 Thus we find
 \beq
r=\frac{\int Dz_{\rm com} 2\sinh(\sqrt{\Lambda_{\rm com}}z_{\rm com})
         \int Dz_{0}(-f'_{\rm Ising}(m_{1},\sqrt{\Lambda_{\rm com}}z_{\rm com}+\sqrt{\Lambda^{\rm Ising}_{0}}z_{0}))
   }{\int Dz_{\rm com} 2\cosh(\sqrt{\Lambda_{\rm com}}z_{\rm com})}
 \eeq

\subsection{Interaction part of the free-energy}

Within the same ansatz, the interaction part of the free-energy \eq{eq-F-int} becomes,
  \beqn
&& \left. \partial_{s}     {\cal F}_{\rm int}[\hat{q}^{1+s}(l-1),\hat{Q}^{1+s}(l),\hat{q}^{1+s}(l)]  \right |_{s=0}
=
\partial_{s} \ln 
\exp \left[ \frac{\Lambda_{\rm com}(l)}{2} \sum_{a,b=0}^{s}\frac{\partial^{2}}{\partial h_{a}\partial h_{b}}\right]
\exp \left[ \frac{\Lambda_{\rm teacher}(l)}{2} \frac{\partial^{2}}{\partial h^{2}_{0}}\right]
\nonumber \\
&& \left.  \prod_{i=0}^{k+1}
 \left. \exp \left[
  \frac{\Lambda_{i}(l)}{2}\sum_{a,b=1}^{s}I_{ab}^{m_{i}}\frac{\partial^{2}}{\partial h_{a}\partial h_{b}}\right]
 \prod_{a=0}^{s} e^{-\beta V(r(h_{a}))} \right |_{\{h_{a}=0\}}  \right |_{s=0}
\nonumber \\
&&= \partial_{s} \ln  \int D z_{\rm com} \int D z_{\rm teacher} e^{-\beta V(\sqrt{\Lambda_{\rm com}(l)}z_{\rm com}+\sqrt{\Lambda_{\rm teacher}(l)}z_{\rm teacher})} \nonumber \\
&& \left. \hspace*{2cm}\int D z_{0}f(m_{1},\sqrt{\Lambda_{\rm com}(l)}z_{\rm com}+\sqrt{\Lambda_{0}(l)}z_{0})  \right |_{s=0}
\label{eq-F-int-k-RSB-teacher-student}
\eeqn
with, for $l=1,2,\ldots,L$. Here we introduced,
\beqn
&&  \Lambda_{\rm com}(l)=r(l-1)R(l)r(l) \nonumber \\
&&  \Lambda_{\rm teacher}(l)=1-r(l-1)R(l)r(l) \nonumber \\
&& \Lambda_{i}(l)=\left \{\begin{array}{cc}
\lambda_{0}(l)-\Lambda_{\rm com}(l) &  (i=0)\\
\lambda_{i}(l)-\lambda_{i-1}(l) & (i=1,2,\ldots,k+1)
\end{array}\right.
\label{eq-def-Lambda-teacher-student}
\eeqn
with $\lambda_{i}(l)$'s defined in \eq{eq-def-lambda} which reads as,
\beq
\lambda_{i}(l)=q_{i}(l-1)Q_{i}(l)q_{i}(l)
\eeq
We also used the family of functions $f(m_{i},h)$ defined recursively
for $i=0,1,2,\ldots,k$ and $l=1,2,\ldots,L$ by \eq{eq-recursion-f} with the initial condition \eq{eq-recursion-f-initial-condition}.
Note that $\Lambda_{0}$ is shifted with respect to that in \eq{eq-def-Lambda} due to $R$ and $r$.

\subsection{Saddle point equations}
\label{sec-sp-teacher-student}

\subsubsection{Variation of $q_{i}(l)$'s}

For the saddle point equations \eq{eq-saddle-q}, we find
formally the same result as \eq{eq-saddle-q-k-RSB} which reads as,
\beq
\epsilon_{i}(l)=-q_{i}(l-1)Q_{i}(l)\kappa_{i}(l)-Q_{i}(l+1)q_{i}(l+1)\kappa_{i}(l+1)
\eeq
with $\kappa_{i}$ defined as \eq{eq-def-kappa} which reads as,
\beq
\kappa_{i}(l) \equiv
\int dh P(m_{i},h,l)(-f'(m_{i+1},h,l))^{2}.
\eeq
The function $P(m_{i},h,l)$ can be also be obtained by the same equations as before \eq{eq-recursion-p} 
but with the initial condition \eq{eq-boundary-P} modified as,
\beq
P(m_{0},h,l)=\frac{
  \int Dz_{\rm com} \int Dz_{\rm teacher} e^{-\beta V(\sqrt{\Lambda_{\rm com}(l)}z_{\rm com}+\sqrt{\Lambda_{\rm teacher}(l)}z_{\rm teacher})}
  \frac{1}{\sqrt{2\pi \Lambda_{0}(l)}}
  e^{-\frac{(h-\sqrt{\Lambda_{\rm com}(l)}z_{\rm com})^{2}}{2\Lambda_{0}(l)}}
}{\int Dz_{\rm com} \int Dz_{\rm teacher} e^{-\beta V(\sqrt{\Lambda_{\rm com}(l)}z_{\rm com}+\sqrt{\Lambda_{\rm teacher}(l)}z_{\rm teacher})}}
\label{eq-boundary-P-teacher-student}
\eeq
Note also that $\Lambda_{0}$ is shifted as in \eq{eq-def-Lambda-teacher-student}.
For the hardcore potential \eq{eq-hardcore} we find,
\beq
P(m_{0},h,l)=\frac{1}{\int Dz
  \Theta \left(
  \frac{
    \sqrt{\Lambda_{\rm com}(l)}z
  }{
    \sqrt{2(\Lambda_{\rm teacher}(l))}
  }
  \right)
}
 \int Dz
  \Theta \left(
  \frac{
    \sqrt{\Lambda_{\rm com}(l)}z
  }{
    \sqrt{2\Lambda_{\rm teacher}(l))}
  }
  \right) 
\frac{1}{\sqrt{2\pi \Lambda_{0}(l)}}e^{-\frac{(h-\sqrt{\Lambda_{\rm com}(l)}z)^{2}}{2\Lambda_{0}(l)}}
\eeq
with $\Theta(h)$ defined in \eq{eq-function-Theta}.

\subsubsection{Variation of $G_{i}(l)$'s}

For the saddle point equations \eq{eq-saddle-Q}, we just need to modify slightly \eq{eq-saddle-point-G},
with $G_{i}$'s defined in \eq{eq-G-Q-teacher-student},
\beqn
\frac{Q_{0}(l)-R^{2}(l)}{G^{2}_{0}(l)}&=&\alpha q_{0}(l)q_{0}(l-1)\kappa_{0}(l)
\label{eq-saddle-point-G-teacher-student}
\\
\frac{1}{G_{i}(l)}-\frac{1}{G_{0}(l)}&=&\alpha \left(\sum_{j=0}^{i-1}(m_{j}-m_{j+1})q_{j}(l)q_{j}(l-1)\kappa_{j}(l)
+m_{i}q_{i}(l)q_{i}(l-1)\kappa_{i}(l)\right)
\nonumber
\eeqn
for $l=1,2,\ldots,L$.

\subsubsection{Variation of $r$}

Here we consider variation of the free-energy 
\eq{eq-F-teacher-student} (see also \eq{eq-S-total}) with respect to $r(l)$
, for $l=1,2,\ldots,L-1$,
\beqn
0&=&\frac{\partial}{\partial r(l)}
\left.  \partial_{s}s_{1+s}[\{{\hat Q}(l),{\hat q}(l)\}]\right |_{s=0} \nonumber \\
&=& \frac{\partial}{\partial r(l)}
\left.  \partial_{s}S_{\rm ent,spin}[\hat{q}^{1+s}(l)]\right |_{s=0}
-\frac{\partial}{\partial r(l)}
\sum_{l'=1}^{L}\left. \partial_{s} {\cal F}_{\rm int}[{{\hat q}^{1+s}(l'-1),\hat Q}^{1+s}(l'),{\hat q}^{1+s}(l')]\right |_{s=0} \qquad
\label{eq-variation-r-teacher-student}
\eeqn

Variation of the entropic part (spin) of the free-energy 
 \eq{eq-S-ent-q-k-RSB-teacher-student}
yields,
\beq
\frac{\partial}{\partial r} 
\left. \partial_{s} S_{\rm ent,spin}[\hat{q}^{1+s}] \right |_{s=0}
=\epsilon_{r}
\eeq
On the other hand, variation of the interaction part of the free-energy 
  \eq{eq-F-int-k-RSB-teacher-student} yields, for $l=1,2,\ldots,L-1$,
  \beqn
&&   -\frac{\partial}{\partial r(l)}\sum_{l'=1}^{L}
  \left. \partial_{s}    {\cal F}_{\rm int}[\hat{q}^{1+s}(l-1),\hat{Q}^{1+s}(l),\hat{q}^{1+s}(l)]  \right |_{s=0} \nonumber \\
  &&
=  -\sum_{l'=1}^{L} \left(
  \frac{\partial \Lambda_{\rm com}(l')}{\partial r(l)}\frac{\partial}{\partial \Lambda_{\rm com}(l')}
  +\frac{\partial \Lambda_{\rm teacher}(l')}{\partial r(l)}\frac{\partial}{\partial \Lambda_{\rm teacher}(l')}
  +\frac{\partial \Lambda_{0}(l')}{\partial r(l)}\frac{\partial}{\partial \Lambda_{0}(l')}
  \right) \nonumber \\
  && \hspace*{2cm} \left. \partial_{s}     {\cal F}_{\rm int}[\hat{q}^{1+s}(l-1),\hat{Q}^{1+s}(l),\hat{q}^{1+s}(l)]  \right |_{s=0} 
  \nonumber\\
&& =r(l-1)R(l)\kappa_{\rm inter}(l)+R(l+1)r(l+1)\kappa_{\rm inter}(l+1)
\eeqn
where we introduced
\beqn
\kappa_{\rm inter}(l) \equiv 
\frac{\int Dz_{\rm com}
  g'_{\rm teacher}(\sqrt{\Lambda_{\rm com}(l)}z_{\rm com})
\int D z_{0} (-f'(m_{1},\sqrt{\Lambda_{\rm com}(l)}z_{\rm com}+\sqrt{\Lambda_{0}(l)}z_{0}))
}{\int Dz_{\rm com} g_{\rm teacher}(\sqrt{\Lambda_{\rm com}(l)}z_{\rm com})
} \qquad
\label{eq-kappa-inter}
\eeqn
with
\beq
g_{\rm teacher}(h) \equiv \int Dz_{\rm teacher} e^{-\beta V(h-\sqrt{\Lambda_{\rm teacher}}z_{\rm teacher})}
\eeq
For the hardcore potential \eq{eq-hardcore},
$g_{\rm teacher}(h)=\Theta(h/\sqrt{2\Lambda_{\rm teacher}})$ with $\Theta(h)$ defined
in \eq{eq-function-Theta}
and $g'_{\rm teacher}(h)=e^{-h^{2}/2\Lambda_{\rm teacher}}/\sqrt{2\pi \Lambda_{\rm teacher}}$.

Using the above results we find, for $l=1,2,\ldots,L-1$,
\beq
\epsilon_{r}(l)=-r(l-1)R(l)\kappa_{\rm inter}(l)-R(l+1)r(l+1)\kappa_{\rm inter}(l+1)
\eeq

\subsubsection{Variation of $R$}

Finally  we consider variation of the free-energy 
\eq{eq-F-teacher-student} (see also \eq{eq-S-total}) with respect to $R(l)$,
 for $l=1,2,\ldots,L$,
\beqn
0&=&\frac{\partial}{\partial R(l)}
\left.  \partial_{s}s_{1+s}[\{{\hat Q}(l),{\hat q}(l)\}]\right |_{s=0} \nonumber \\
&=& \frac{\partial}{\partial R(l)}
\frac{1}{\alpha} \left.  \partial_{s}S_{\rm ent,bond}[\hat{Q}^{1+s}(l)]\right |_{s=0}
-\frac{\partial}{\partial R(l)}
\left.  {\cal F}_{\rm int}[\{{\hat Q}^{1+s}(l),{\hat q}^{1+s}(l)\}]\right |_{s=0}
\label{eq-variation-R}
\eeqn

Variation of the entropic part (bond) of the free-energy 
\eq{eq-S-ent-bond-teacher-student}
yields,
\beq
\frac{\partial}{\partial R} 
\left. \partial_{s} S_{\rm ent,bond}[\hat{Q}^{1+s}] \right |_{s=0}
=-\frac{R}{G_{0}}
  \eeq
On the other hand, variation of the interaction part of the free-energy 
  \eq{eq-F-int-k-RSB-teacher-student} yields,
  \beqn
&&  - \frac{\partial}{\partial R(l)}
\left. \partial_{s}     {\cal F}_{\rm int}[\hat{q}^{1+s}(l-1),\hat{Q}^{1+s}(l),\hat{q}^{1+s}(l)]  \right |_{s=0} \nonumber \\
&& =
-\left ( \frac{\partial \Lambda_{\rm com}(l)}{\partial R(l)}\frac{\partial}{\partial \Lambda_{\rm com}(l)}
+ \frac{\partial \Lambda_{\rm teacher}(l)}{\partial R(l)}\frac{\partial}{\partial \Lambda_{\rm teacher}(l)}
+ \frac{\partial \Lambda_{0}(l)}{\partial R(l)}\frac{\partial}{\partial \Lambda_{0}(l)}
\right ) \nonumber \\
&& \hspace*{2cm}\left. \partial_{s}     {\cal F}_{\rm int}[\hat{q}^{1+s}(l-1),\hat{Q}^{1+s}(l),\hat{q}^{1+s}(l)]  \right |_{s=0} \nonumber \\
&& =r(l-1)r(l)\kappa_{\rm inter}(l)
\eeqn
with $\kappa_{\rm inter}(l)$ defined in \eq{eq-kappa-inter}.
Using these results in \eq{eq-variation-R} we find,
\beq
R(l)=\alpha G_{0}r(l-1)r(l)\kappa_{\rm inter}(l)
\eeq

\end{appendix}
\end{document}